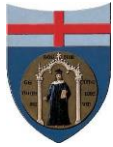

UNIVERSITY OF GENOVA

PHD PROGRAM IN BIOENGINEERING AND ROBOTICS

# Fabrication and characterization of plasmonic nanopores for Raman detection of biomolecules

**Marzia Iarossi**

Thesis submitted for the degree of *Doctor of Philosophy* (33° cycle)

**Thesis Jury:**  **Supervisor**

Prof. Antonio De Luca    Dr. Francesco De Angelis

Prof. Ricardo Martín Abraham-Ekeroth

**Head of the PhD program**

Prof. Giorgio Cannata

Dibris

Department of Informatics, Bioengineering, Robotics and Systems Engineering

# Table of Contents

# Introduction

In the last two decades the development of biosensing platforms has received great attention and resulted in substantial advances in diagnostics for biomedical applications. Nowadays, due to the great effort that has been invested to improve the sensitivity of biosensing devices, based on both electrical and optical methods, it is possible to probe molecules in liquid at the single-molecule level, study the interactions between molecules that play a fundamental role in biological processes and resolve their structural composition[1,2]. Amongst others classes of biosensors which generate an electrical readout, solid-state nanopores have been widely investigated due to the fact that they can be miniaturized to the nanoscale with dimensions similar to those of a molecule and be used to probe molecules directly in liquid when they enter in the sensing volume delimited by the nanopore channel[3]. The molecules are detected in real-time by monitoring the ionic current trace over time following the principle of resistive pulse sensing, that was conceived for the first time in the 1950s by Coulter to count particles suspended in an electrolyte solution. The fast progress of nanotechnology has enabled the realization of solid-state nanopores with dimensions of few nanometers in a variety of different materials in order to confer them new functionalities, which are related to their specific surface chemistry[4]. This aspect is important because it strongly affects the interactions between the molecules and the nanopore walls and as a result also the overall process of the passage of the molecule through the nanopore, changing the average residence time. Indeed, solid-state nanopore-based devices have been widely investigated not only for biomolecule identification but also for the potential to use them for sequencing of nucleic acids, such as DNA and in prospective for protein sequencing[5]. In fact, not only can several examples be found in literature of single solid-state nanopore devices which enable DNA sequencing but also commercial devices for sequencing technology, such as from Oxford Nanopore Technology[6]. For single molecule detection and sequencing it is important to improve the spatial and temporal resolution of single solid-state nanopores. In fact, the size and the thickness of the nanopore determines the sensing volume which should be minimized to improve the sensitivity: the diameter of the pore has to be comparable to the size of the molecule to allow only one molecule per time to reside in the nanopore and the length of the pore should be as small as possible to allow only a small portion of the molecule to reside in the sensing volume at a given time.  These requirements on the geometry of the nanopores are necessary to improve the spatial resolution of the system. Nowadays, nanopores with a sub-nanometer dimension and a length of 1-2 nm can be produced due to the technological advances of the equipment used for fabrication and to the use of new materials, such as the so called 2D materials. Nevertheless, improvements on the control of the geometry of the nanopore and its surface chemistry are still required to develop devices which enable us to control the motion of molecules during the translocation and to perform selective sensing with the aim to extend their use not only for proof-of-concept but also for real-life applications. Generally speaking, the design and implementation of new architectures made of new materials and/or with different geometries is still in demand in order to overcome some of the main limitations of the current solid-state nanopores devices, which are the limited spatial resolution, the impossibility to perform high-throughput analysis and the lack of specificity.

Indeed, solid-state nanopore devices need to be integrated with additional sensing modalities because often the information that it is possible to extract from the ionic current trace is not sufficient for some applications in which it is required to differentiate molecules in complex biological media. With this aim, solid-state nanopores have been coupled to various optical sensing techniques to combine the confinement of the molecules in the sensing volume delimited by the nanopore and the ionic current readout to the additional information provided by an optical readout[7]. Among others approaches, metallic nanostructures have been integrated with conventional nanopores to take advantage of the strong localized electromagnetic field that it is possible to excite upon illumination at the resonant wavelength, also referred to as plasmon resonance. In general, various plasmonic nanostructures, such as gold nano bowties, metallic nanoslits and, more in general plasmonic nanocavities, have been fabricated near a solid-state nanopore for electro-optical detection of molecules, even at the single-molecule level, with various techniques based on scattering and transmitted light, fluorescence enhancement and Surface-Enhanced Raman Spectroscopy (SERS)[7,8]. In fact, the coupling of the electrical and the optical readout has provided signals that are fingerprints of the target molecule, achieving in some cases both high sensitivity and specificity. Nevertheless, plasmonic nanopores also still face challenges related to their limited spatial and temporal resolution. Concerning the temporal resolution, a strategy to slow down the passage of the molecule in the plasmonic sensing volume is necessary because a residence time of at least few milliseconds is often necessary to allow optical detection. Therefore, various mechanisms for trapping or significantly slow down the molecules during their passage through plasmonic nanopores have been successfully proposed. Whereas concerning the spatial resolution, in analogy to common solid-state nanopores, the proper design of the plasmonic nanostructure is fundamental to enable the enhancement of the electromagnetic field only in a narrow sensing volume. Ideally, this means that the enhanced electromagnetic field has to be extended only a few nanometers both in the imaginary plane containing the plasmonic nanopore and in the z-direction perpendicular to this plane. For this purpose, the plasmonic nanopore has to be properly engineered to restrict the plasmonic sensing volume as much as possible and in turn, increase the sensitivity.

*Goals: In this thesis, new classes of materials and designs of plasmonic nanopores are explored with the aim to develop biosensing platforms with high sensitivity and confined sensing volumes that can be eventually integrated with conventional solid-state nanopores for electro-optical detection of molecules. Therefore, the long term scope of this work is to develop multifunctional nanopores capable of detecting biomolecules by means of both electrical and optical methods.*

In particular, two types of plasmonic nanostructures have been studied. The first one consists of concentric nanoholes in a multilayer film made of alternating layers of metallic/dielectric units and forming a so-called hyperbolic metamaterial (HMM). HMMs are a fascinating class of artificial materials exhibiting anisotropic optical properties and supporting highly confined propagating electromagnetic modes. Beside the choice of the plasmonic material also the design of the plasmonic nanostructure is important to excite plasmonic modes localized in very narrow volumes. The second type of plasmonic nanostructures that have been investigated are V-shaped gold nanopores with ultrathin tips to confine the electromagnetic field at the edges of the nanopores.

In Chapter 1 an overview of solid-state nanopores and their capabilities as functional sensing elements are discussed. In particular, the first part of Chapter 1 is focused on the transport properties of nanopores and on how molecular sensing is typically performed from an electrical readout, namely monitoring the ionic current over time and checking the events related to the passage of the molecules through the nanopore. The second part of Chapter 1 is dedicated to the description of plasmonic nanopores and how plasmonic nanostructures have been integrated with solid-state nanopores to couple the electrical readout with an optical readout in order to provide more information on the target molecules and overcome some limitations, such as the lack of specificity, of common solid-state based nanopore devices. Chapter 2 describes briefly why HMMs could be interesting as a class of materials to be engineered for the development of plasmonic nanoholes/nanopores and the fabrication procedure implemented for the realization of periodic nanoholes in HMM films. In Chapter 3 a discussion on the optical properties of HMM nanohole/nanopores, their potential and limitations for biosensing is provided. Chapter 4 describes briefly the common approaches used for the fabrication of solid-state nanopores and the strategy used for the fabrication of V-shaped plasmonic nanopores in a free-standing plasmonic nanoassembly. Chapter 5 discusses the optical properties of V-shaped plasmonic nanopores and how the V-shaped nanopores in a free-standing plasmonic nanoassembly can be used to probe biomolecules with SERS.

# 1 Solid-state nanopores for biological applications: an overview

Nanopore-based sensors have been widely explored in the recent past since they have shown great potential as a tool for biological applications, such as detection and sequencing of the building block of life, namely proteins and nucleic acids. Typically, the predominant method for the molecular readout is based on monitoring the ionic current trace across the nanopore over time. Solid-state nanopores are artificial nanopores that have been developed over the last decades taking in mind their biological counterparts, such as protein channels, which are involved in intra and extra cellular transport in a variety of fundamental biological processes. Recent advances in nanotechnology and in surface chemistry have enabled the development of robust devices improving the sensitivity, the selectivity and the throughput of solid-state nanopore based on an electrical readout. More recently, others sensing strategies based on optical methods have received great interest. In fact, plasmonic nanostructures have been integrated with solid-state nanopores to improve the sensing capabilities of these devices by boosting the optical signals with plasmonic effects, as in the cases of fluorescence enhancement and Surface-Enhanced Raman Spectroscopy. Nowadays, various examples of devices based on plasmonic nanopores which allow for simultaneous electrical and optical readouts have been reported in literature. Indeed, the choice to carry out both optical and electrical measurements has shown to guarantee a certain control on the motion of the molecules during the sensing experiments.

In this Chapter, the main features of solid-state nanopores as functional sensing elements for both electrical and optical readout are addressed. Section 1.1 is dedicated to a description of the transport properties of solid-state nanopores discussing the role of the ionic distribution at the interface between the charged walls of a nanopore and a liquid. The main electrokinetic phenomena generated in the presence of an external electric field are described. Furthermore, it is discussed how to characterize nanopores with electrical measurements and detect molecules by monitoring the ionic current over time during their passage through the nanopore. Section 1.2 describes the optical properties of nanopores starting from a brief summary of the plasmonic properties of both a single plasmonic nanohole and periodic array of plasmonic nanoholes. Then, the principal techniques used for optical and electro-optical sensing are detailed focusing on some relevant results achieved until now for what concerns detection at single-molecule level and sequencing. Finally, even if the majority of the examples reported regards single plasmonic nanopores some examples of plasmonic arrays of nanopores are reported in Section 1.3 since these platforms have potential for real time analysis of biological samples for diagnostic applications.

## 1.1 Nanopores as functional sensing nanodevices for electrical readout

It is well-known that when objects are scaled down to the nanometer scale the surface-to-volume ratio increases and surface effects play an important role. Among other nanosystems that have been investigated in the last two decades, nanopores and nanochannels, namely apertures with characteristic dimensions on the nanometer scale and connecting two fluidic compartments, have received great interest. In fact, unique properties and peculiar phenomena of transport have been observed from artificial nanochannels/nanopores which are strongly dependent on the characteristic dimensions of these nanosystems[9]. In general, we can distinguish between nanochannels and nanopores from their length: the first have a length that is significantly longer than the size of their opening whilst the latter have a length within an order of magnitude of the size of their opening. Among others transport properties, the high permeability and selectivity for specific ion types have been widely investigated since they have some similarities with functionalities that are typical of biological nanopores[10]. The permselectivity of nanochannels offers the possibility to create devices for chemical filtration, desalination and energy conversion. Importantly, nanochannels are responsive to various external stimuli, such as pH, concentration gradients, pressure gradients, electric fields and temperature. Because of this, many strategies have been explored in order to manipulate and control ionic transport properties of responsive nanochannels. Furthermore, in the last two decades nanochannels have been widely designed and tested for their great potential in the field of analysis of biomolecules[11]. The mechanisms of interaction between a biomolecule and a nanopore are complex and depend on the intrinsic features of both the nanochannel and the molecule, such as their size and shape, the superficial charge distribution and functionalization, and on some parameters that can be tailored, such as the medium properties (type and concentration of electrolytes, pH and so on) and the application of external forces. In this regard, external electric fields have been employed to regulate molecular transport through the control of electrokinetic phenomena, such as electrophoresis and electro-osmosis. The capture and transport of biomolecules through a nanopore can be used for fundamental studies of the physiochemical properties of such molecules and for many biological applications, such as biomolecule separation, identification and sequencing. In the next section an overview of the electrostatic interactions at the interface between the nanopore walls and a liquid, as well as the definition of some important physical quantities, such as the electrical double layer and the $\zeta$ potential, are reported following the review of Schoch et al.[12]

## 1.1.1 Electrostatics at the interface between a solid surface and a liquid

A solid surface in contact with a liquid exhibits a surface charge as a consequence of the dissociation of specific functional groups through acid-base reactions with the ions of the liquid or from the adsorption of ions from the liquid to the surface. The sign of the surface charge depends on the numbers and types of ions in the liquid. Indeed, even materials that don't have ionizable surface groups, such as Teflon, when in contact with water exhibit a surface charge due to the adsorption of hydroxyl ions. This surface charge determines the distribution of the ions in the liquid in proximity to the solid-liquid interface. In order to guarantee the electroneutrality, ions in the liquid with opposite charge, termed counterions, accumulate close to the surface charge and screen it. Conversely, ions with equal charge, termed co-ions are repulsed by the surface charge and diffuse into the liquid interacting electrostatically with the counterions.

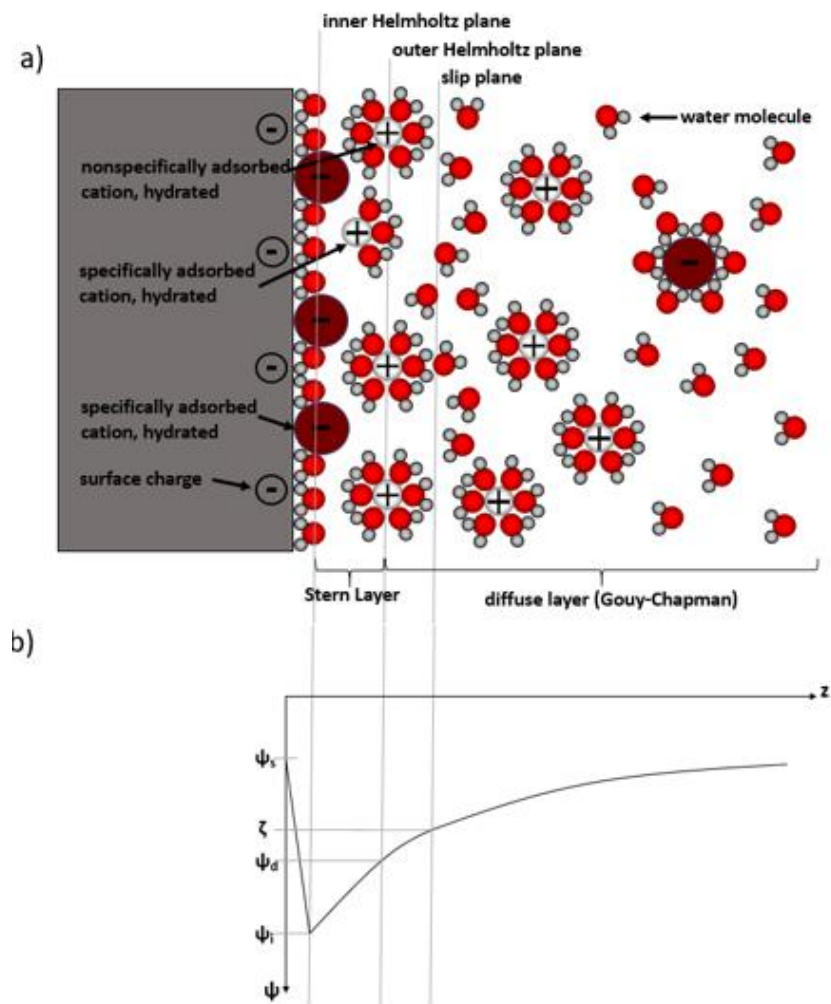

**Figure 1.1.** a) Schematic illustration of the Gouy-Chapman-Stern model for a negatively charged solid surface in contact with a liquid. In the inner Helmholtz plane only non-solvated counterions and co-ions are present whilst in the outer Helmholtz plane the ions in solutions are bound to the surface but solvated. Further from the slip plane ions are mobile and form the diffuse layer. b) Potential distribution as a function of the distance from the surface. Adapted with permission from reference[12].

The screening region composed of the charged surface and the surplus of counterions and co-ions around the solid-liquid interface is referred to as the *electrical double layer* (EDL) and can be properly described in the Gouy-Chapman-Stern model by introducing three virtual layers, as schematically illustrated in Figure 1.1a. Let's consider the case of a solid surface with a negative surface potential $\psi_s$ [12]. The first layer consists of the co-ions and counterions of the liquid adsorbed on the surface and is delimited by the inner Helmholtz plane with potential $\psi_i$. The second layer, also known as *Stern layer*, includes the counterions that are attracted and bound to the solid-liquid interface by electrostatic forces and is delimited by the outer Helmholtz plane with potential $\psi_d$. The third layer is a *diffuse layer* in which co-ions and counterions are still affected by the charged surface but are mobile. The imaginary plane that divides those bound at the interface and the ones that are mobile in the solution is named the slip plane and the value of the electric potential at this plane is known as $\zeta$ *potential*. Figure 1.1b reports the electric potential as a function of the distance from the surface; it has a maximum at the surface and then drops as the distance from the surface increases until its value is equal to 0 at the boundary of the diffuse layer within the EDL.

### 1.1.2 $\zeta$ potential

In section 1.1.1 the $\zeta$ potential has been defined as the value of the electric potential at the shear plane, an imaginary plane that divides the region where the ions are bound on the charged surface from the region where they are mobile in the medium. In the field of colloid and surface chemistry the $\zeta$ potential is an important parameter that can be measured experimentally for solid macroscopic surfaces as well as for colloids, nanoparticles and biomolecules, whilst the surface potential $\psi_s$ is difficult to measure[13]. The properties of the medium, such as the pH, the electrolyte composition and concentration, strongly affect the EDL formation and as a result the $\zeta$ potential. Usually, $\zeta$ potential decreases as the pH increases and a charge reversal is observed after the isoelectric point. The isoelectric point is defined as the value of the pH at which the $\zeta$ potential is equal to 0.

### 1.1.3 Exclusion–enrichment effect

As discussed in section 1.1.1, at the interface between a charged wall and a liquid the EDL is established to guarantee the overall electroneutrality. In the case of nanochannels, two EDLs are formed at the interface between the liquid and each charged wall. Under proper geometrical conditions and at high ionic strength, the EDLs can fully extend to cover the region between the two walls, as happens in microchannels and thus, the electric potential in the middle of the channel is equal to 0. Otherwise, when the size of the channel is comparable with the Debye length or at low ionic strength conditions, the diffuse layers of the EDLs overlap and due to their interaction, the electric potential in the middle of the channel differs from 0. In these conditions, the channel allows the transport of ions selectively. In fact, in microchannels and nanochannels in which the EDLs are not overlapped, the counterions and co-ions concentrations have an exponential profile in the EDL and they reach the bulk concentration in the proximity of the middle of the channel where the surface potential is equal to 0. On the other hand, when the nanochannel size is of the order of the Debye length the counterions and co-ions concentrations are different on average from the corresponding bulk concentrations due to the presence of the surface potential, that differs from 0 even in the middle of the channel (see Figure 1.2a-b). In this case, an enrichment of counterions inside the channel is observed due to the overlap of the EDLs and an exclusion of co-ions occurs as well. This phenomenon is known as the exclusion-enrichment effect and has been widely investigated in nanoporous membranes[14]. Because of it, charged nanochannels have shown great potential for chemical separation and filtration. As examples, Sint et al. designed a fluorine−nitrogen-terminated nanopore in a graphene monolayer in which cations such as $Li^+$, $Na^+$ and $K^+$ can pass and anions are blocked, while with a hydrogen-terminated nanopore anions, such as $F^-$, $Cl^-$ and $Br^-$ can pass and cations are blocked[15].

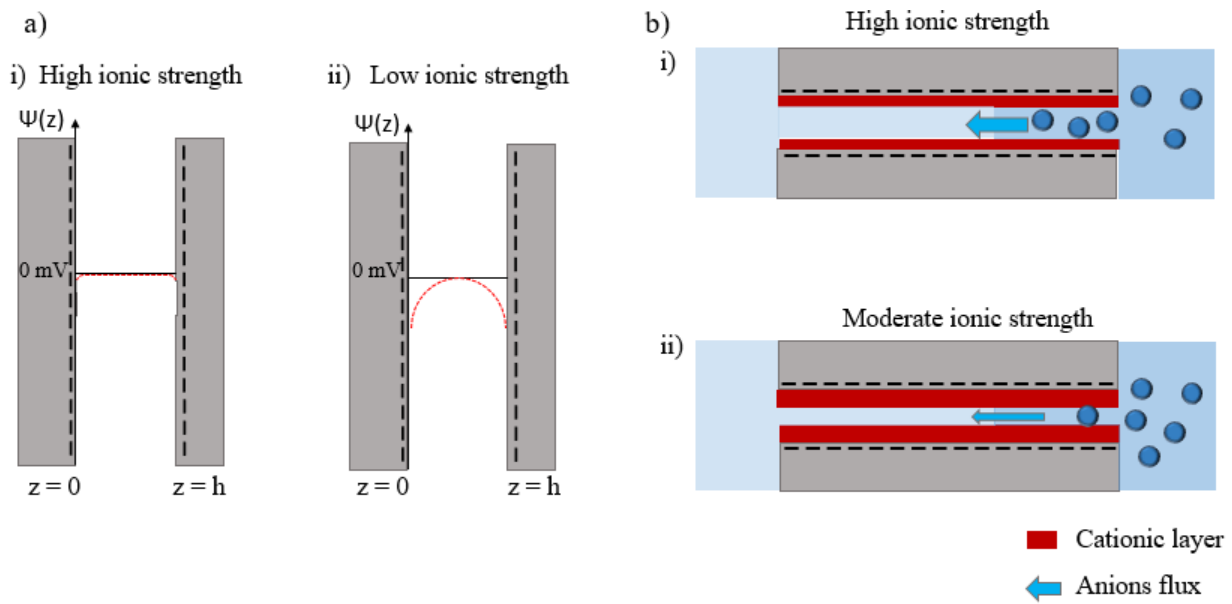

**Figure 1.2.** a) Electric profile in the nanochannel at i) high ionic strength (100 mM) and at ii) low ionic strength (1 mM) respectively. In the first case the potential drops fast to zero whilst in the second case the potential decays slower and does not reach 0 in the middle of the channel. b) Sketch of the extension of the EDL at i) high and ii) low ionic strength respectively. Adapted with permission from reference[14].

While Wang et al. demonstrated that a nanoporous polymer membrane allows the passage of $K^+$ with a high transport rate up to 14 mol h$^{-1}$ m$^{-2}$ maintaining at the same time a high selectivity over alkali metal ions[16]. Moreover, biomimetic nanopores, that mimic the permeability and selectivity of protein channels, have been studied to discriminate between the cations $K^+$ and $Na^+$, as this can be achieved with oxygen doped graphene nanopores in which the distance between oxygen atoms is precisely controlled[17].

### 1.1.4 Electrokinetic phenomena in nanochannels

So far in this chapter we have discussed the interfacial forces at the interface between a solid surface and a liquid and explained that when the surface and the medium composition are fixed a corresponding surface charge, EDL and potential distribution are created. Here, we are going to focus the attention on the principal transport mechanisms, on the interaction between ions within the EDL and external forces that gives rise to many effects, known as electrokinetic phenomena[18]. Generally speaking, the nature of such external forces can be electrical, pressure gradients, gravity and concentration gradients and they can drive the motion of charged particles in a fluid or the ions in solution at the interface between a charged surface and a liquid. Here, we describe the main electrokinetic phenomena occurring in the presence of an applied electric field, concentrating on electrophoresis and electro-osmosis.

Electro-osmosis occurs at the interface between a charged solid wall and a liquid in the presence of a tangential electric field. The ions in the diffuse layer of the EDL are driven into motion by the tangential electric field and because of viscous interactions also the bulk liquid moves, generating a flow. By definition the velocity of the ions at the slip plane is equal to 0, then it increases with the distance from the wall to a maximum value, known as the electro-osmotic velocity $v_{eo}$, beyond this its value is constant and equal to $v_{eo}$. At the outer boundary of the EDL this flow has the constant velocity $v_{eo}$, that was calculated for the first time by Smoluchowski, and can be written as:

$$v_{eo} = -\frac{\varepsilon_0 \varepsilon_r \zeta}{\eta} E_0,$$

(1. 1)

where $E_0$ is the strength of the electric field applied along the surface, $\varepsilon_0$ is the dielectric constant of the free space, $\varepsilon_r$ is the dielectric constant of the medium and $\eta$ the dynamic viscosity of the fluid. The electro-osmotic velocity is usually described by the product of the electric field and the electro-osmotic mobility $\mu_{eo}$, the latter defined as $\mu_{eo} = \frac{\varepsilon_0 \varepsilon_r \zeta}{\eta}$.

On the other hand, the motion of a charged particle in an electrolyte solution in the presence of an applied electric field is known as electrophoresis. Due to the presence of the electric field, the counterions in the diffuse layer move in the opposite direction to that which the particle would move if it was not screened. The fluid motion around the particles due to electro-osmosis also causes the motion of the particle in the opposite direction through a viscous stress. The velocity of the particle can be determined by one of two different limits depending on the thickness of the EDL, that are defined by the value of the product between the Debye-Hückel parameter $\kappa$ and the particle radius $a$. In particular, an estimation of the extension of the EDL is given by the Debye length $\lambda_D = \kappa^{-1}$, which corresponds to the distance from the charged surface where the electrostatic interactions between the ions in solution and the solid surface are reduced to a factor $e^{-1}$ of their initial value. For a thin EDL, namely when the EDL thickness is smaller than the particle radius ($\kappa a \gg 1$), electrophoresis is the converse of electro-osmosis. This means that the electrophoretic velocity of the particle $v_{ep}$ is given by the Helmholtz-Smoluchowski equation and is proportional to the electric field $E_0$ via the electrophoretic mobility $\mu_{ep} = \frac{\varepsilon_0 \varepsilon_r \zeta}{\eta}$:

$$v_{ep} = \mu_{ep} E_0 = \frac{\varepsilon_0 \varepsilon_r \zeta}{\eta} E_0. \tag{1.2}$$

For a thick EDL ($\kappa a \ll 1$), Hückel found the following expression for the electrophoretic mobility:

$$\mu_{ep} = \frac{2\varepsilon_0 \varepsilon_r \zeta}{3\eta}. \tag{1.3}$$

In the previous paragraph, we have introduced the phenomenon of electro-osmosis for a charged flat surface in contact with a liquid and in the presence of an external electric field, as illustrated. Electrokinetic phenomena play an important role in micro/nanochannels with charged walls when an electric field is applied along the channel[19]. The transport properties of the channel are strongly dependent on the surface charge distribution and on the EDL's extension, that in turn are affected by the size of the channel[20]. Under these conditions, the EDL, which screens the walls' surface charge, is in motion and thus, also the fluid is dragged along through viscous forces. Therefore, the Helmholtz-Smoluchowski equation (1.1) predicts the electro-osmotic velocity of the fluid if the Debye length $\lambda_D = \kappa^{-1}$ is smaller than the channel width $h$, that means $\kappa h \gg 1$. From equation (1.1), we deduce that within a nanochannel with charged walls, and in contact with a liquid, the velocity of electro-osmotic flow (EOF) generated in the presence of a tangential electric field does not depend on the dimensions of the nanochannel. However, the thickness of the EDL plays an important role on the transport properties of a nanochannel, and the EOF is also affected from the extension of the EDL inside the channel. In fact, equation (1.1) is valid only at high ionic strength or more in general when the EDLs at the opposite walls of the channel are fully extended and do not overlap, because in this case the total electric potential in the middle of the nanochannel is 0. Otherwise, in the regime $\kappa h \ll 1$ the EDLs formed on the two sides of the nanochannel overlap, as occurs for solutions of low ionic strength, and the EOF is affected by the electric potential $\psi(z)$. As a result, equation (1. 1) is modified as follows:

$$v_{eo} = -\frac{\varepsilon_0 \varepsilon_r \zeta}{\eta} E_0 \left(1 - \frac{\psi(z)}{\zeta}\right). \tag{1.4}$$

Haywood et al. measured the electro-osmotic mobilities from micro/nanochannels of different depths and compared their results with theory[21]. They studied the Smoluchowski regime ($\kappa h \gg 1$) and the one in which the EDL thickness is comparable with the width of the channel ($\kappa h \ll 1$), obtaining that in this latter case the electro-osmotic mobilities significantly decrease due to the confinement within the channel caused by the EDL extension. In general, both electrophoresis and electro-osmosis play an important role during detection experiments of biomolecules with electrical measurements. In the next section it is briefly discussed how nanopores/nanochannels are characterized with electrical measurements and how the ionic current trace can be monitored during sensing experiments to detect biomolecules.

### 1.1.5 Characterization of nanopores with electrical measurements

Typically, electrical measurements of the current flowing through solid-state nanopores are performed with conventional patch-clamp amplifiers used in electrophysiology. The chip with the nanopore is encapsulated between two compartments filled with an electrolyte solution and two electrodes are placed into these two compartments, close to the nanopore's openings (see the inset in Figure 1.3.a). It is possible to show that the conductance of the nanochannel $G$ is obtained by summing the bulk conductance and the excess ion conductance[22]:

$$G = (\mu_{K^+} + \mu_{Cl^-}) c N_A e \frac{wh}{d} + 2 \mu_{K^+} \sigma_s \frac{w}{d}, \tag{1.5}$$

where $w$, $d$ and $h$ are the width, the length and the height of the channel respectively, $\mu_i$ is the mobility of ions $i$, $e$ is the charge of the electron, $N_A$ is the Avogadro constant, $c$ is the concentration of the electrolytes and $\sigma_s$ is the surface charge density. At electrolyte concentrations lower than $c_e = 2 \frac{\sigma_s}{h N_A e}$ the dominant contribution to the conductance is given by the excess ion conductance that depends on $\sigma_s$ and is independent of the height of the channel. At electrolyte concentrations higher than $c_e$ the conductance is affected by the geometry of the channel. The conductance plateau is reached at the electrolyte concentration $c_t$ for which the bulk conductance and the excess ion conductance are comparable. It is important to point out that by decreasing the electrolyte concentration, the absolute value of the $\zeta$ potential increases and this implies that the number of counterions near the surface increases and that the surface charge density decreases.

Thus, the increase of the ζ potential and the decrease of the surface charge are two competitive factors for the conductance of the nanochannel at low electrolyte concentrations. For this reason, it is not the surface charge that appears in equation (1. 5) but the surface charge in the Stern layer and in the diffuse double layer $\sigma_s$ that take into account of these factors. In general, the voltage-current curve for a symmetric nanochannel immersed in a symmetrical electrolyte solution is linear and the absolute values of the current that are measured by applying opposite voltages with the same amplitude are similar[23].

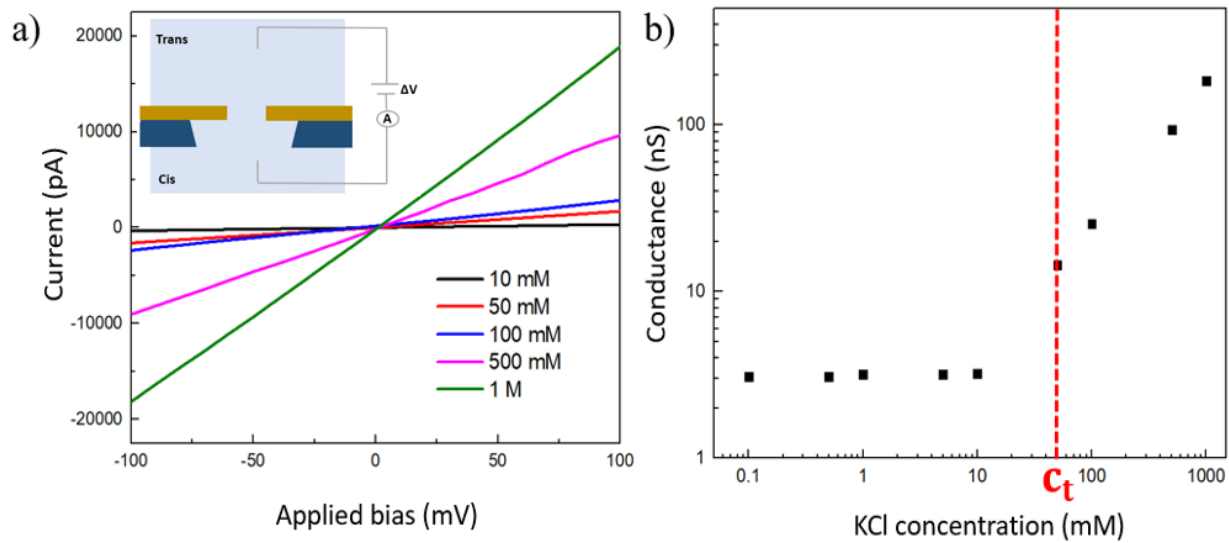

**Figure 1.3.** a) I-V curves obtained from a nanopore with a diameter of 50 nm at various KCl concentrations (experimental data obtained from the patch clamp amplifier of our lab). The inset shows a schematic representation of a solid-state nanopore encapsulated between two compartments to enable the ionic current recording under an applied voltage. b) Corresponding conductance as a function of KCl concentration.

As an example, a plot of the experimental I-V curves acquired from a nanopore of 50 nm on a silicon nitride membrane at various KCl concentrations is reported in Figure 1.3a. At low electrolyte concentrations, the slope of the I-V characteristics does not change if the electrolyte concentration is increased and the conductance is constant. Then, at concentrations higher than $c_t$ the slope of the I-V characteristics, that represents the conductance, depends linearly on the electrolyte concentration (see Figure 1.3b).

### 1.1.6 Detection of molecules from electrical recording

During the passage of a molecule inside a nanopore it is temporarily partially blocked leading to a reduction of the ionic current compared to the one which flows in the unblocked nanopore[24,25]. This has been widely observed and investigated for both biological and solid-state nanopores. Moreover, the study of the modulation of the ionic current provides information on the molecule, its interactions with the charged walls and the translocation process. An example of an ionic current trace showing the modulation of the current after the introduction of the analyte, in this case DNA molecules, is reported in Figure 1.4. Each translocation event corresponds to a blockage of the ionic current, this is characterized by a change from the mean ionic current value in the absence of the molecules $<i_o>$ to the mean-blockade pore current $<i_B>$ and by the duration of the event $t_D$ [26]. Information about the molecule and its translocation through the nanopore is recovered by the analysis of the fractional amplitude $<i_B/i_o>$ and the dwell time $t_D$.

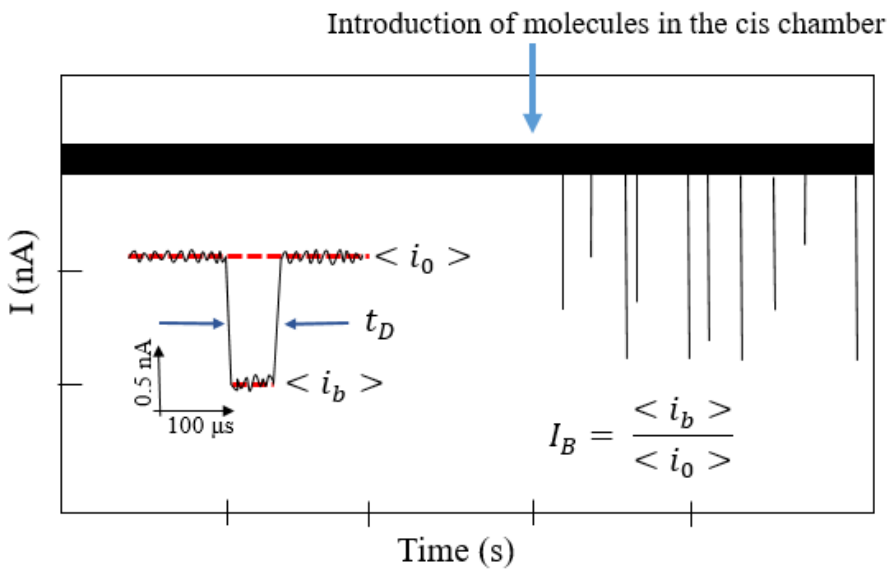

**Figure 1.4.** Typical ionic current trace from a nanopore (4 nm in diameter) before and after the injection of the analyte (in this case linear DNA fragments, length: 400 bp, concentration: 5 nM) in the cis chamber. The detection of single molecules of DNA correspond to the blockade-current events observed in the current trace. The inset shows a magnification of one of the detection events where the main parameters, that are the duration time $t_D$, the mean-blockade pore current $<i_B>$ and the mean current in absence of the molecule $<i_o>$, are illustrated. Adapted with permission from reference[26].

Indeed, from the analysis of the current trace a distribution of duration times is found, which results from the interplay between several phenomena occurring during the DNA translocation, namely electrophoretic force, friction and thermal motion[27]. Moreover, the development of a device based on a single nanopore able to recognize the sequence of an RNA molecule or a single/stranded DNA from the analysis of the ionic current trace is extremely challenging. In fact, for a nanopore on a 30 nm thick $SiN_x$ membrane the translocation of a DNA molecule implies that a portion of about 30 nm of the molecule resides inside the channel at a given time. As a result, about 100 base pairs (bps) are contemporarily located in the channel at a given instant and thus, the resolution of single bases is not possible[5]. Recently, 2D materials, made of a few layers of exfoliated graphene, boron nitride (BN) or molybdenum disulphide (MoS$_2$) have attracted great interest due to the possibility to reduce both the noise level and the sensing length. Graphene-based nanopores with a thickness of 0.3 nm and a diameter comparable with the size of a DNA molecule could be used to discriminate the nucleotides of a DNA sequence. However, there are some issues, such as higher noise than standard solid-state nanopores, sticking of DNA molecule/clogging due to the strong interactions between DNA molecules and the nanopore, and the speed of the DNA molecules are still difficult to overcome during translocation experiments. Others 2D material-based nanopores, such as BN nanopores have provided an alternative to reduce the interactions with DNA molecules and avoid sticking/clogging. Furthermore, Liu et al. used a transferred flake of MoS$_2$ on a pre-etched opening in a SiN$_x$ membrane to study the translocation of DNA through the subnanometer thick MoS$_2$ flake to improve the transverse detection and significantly decrease the interactions between the pore and the DNA molecules[28]. Nonetheless, the identification of nucleotides during translocation experiment is still challenging due to the speed of the translocation events under the applied voltage. The average translocation speed of DNA molecules is in the range of 30$^{-1}$- 0.3 bp per μs, which is too fast to detect the nucleotides. The temporal resolution is a limitation for the sequencing of DNA molecules and thus, strategies to slow down the DNA to a velocity of about 1 ms per μs are required. In this regard the control over the electrophoresis and the electro-osmosis is fundamental to achieve a better control on the motion of the molecules and even to attempt to slow them down during the translocation.

### 1.1.7 Role of electrokinetic forces in electrical detection of biomolecules

The translocation behavior of a molecule strongly depends on both electrophoresis and electro-osmosis phenomena and in turn, on the $\zeta$ potential of the molecule $\zeta_p$ and the one of the charged walls of the pore $\zeta_{wall}$ (see Figure 1.5a). Firnkes et al. studied the translocation of avidin through a SiN membrane with a width of 20 nm and a length of 30 nm varying the pH of the electrolyte solution in order to change the net charge of both the protein and nanopore walls[29]. Then, measuring $\zeta_p$ and $\zeta_{wall}$ as a function of pH and comparing these values to the corresponding translocation rate events of avidin, they found that both the translocation direction and rate depend on the difference in zeta potentials of the protein and the nanopore walls $\zeta_p - \zeta_{wall}$. In order to explain their results, they proposed that the effective velocity of avidin inside the pore is given by the sum of the electrophoretic and electro-osmotic transport:

$$v_{eff} = \frac{\varepsilon E}{\eta}(\zeta_p - \zeta_{wall}). \qquad (1.6)$$

From equation (1. 6) it is clear that the EOF can enhance or reverse electrophoretic transport depending on both the signs and magnitudes of $\zeta_p$ and $\zeta_{wall}$. This means that the surface charge of the pore and the protein charging state should be controlled properly during translocation measurements. Moreover, since proteins have different charge distributions, the control on the electro-osmotic effects could potentially represent a key factor for the discrimination of proteins.

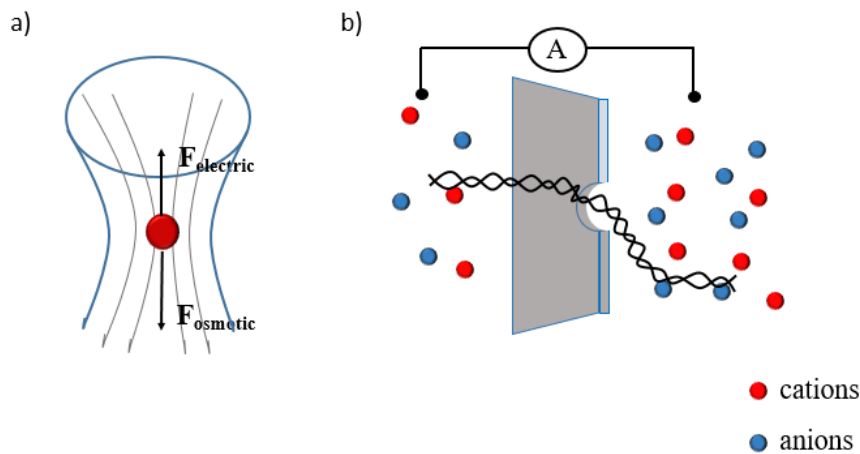

**Figure 1.5.** a) A scheme of the principal electrokinetic forces acting on a molecule into a nanopore. b) Illustration of DNA translocating from the cis chamber to the trans chamber under an applied salinity gradient.

Another strategy to control the translocation events manipulating the EOF is achieved by an ionic concentration gradient between the cis and trans compartments, as achieved by McMullen et al.[30]. To begin with the cis and trans chambers are filled with the same electrolyte and the molecules of interest, then λ-DNA molecules, are introduced into the cis compartment. Then, the electrolyte solution in the trans compartment is varied by adding an electrolyte solution at higher concentration (see Figure 1.5b). At this point, the ionic current is measured and from the analysis of the current traces an enhancement of the translocation rate is obtained compared to the one that is found when the two compartments are filled with an equal and symmetric electrolyte concentration. They showed that the osmotic effect originating from the salinity gradients is able to enhance translocation of λ-DNA molecules even in absence of the applied bias. Moreover, they observed that the transport of λ-DNA molecules is promoted by this osmotic effect even acting against an applied bias of tens of millivolts[31].

Indeed, the conformation of DNA during translocation strongly affects the current traces and in fact, different signals have been observed experimentally from linear and folded states of double stranded DNA. In order to encode information from the current drop in a reliable and reproducible way, it is often required to control the conformation of DNA during translocation and avoid translocations in the folded states. Ermann et al. enhanced single-file DNA translocations through conical pores of 12.5 nm in diameter by changing the electrolyte concentration and pH of the buffer[32]. They observed an enhancement of translocations in a linear state at low salt concentrations to above 90 % (see Figure 1.6a-b) and gave as explanation the occurrence of a stronger EOF at low salt concentrations by using the fact that the longitudinal flow velocity is inversely proportional to the square root of the electrolyte concentration.

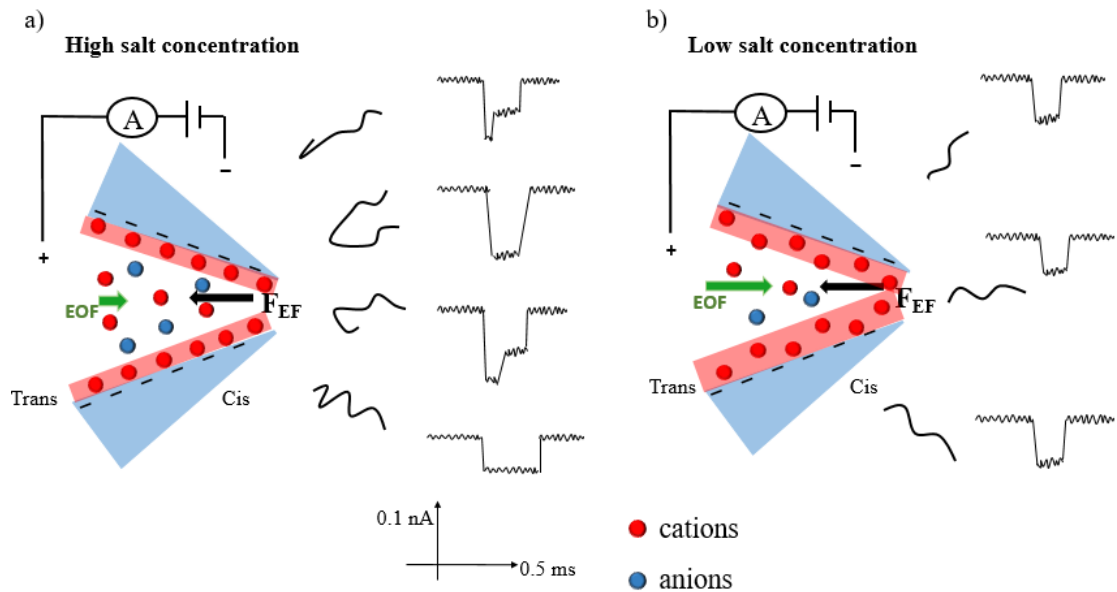

**Figure 1.6.** Schematic illustration of a conical nanopore under an applied bias during the translocation of DNA molecules at a high salt concentration in a) and a low salt concentration in b). Examples of the corresponding current traces are also reported. Adapted with permission from reference[32].

An important constraint during the translocation of biomolecules is the limited range of voltage bias, from tens to hundreds of mV, which can be applied. Therefore, a low bias voltage leads to a small current and a low capture rate whilst a high voltage bias leads to too fast DNA translocations. A robust current signal is necessary for the analysis of the translocation events and thus, instead to decrease the applied voltage bias some tricks to slow down the DNA translocation are required. Hoogerheide et al. developed a DNA molecule trap near an electrolyte-filled solid-state nanopore by applying a voltage bias and a pressure bias across the pore[33]. In this approach, the net force on the DNA molecule is close to 0 because the electric force and the viscous flow induced by the pressure are nearly balanced. In this condition, a "pressure-voltage trap" forms at the entrance of the pore and multiple-attempt events for the DNA translocation are acquired. As a result, it is even possible to study fluctuations of DNA motion during translocation. In general, many studies about DNA translocations through solid-state nanopores with a diameter in the range 10-20 nm have shown that the average translocation speed is remarkably fast, about 10 nm/µs or 30 nm/bp. Due to the instrumental bandwidth limit the temporal resolution is often not sufficient to detect the nucleotides during a translocation event. In fact, a low-pass filter of 10 KHz is usually employed to attenuate noise in electrical measurements with a temporal resolution that is around 50 µs. This means that in electrical sensing with a nanopore-based device it is not possible to resolve single nucleotides. For this reason, many approaches have been proposed to slow down the translocation process by modifying the nanopore shape, charge and surface properties[34]. Fologea et al. studied the role of viscosity, electrolyte temperature and salt concentration on the translocation speed of a DNA molecule through a silicon nitride nanopore with a diameter of 4-8 nm[35]. They found that adding 20% of glycerol in an electrolyte solution of 1.6 M KCl and performing the experiment at 4 °C with an applied voltage bias of 40 mV increases the translocation time of one order of magnitude more than the one recorded in standard conditions (room temperature, no glycerol, applied voltage bias of 120 mV and KCl concentration of 1M). Importantly, the interactions of DNA with the nanopore walls have a strong impact on the translocation speed. As example, the hydrophobic interactions between a single stranded DNA and a 2D nanopore in graphene with a diameter of 3.3 nm are produce translocation times in the range of 50-400 μs[36]. In the case of proteins, the translocation speed of streptavidin has been reduced by covering with a lipid coating a nanopore in a silicon nitride window due to the viscosity of the fluid coating[37].

## 1.2 Plasmonic nanopores for optical sensing

From the analysis of the ionic current traces, and specifically from the characteristic current blockage signatures, it is possible to reconstruct information about the molecules during the translocation. However, this approach is limited by the signal to noise ratio and thus, the detection of very small molecules or the identification of portions of the molecules, as in the case of DNA and proteins sequencing, is not trivial. Indeed, plasmonic nanopores, namely nanopores in metallic films or plasmonic nanostructures integrated with common nanopores are of great interest because they combine the transport properties of nanopores and plasmonic effects originating from metallic nanostructures, which are usually exploited for optical detection. Among others techniques, Surface Enhanced Raman Spectroscopy (SERS) is a label-free technique which allows for the identification of molecules with high specificity since the information about them is extrapolated from the characteristic signature of the collected Raman spectra. In this sense, plasmonic nanopores coupled with SERS measurements have great potential for biological sensing even at the single-molecule level. Other techniques based on monitoring the transmission signal or the fluorescence signal have been widely explored from various metallic nanostructures. Depending on the application, it could be convenient or not to realize a single plasmonic nanopore or a periodic array of plasmonic nanopores: the first ones are suitable for the identification of target molecules at low concentrations or even at single molecule level and in prospective to probe the entire translocation process, while the latter is a better choice for the detection of biomolecules at higher concentrations, filtration and in prospective for parallel sensing. However, the optical properties of single plasmonic nanostructures and ordered arrays can be completely different because the nature of the modes that can be excited in these structures are different. In order to give an overview on the optical properties of plasmonic nanopores and simultaneously on the related optical techniques that are used in sensing experiments, the next section briefly describes why plasmonic nanostructures are usually employed as biosensing elements probing local refractive index changes related to the presence in the active sensing area of target molecules.

### 1.2.1 Plasmonic Biosensing Platforms: how they work

In the last decades the scientific community has dedicated great attention and effort in the development of biosensing platform based on plasmonics. Generally speaking, plasmonic biosensors can be divided in two categories: the ones based on metallic thin films which rely on surface plasmon resonance (SPR) and the ones based on single metallic plasmon resonant nanostructures, also referredo to as localized surface plasmon resonance (LSPR)-based sensors. There are various sensing modalities in which plasmonic biosensors can operate and even examples of biosensing platforms in which the two plasmonic biosensors categories are combined together[38,39].

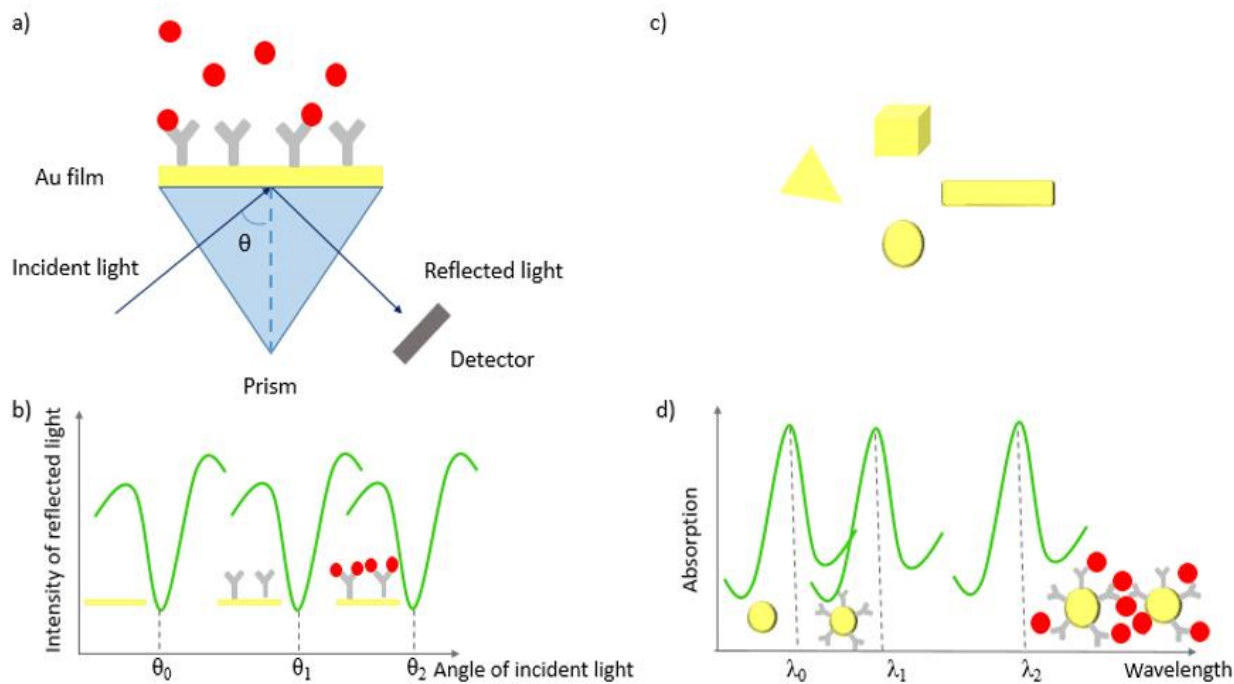

**Figure 1.7**. a) Scheme of a SPR-based sensor: the SPR resonance is sensitive to refractive index changes within the sensing area (hundreds of nanometers). In this case the surface plasmon propagation at the metal-dielectric interface is allowed by a light coupling mechanism obtained by illumination through a prism. b) Plot of the reflection as a function of the angle of incidence of the incoming light. At a certain angle of the incident light the SPR has a minimum in the reflection signal observed. The position of this angle changes each time that some molecule is in the sensitive area above the metallic film: as a result, a different angle at which the SPR is observed is found

after both the immobilization of antibodies on the Au surface and the capture of target molecules. c) A few examples of metallic nanoparticles of different shapes. d) Plot of the absorption as a function of the wavelength. Metallic nanoparticles exhibit a maximum in the absorption spectrum as a result of the excitation of the LSPR. Metallic nanoparticles act as local probes because they are sensitive to local refractive index changes. In fact, a shift of the resonance is observed after the functionalization of the surface with some molecules, such as antibodies. For functionalized nanoparticles in solution it is possible to induce aggregation of the nanoparticles by introducing the target molecules: the corresponding shift in the absorption spectrum is usually of several tens of nanometers and thus, LSPR-based sensors can be used for colorimetric sensing.

### 1.2.2 Propagating Surface Plasmon Resonance Sensors

Thin metallic films support surface plasmons, which are collective oscillations of the surface conduction electrons. Among others metals, gold (Au) films have been widely investigated due to the fact that the SPR is in the visible range and thus, can be excited by using visible light[40]. Furthermore, Au does not oxidize and can be easily functionalized with thiol terminated molecules by adsorbing on the Au surface due to the covalent Au-S bond. In order to excite SPRs it is fundamental to have a light coupling mechanism to match the momentum of the photons of the light beam and the free surface electrons of the film[41]. For instance, in the Kretschmann configuration the light coupling mechanism is typically achieved with a prism to generate total internal reflection of the light beam at the interface between the Au film and surrounding medium. These sensors work in angle-resolved or in wavelength-resolved configurations. In the first one, the beam source is a laser and the angle between the source and the film is adjusted to find the one at which there is coupling between the light and the film (Figure 1.7a). In the second configuration, the light source is white light at a fixed angle and there is only a certain wavelength at which the SPR occurs. In both configurations the SPR peak corresponds to a minimum in the reflection signal at a certain angle or wavelength because the transmitted light is lost at that specific angle/wavelength due to the coupling with the superficial electrons of the metal. Therefore, the SPR dip in reflection is monitored over time because its position depends on the refractive index of the medium that is in contact with the film. Various strategies have been used to form a monolayer of molecules on the metallic surface in order to detect them from the shift caused by the change of the refractive index induced by the molecules[42]. As example, Pernites et al. developed a method based on non-covalent electropolymerized molecular imprinted polymers to form a thin film with high selectivity for the detection of naproxen, paracetamol, and theophylline on the surface of a Au film[43]. Usually receptor molecules, such as antibodies or aptamers, are immobilized on the surface by using various surface attachment immobilization techniques in order to improve the selective capture of the analyte molecules (Figure 1.7b)[44–46]. This step is crucial and several groups have worked to realise assays based on SPR sensing with high specificity, often using functionalized nanoparticles which are key elements in the process of capture of the analyte and, at the same time, can improve the performance of the SPR sensing, boosting their plasmonic response. After the immobilization of the receptors on the surface of the film the baseline of the SPR sensor is determined. Then, the analyte is introduced over time and the binding kinetics of the receptor-analyte interactions are monitored from the position of the SPR peak. Nowadays, SPR-based biosensing is so widespread that it is possible to find on the market commercial SPR modules integrated in ellipsometer systems to perform biosensing experiments from thin metallic films. There are also examples of SPR-based platforms that are designed in order to have multiple independent channels and thus, can be used for parallel sensing. As example, Vala et al. developed a prototype based on a Au-coated diffraction grating with ten independent fluidic channels and tested the capabilities of the device for multiple detection in the presence of oligonucleotides of different concentrations[47]. Nevertheless, one limitation of SPR-biosensors is that they require bulky instrumentation due to the necessity to couple the light with the superficial free electrons of the film and thus, they are not easily incorporated into portable devices.

### 1.2.3 Localized Surface Plasmons Sensors

Not only thin metallic films but also metallic nanostructures with characteristic dimensions smaller than the wavelength of the incoming light exhibit remarkable optical properties due to the collective oscillation of free surface electron upon illumination (Figure 1.7c). In this case the collective oscillations are confined within the nanostructure and for this reason they are referred to as localized surface plasmon resonances (LSPRs). Since this category of plasmonic sensors is based on plasmonic nanoparticles it is easier to integrate them into lab-on-a-chip devices compared to the SPR-based sensors. Furthermore, these plasmonic nanosensors exhibit resonances that can be tuned by changing their geometric features, such as size, shape and aggregation state[48,49]. The position of the LSPR depends on the refractive index of the surrounding medium and thus, they act as probes of the local environment. Compared to SPR-based sensing devices which are sensitive to hundreds of nanometers far from the metallic film, LSPR-based sensors show smaller sensing volumes which extend no more than tens of nanometers from the nanostructure. For this reason, LSPR-based sensors often have a smaller limit of detection than their SPR counterpart and show higher sensitivities for molecular sensing than SPR sensors whilst the latter are more suitable for bulk refractive index sensing. Often the modality for sensing biomolecules in LSPR-based devices involves monitoring the transmission of light over time from functionalized nanoparticles, which are immobilized on a substrate. This can be accomplished with commercial standard spectrophotometers and even visually because often the LSPR is in the visible range for many metallic nanostructures giving rise to a colorimetric readout. Many LSPR- based devices with a positive/negative colorimetric indication in the presence/absence of the analyte of interest have been produced (Figure 1.7d)[50–53]. It is worth mentioning that the colorimetric readout is cost-effective and simple but often is not quantitative if not integrated with the spectrophotometer readout. Plasmonic nanoparticles have been used also in SPR sensing because they can be useful elements that helps the selective capture of antigens and, importantly, they enhance the signal intensity of SPR sensors[54,55]. As example, Matsui et al. placed on top of a Au film a molecularly imprinted polymer gel with embedded gold nanoparticles and used the swelling of the polymer induced by the binding of the analyte to change the distance between the Au nanoparticles and the plasmonic film, which n turns leads to a shift of the SPR dip[56]. In fact, more complex designs involve nanoparticles with many biological recognition elements in order to improve the performance of simple SPR sensors.

### 1.2.4 Optical properties of plasmonic nanoholes and nanopores

Among others biosensing platforms plasmonic nanohole arrays and more recently, plasmonic nanopores, which are sub-100 nm apertures connecting two compartments, have attracted great attention due to their potential not only for biological sensing but also for single-molecule detection and sequencing. For what concerns the optical properties, plasmonic nanopores exhibit a behavior that is the equivalent of plasmonic nanoholes, which are holes in a metallic film with a substrate as support[57,58]. Plasmonic nanoholes in thin metallic films exhibit both SPR and LSPR modes. Indeed, single plasmonic nanoholes show similarities with plasmonic nanoparticles because they also support localized plasmon modes which are strongly affected by the shape of the aperture, as in the case of nanoparticles. Furthermore, for single apertures in thick (hundreds of nanometers) metallic films it is possible to excite plasmonic modes which propagate along the aperture in the direction perpendicular to the film plane. For ordered periodic arrays of nanoholes in a thick metallic film the coupling of the incident light and SPPs are responsible for a phenomenon known as extraordinary optical transmission (EOT). In 1998 for the first time Ebbesen et al. observed EOT from a periodic nanohole arrays in a Ag film of 200 nm[59]. They measured the transmission signal and found the usual peak at 326 nm related to the Ag plasmon peak and many other gradually stronger peaks at longer wavelengths. In fact, they observed an additional minimum related to the quartz-metal interface at $\lambda = a_0\sqrt{\varepsilon}$, where $a_0$ is the periodicity of the array and ε is the dielectric constant of the substrate and the strongest peak at 1370 nm with a transmission efficiency higher than 2, meaning that the transmitted light is two times the incident light. Furthermore, they showed that the positions of the peaks shifted in opposite directions as a function of the angle of the incident light and that splitting of peaks was observed increasing the angle of incidence. This suggested that this new phenomenon had similarities to the coupling of SPPs with a reflection grating. Starting from the same year on, theoretical explanations of EOT were proposed showing that the origin of the selective transmission of periodic nanohole arrays at certain wavelengths with efficiencies even 1000 times higher than the one calculated for single nanoholes is due to the coupling of SPPs excited at the film interface with the incident light[60]. Afterwards, many groups devoted great attention to SPR sensors based on ordered arrays of nanoholes on plasmonic films from a fundamental point of view to understand the role of SPPs in the EOT phenomenon[61] and for their prospective to be used as miniaturized SPR-based sensing elements overcoming the main challenge of common SPR sensors, namely the bulky required optics to excite propagating SPPs. Plasmonic devices based on ordered nanohole arrays usually operate in transmission mode to take advantage of the EOT for many applications, such as refractometric sensing, study of biomolecular interactions and monitoring spatial microfluidic concentration gradients[62–64]. Among others, Couture et al. showed that properly designed nanohole arrays working in the Kretschmann configuration have a better performance than SPR sensors both for refractive index sensing and molecular sensing, as they experimentally prove in the case of the detection of IgG[65]. Indeed, the arrangement of the periodic nanohole arrays affects the performance of the platform as refractive index sensor. Among others configurations, the most commonly explored are square and hexagonal arrays of nanoholes (Figure1.8a-b respectively) because they can be produced with expensive top-down fabrication techniques, such as electron beam lithography, as well as with cheap approaches with high throughput such as nanosphere lithography. Square and hexagonal arrays of nanoholes exhibit similar sensitivities for similar wavelength resonances but the transmission spectra of hexagonal arrays show sharper EOT peaks than the ones corresponding to the square arrays and thus, the former supports a larger figure-of-merit[66]. Recently, Wang et al. reported a new asymmetric nanohole array obtained combining nanosphere lithography and oblique angle deposition, in order to realise a periodic nanohole array with a rod structure inside the hole[67]. They proved that the array exhibits an EOT mode in the near-infrared region and strong polarization-dependent optical properties. Furthermore, Tavakoli et al. studied the EOT from an ordered plasmonic nanohole array with an elliptical shape and the dependence of the quality factor of the main peaks in rotation in order to develop a platform useful for rotational sensing[68].

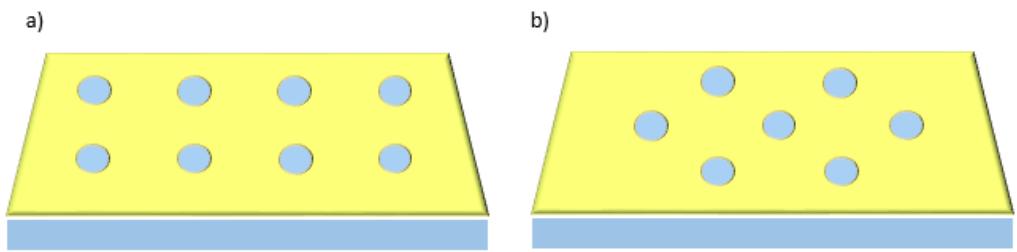

**Figure 1.8.** Sketch of periodic nanohole arrays with two different arrangements of the holes: a) a square lattice and b) a hexagonal lattice.

Apart from the geometrical features of the nanohole array also the choice of the material plays an important role because it strongly affects the spectral features. From periodic arrays of nanoholes on thin films the opposite phenomenon can be observed, namely the suppression of transmission or equivalently the enhancement of absorption at certain wavelengths[69]. Even in this case the origin of the absorption enhancement resides in the excitation of SPPs at the interface of the thin metallic film. For ultrathin films (with a thickness not larger than 20 nm) a single hole exhibits a LSPR mode inside the hole itself and a dominant SPP mode elsewhere in the surrounding film, as confirmed by scanning near field optical microscope (SNOM) measurements[70]. On the other side, for an ordered array of nanoholes the LSPR mode and the SPP mode are coupled together and give rise to localized optical modes, whose features depend on the geometrical parameters of the array, such as the periodicity and the diameter of the holes. Although ordered nanohole arrays have been widely investigated from both a theoretical and an experimental point of view, many explanations and theories have been discussed to predict their optical properties and the debate is still open since an exhaustive understanding of the mechanism responsible for the origin of their optical behaviour is still missing.

### 1.2.5 Plasmonic nanopore as enhanced resonance shift assays

Typically, plasmonic nanopores/nanoholes have been used as sensing platforms by exploiting the red-shift of the plasmonic resonance due to a local increase of the refractive index as a result of the molecules binding on the surface in analogy with SPR-based sensors. However, the main difference between SPR-based sensors and plasmonic nanoholes/nanopores-based devices is that the signal that is usually monitored, the transmission, can be easily acquired with a spectrophotometer because there is no need to implement a set-up which satisfies the matching of the momentum between the incoming photon of light and the conduction electrons of the metallic film, as in the case of the Kretschmann configuration. As a result, platforms based on plasmonic nanohole arrays have potential to be used as label-free sensors for real-time measurements. Sensors based on nanohole arrays, as in general SPR-based sensors, are used to investigate molecular interactions for fundamental studies, but are not suitable for real biological applications due to the non-specific binding of molecules present in biological media, which causes drastic increases in the detection limit. Nevertheless, a few examples of detection of molecules in real biological media can be found[71]. Recently, Jia et al. developed a method to transfer a plasmonic pattern on a tape and obtain a transparent tape with a plasmonic pattern on it[72]. Furthermore, the transferred nanohole array showed the highest sensitivity in the refractive index sensing experiment, thus confirming its potential to be used as a plasmonic flexible and stretchable sensor. Furthermore, the performance of plasmonic nanohole arrays for refractometric sensing can be improved by using additional boosting EOT with index matching. Among others, this has been studied by Vala et al.

by using a water-index-matched polymer layer (Cytop) underneath the nanohole array and measuring the reflected light from the backside[73]. They obtained a high extinction amplitude and high Q-factor from the device. Although there are several examples in literature of refractometric sensing from nanohole arrays platforms for detection of proteins and studies of protein-membrane interactions, it is not easy to find strategies to take advantage of the nanoholes due to the presence of the underneath substrate. For this reason, the attention of the scientific community in the last decade has been mainly focused on the development of sensing platforms based on plasmonic nanopores. Historically, nanopores have been developed for electrical sensing which usually suffers from poor signal to noise ratio and a limited bandwidth amplitude. These constraints and the fact that the electrical sensing is not specific, namely needing advanced statistical analysis to extract information, are the main reasons why optical detection from plasmonic nanopores has attracted such great interest[7].

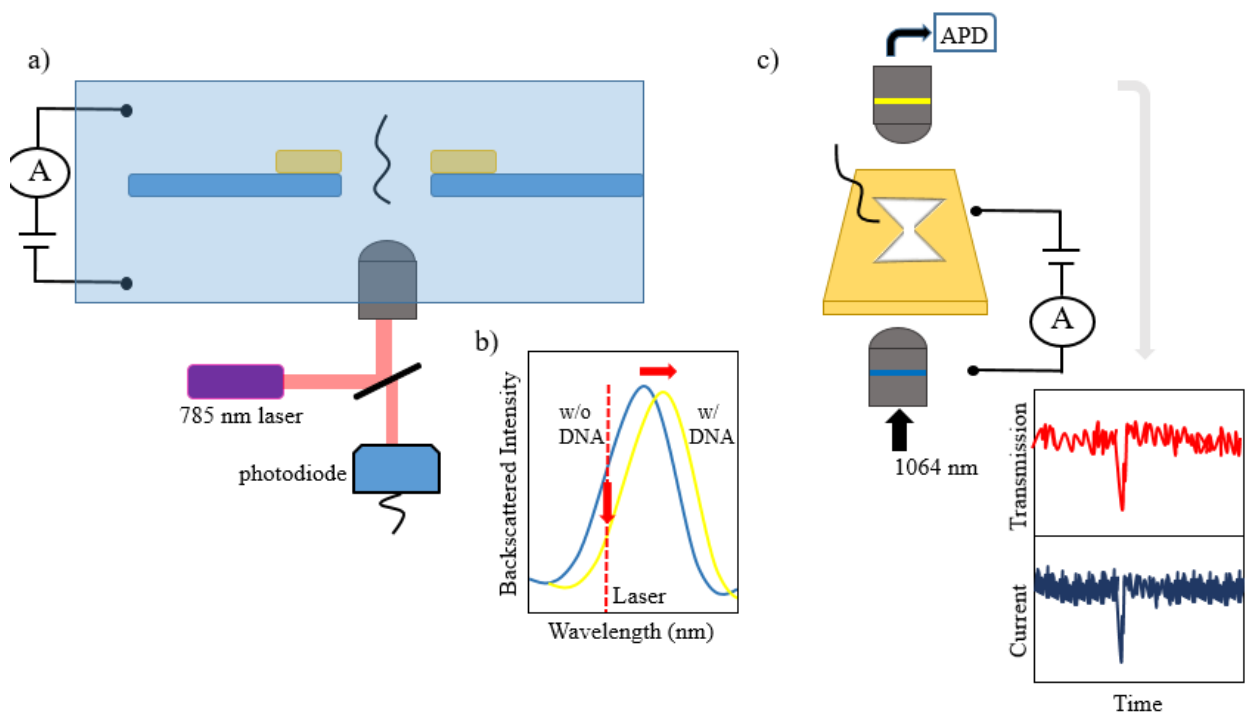

**Figure 1.9**. a) Scheme of a DNA molecule that is driven through a plasmonic nanopore with an external electric field and detected by optical backscattering from the plasmonic antenna. b) The presence of a target molecule in the hotspot area of the plasmonic nanopore causes a red-shift of the plasmonic resonance. As a result, the scattering intensity that is detected at the excitation laser wavelength decreases[74]. c) Schematic of the experimental setup used to monitor the transmission signal from an inverted bowtie plasmonic nanopore. The ionic current trace is monitored over time to check the simultaneous detection of the molecules in the ionic current and the light transmission signal[75].

As discussed in section 1.2.4, the optical transmission of nanoholes/nanopores can be monitored over time for sensing applications since a shift of the resonance is expected in the presence of target molecules, which induce a change of the local refractive index (see Figure 1.9a-b). Therefore, as a result of the resonance shift an increase of the transmission of scattered light through the nanopore is observed in the presence of a molecule within the nanopore. This approach has enabled the development of plasmonic-nanopore-enhanced resonance shift assays which are usually made of large nanopores (at least 100 nm in diameter) in thick metallic films. These devices have shown good capabilities by sensing of polystyrene nanoparticle of about 70 nm with an increase of the scattering intensity 20–50%[76]. Anyway, biomolecules are usually smaller and thus, produce smaller refractive index changes. For this reason nanopore-enhanced resonance shift assays are often not sensitive enough for biomolecule sensing. Nevertheless, some improvements have been shown from the use of plasmonic inverted bowtie nanoantennas, which allowed the detection of DNA molecules (see Figure 1.9c)[75]. Therefore, other optical techniques coupled to the plasmonic effects generated from plasmonic nanopores have been investigated. In sections 1.2.6-7, remarkable examples of plasmonic nanopores and their application for biosensing, even at the single molecule level, are discussed, as well as the optical detection mechanisms which underlie their performances and capabilities.

### 1.2.6 Fluorescence enhancement from zero-mode waveguides and plasmonic nanopores

Among others optical sensing methods, fluorescence has been widely explored due to the fact that nowadays commercial dyes can be easily attached to specific positions of proteins and nucleic acids. The first experiments have been carried out on large zero-mode waveguides (ZMWs) due to some technological challenges in the fabrication of small nanopores in thick materials[77–79]. Typically, the target molecules are tagged with a suitable dye so that they are detected from the fluorescence enhancement arising from their passage through the plasmonic nanopore. A ZMW is a nanometric well, typically made of a nanohole in a thin Al film, designed to confine the electromagnetic field at the bottom of the well. ZMWs have been employed for fluorescence enhancement because fluorophores are excited only in the zeptoliter excitation volume formed inside the well whilst the other ones present in the reservoir are screened by the film[80]. Indeed, Al is also a plasmonic material which provides a high confinement of the electromagnetic field inside the nanowell but also a poor enhancement of the fluorescence. In this regard, other materials such as Au or Ag represent a better choice to produce a higher enhancement and even the shape of the well can be optimized to produce a stronger signal. However, the fluorescence enhancement (EF) observed from smaller nanopores and smaller analytes is usually very small.

Various designs of ZMWs, for instance with a rectangular shape instead of a circular one, have been proposed both theoretically and experimentally to improve the enhancement of the signal from the pore[81,82]. Indeed, the optimization of ZMWs is focused also on the extension of the working range they operate in with the aim to use them in the UV spectral region because it could enable label-free detection of various biomolecules which have intrinsic fluorescence in that range[83]. ZMWs can be integrated with solid-state nanopores to enable optical sensing in flow-through configuration. As example, Assad el al. proposed a ZMW on a thin silicon nitride membrane which was placed close to a 10 nm nanopore drilled by a transmission electron microscope[84]. In this configuration, they showed that the system successfully worked for single-molecule fluorescence detection. Furthermore, in order to improve the performance of ZMWs integrated with single solid-state nanopores it is important to take into account the electrokinetic phenomena occurring during translocation. With this purpose, Larkin et al. showed that applying an electric bias to drive DNA molecules with a sub-nanogram concentration to the zeptoliter excitation volume of the nanocavity leads to an efficient delivery and improves the performance of the ZMW for electro-optical detection of biomolecules[85].

### 1.2.7 SERS platforms based on plasmonic nanopores

Among other techniques based on optical detection boosted by plasmonic effects, Surface-enhanced Raman Spectroscopy (SERS) has been widely used for biosensing and molecular interactions studies. In SERS even if the sensing platform is label-free the detection is highly specific because upon laser illumination each molecule exhibits characteristic Raman peaks which are related to their specific vibrational modes uniquely related to its chemical bonds[86,87]. The plasmonic platform provides an enhancement of the local electromagnetic fields, which in turns gives an increase of several order of magnitude of the scattering signal collected by the detector.

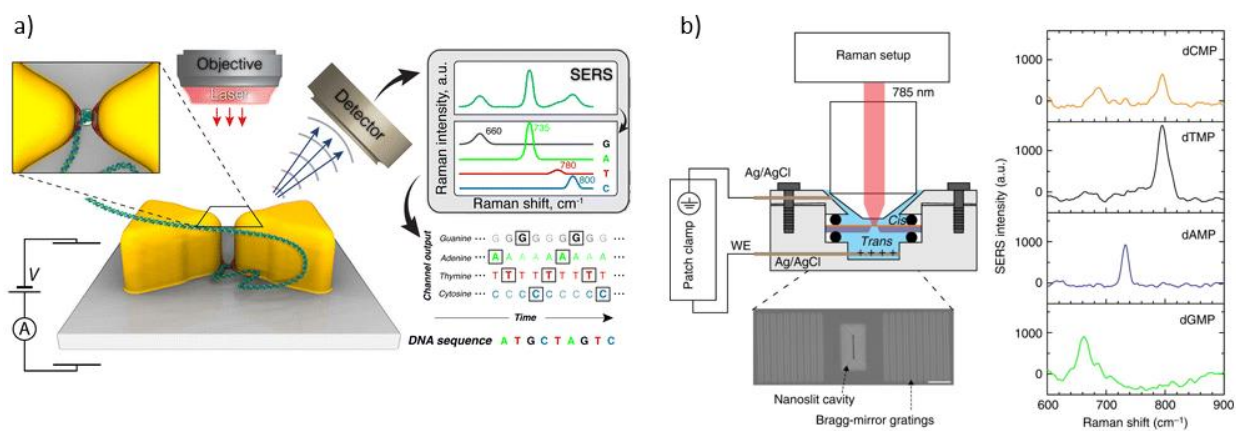

**Figure 1.10.** a) Scheme of a metallic nano bowtie formed by two gold triangular prisms integrated with a solid-state nanopore in the gap of the nano bowtie. An electric field can be applied to drive molecules, such as DNA through the nanopore. From theoretical simulations, switching the laser

on excites the plasmonic hotspots that can be used for SERS measurements and from the analysis of the spectra it is possible to reconstruct information on the nucleotides composing the DNA sequence. Reproduced with permission from reference[88]. b) Scheme of the setup used for a plasmonic nanoslit sealed in a fluidic cell and excited with a 785 nm laser for SERS. Applying an electric field it is possible to deliver DNA molecules and discriminate between individual nucleotides and the ones incorporated in DNA strands. Reproduced with permission from reference[89].

Among other nanostructures, nanopores integrated with plasmonic nanostructures such as metallic bow ties[90–92], nanogaps[93] and nanotips[94] have been used since these are excellent nanoprobes for SERS. This is because within the local area around the gap (< 10 nm) in which the electric fields are highly enhanced, also known as plasmonic hotspots. Indeed, the combination of plasmonic nanopores with SERS has attracted great interest due to the possibility to develop sensing platforms in flow-through configuration[95]. This overcomes some issues of surface-based sensors, such as the long detection times at low concentrations of the analyte which results from the fact that the mass transports affects the binding kinetics. Nevertheless, the main limitation of SERS is the long acquisition times, typically in the range of several milliseconds, when compared with the short time that molecules spend close to the hotspot, which for free diffusion is about 1 µs. For this reason, when plasmonic nanopores are combined with SERS it is fundamental to implement a strategy to trap the analyte in the hotspot for a time that is long enough to acquire SERS spectra even multiple times in order to ensure reliability and attempt to collect more information from different portions of the molecule (see Figure 1.10a). Among other strategies, optical trapping phenomena have been widely explored in order to trap molecules close to a plasmonic hotspot. Recently, Chen et al. succeeded in the combination of SERS measurements performed on a plasmonic nanoslit, namely an elongated nanopore structure, and an external electric field to encourage DNA molecules to pass through the nanoslit for real-time molecular sensing in fluid (see Figure 1.10b)[89]. At low concentrations of DNA oligonucleotides they studied the single molecule regime and were able to discriminate adjacent bases when DNA oligonucleotides were temporarily adsorbed on the nanoslit surface showing that this platform has sub-nanometer spatial resolution. Nevertheless, the main limitation of this approach is the fact that it is not easy to combine these nanoslits for SERS with electrical recording due to the high ionic current background corresponding to the large area of the nanoslit. More in general, the use of an external electrical voltage to drive molecules to the pore entrance is useful especially for polymeric molecules, such as DNA, providing a way to stretch the molecule from a folded configuration to a linearized one. In fact, flow-through sensing offers the advantage that molecules can be directed through the pore very close to the plasmonic sensing area. In particular, the motion of the molecules through the pore can be controlled by exploiting electrokinetic phenomena, such as electrophoresis and electro-osmosis, occurring at the interface between the liquid medium and the nanopore walls. In this way, it is possible to control the transport rate of molecules through the pore and partly their motion too, even if the microscopic control over the molecules' trajectory during the motion through the pore is still missing, and this constitutes one of the current limitations of nanopores-based devices developed for the analysis of molecules constituents, such as for sequencing of DNA or the identification of proteins' aminoacids. Indeed, a flow-through configuration offers many advantages compared to surface-based sensors because there are many effects that can be exploited to concentrate the analyte in the sensing area. If the molecules are free to diffuse through the pore, after a while the system reaches equilibrium: as a result, the rate of binding and unbinding are the same and thus, there is no enhancement in the signal even in this configuration because it is not possible that other molecules bind the active surface. In nanopore-based devices the sensitivity is higher than surface-based sensors because it is possible to accumulate more molecules close to the pore entrance and concentrate the target molecules within the sensing surface. As already mentioned, this is due to the fact that it is possible to apply an external electric field to drive the molecules close to the walls through the nanopore. This is an important aspect not only for devices based on single plasmonic nanopores that aim for single-molecule detection but also for the ones based on arrays of nanopores that are more suitable for filtration after a proper functionalization. In general, since the walls of a pore (any type of pore, not only the plasmonic ones) in contact with a liquid exhibit a surface charge it is important to understand which are the transport properties they show and how it is possible to control them. For this reason, platforms based on plasmonic nanopores can be extended for electro-optical detection with the idea to combine optical techniques boosted by plasmonics and electrical forces to govern the transport of the molecules in a controlled way.

Besides the issue of the temporal resolution of SERS, another problem is the design of plasmonic nanopores that are able to highly enhance the SERS signal with a good reproducibility. This is an important aspect and usually also the reason why an ordered array of plasmonic nanopores is preferable than a disordered one in order to avoid the presence of random hotspots which can compromise the reproducibility of the SERS spectra. With this purpose, many approaches have been developed to create arrays of nanopores from cheap and high throughput lithographic techniques, such as nanosphere lithography and self-organized nanoporous templates of anodized aluminum oxide (AAO). For instance, Choi et al. realised a self-aligned SERS substrate by depositing a Au film on a self-organized hexagonal AAO nanopore array with channel diameters of about 30 nm fabricated by the two-step anodization of aluminum[96]. From SERS measurements the estimated enhancement factor is $10^7$ with a good degree of reproducibility and thus, these platforms have great potential as biosensing platforms but they are limited by the fact that they are not flow-through. Until now, few works have been found in literature which demonstrate the possibility to use ordered array of nanopores as a vertical flow assay. As example, Chen et al. employed nanoporous AAO as a sensing membrane for SERS and functionalized the pores' walls with multiple antibodies which enable the capture of inflammatory biomarkers. In this work the optimal structures had a diameter of 350 nm and the detection limits obtained for different biomarkers are in the range of tens of fg mL$^{-1}$ [97]. In general, the fabrication of ordered arrays of plasmonic nanopores is still challenging from a technological point of view because it cannot rely on the same equipment that are usually employed for the realization of single pores in free-standing insulating membranes, such as focused ion beam systems or transmission electron microscopes, due to the costs and the long time required to drill the pores one by one. Few examples of ordered nanopore arrays fabricated with alternative methods, which have good reproducibility and control over the dimensions of the pores, are reported in literature. For instance, Dahlin et al. succeeded in the development of nanopores on metal-insulating metal (MIM) films by self-assembling colloidal spheres on a silicon nitride membrane and using them as a mask for the deposition of a Au layer[98].

Until now the main challenge of biosensing performed with SERS both from single and ordered arrays of plasmonic nanopores is the control over the molecules' motion towards the pores. In the first place, the molecules have to be driven near the hotspot of the plasmonic nanopore because the diffusion of the molecules requires long times. Then when the molecule is in the plasmonic hotspot, the interactions between the molecules and the pore walls affect the average time that they spend near the hotspots determining if the temporal resolution of SERS is enough or not to collect information on the molecules. Therefore, in the first place it is important to deliver the molecules close to the hotspot, then provide a strategy to be sure that they reside in the hotspots for a time sufficient to collect Raman spectra and finally, move them away to avoid the occurrence of clogging. Furthermore, even the stochastic movements of the molecules in the hotspot affect the SERS signals and often can be responsible for a lack of reproducibility. For these reasons the microscopic control over the molecules during their passage through the pore is a fundamental aspect. For what concerns the delivery of the molecules through the hotspot, this is usually achieved applying an external electric field taking advantage of the fact that biomolecules in solution often exhibit a superficial charge. In general, this is a well-known approach for charged biomolecules, such as DNA, but it is not trivial for other biomolecules, such as proteins, which expose different charges, positive, neutral or negative, depending on the ionic strength and the pH of the solution in which they are.

### 1.2.8 Electro-optical detection from plasmonic nanopores

In the recent past, many examples of plasmonic nanopores in free-standing membranes have been reported as efficient biosensing platform to probe molecules even at single-molecule level by using the transmission signal, EF or SERS. However, the integration of plasmonic nanostructures with a solid-state nanopore in a freestanding membrane turned out to be a key factor for the development of biosensors at the single molecule level. Among other plasmonic nanostructures, nano antennas have been implemented for electro-optical detection by monitoring the resonance shift during the passage of DNA or other molecules and simultaneously the ionic current trace which tracks the molecules delivered through the pore with an external electric field from the temporary current blockages. Depending on the composition of both the charged walls of the nanopore and the target molecules their reciprocal interactions can be weak or strong. In the first case, the molecules pass through the pore very fast and sometimes this is a problem for the limited temporal resolution of optical detection techniques. Otherwise, the residence time of the molecule in the nanopore is long enough to allow their detection, but, unfortunately, strong interactions can lead to clogging of the pore or denaturation, as in the case of proteins. In this case the use of an external electric field is useful to facilitate the passage of the molecules through the pore. As example, Verschueren et al. used an electrical bias applied across a plasmonic nanoantenna to deliver nanoparticles and proteins in the plasmonic hotspot and control their average residence times[99]. In this way, optical nanoantennas allows the detection of proteins by monitoring the transmission signal over time and the electrical field encourages them to move away from the plasmonic hotspot preventing clogging. Recently, Cao et al. developed a protocol to produce a plasmonic nanopore on the tip of a nanopipette by assembling Ag nanoprisms on a glass nanopipette through a functionalization for silanization. Thus, they performed sensing experiments by delivering rhodamine 6G from the pipette to the plasmonic pore by applying an electric field and then, they extended the use of this plasmonic nanopore as sensing device to the detection of the four nucleotides testing at high concentrations ($10^{-5}$ M)[100]. Indeed, also Yang et al. studied a similar platform composed by a glass nanopipette with the tip coated with Au nanopores produced by in situ reduction of gold on the tip of the nanopipette[101]. In this case the nanopipette integrated with a matrix of disordered Au nanopores was successfully tested for the detection of oligonucleotides showing that it is possible to discriminate between oligonucleotides that differs for a single nucleobase from the analysis of the Raman trajectories. Also, in this case the molecules are driven to the hotspots of the plasmonic material by an external electric field and are not adsorbed on the surface. So far it is clear that the presence of an external electric field can be very useful to electrokinetically deliver molecules one at time very close to plasmonic nanostructures promoting their optical detection boosted by plasmonic effects. However, electro-optical detection still faces two important challenges: the first one is still the limited temporal resolution of the commonly employed optical techniques and the second one is the lack of molecular selectivity of solid-state nanopore based devices for real biological applications. The limited temporal resolution can be solved by slowing down the molecules in order to increase the average time which the molecules spent in the plasmonic hotspot. Indeed, both simulations and experiments on plasmonic nanostructures have shown that it is possible to have plasmonic trapping effects when a molecule is in the plasmonic hotspot which enables you to trap a small portion, a few nucleotides off a DNA molecule, for long times, even several minutes. In the case of SERS, the design of the plasmonic nanostructures is important in order to improve the microscopic control over the translocation of biomolecules through the pore because a lack of the control of their motion causes a lack of reproducibility of the SERS spectra since the translocation is a stochastic process. For this reason, it can be useful to use a trapping mechanism during the passage of molecules which has the double role to trap the molecule close to the hotspot in order to overcome the issue of the limited temporal resolution of optical detection techniques and to limit the freedom of the molecule during the translocation which stochastically moves and causes different fingerprints in the spectra. Optical trapping of small molecules can be achieved by plasmonic nanocavities[102]. Among others effects, in self-induced back-action (SIBA) optical trapping the trapped nano-object affects the enhancement of the local electromagnetic field because the resonance of the cavity shifts as the position of the particle changes due to a variation of the local refractive index and this, in turn, gives ride to a change of the electromagnetic field intensity. Indeed, plasmonic trapping of small nanoparticles based on SIBA (about 20 nm in diameter) has been observed from a resonant metallic rectangular aperture due to the existence of a confined gap mode which strongly enhances the intensity of the electromagnetic fields generating a dynamic trapping effect and enhances the far-field light transmission too[103]. Recently, Huang et al. have shown that it is possible to have plasmonic trapping between a Au nanostar and the wall of a Au nanopore (see Figure 1.11.a)[104,105]. The nanogap generated between the nanostar and the wall of the pore creates a huge enhancement of the electromagnetic fields and thus, functionalizing the nanostar with the molecule of interest, in this case DNA, enables the detection of single nucleotides. Indeed, since a part of the light is absorbed by plasmonic nanostructures, thermal effects play an important role and thus, thermophoretic forces act on the nanoparticles/molecules which are in plasmonic hotspots. As recently shown by Jiang et al., some parameters, such as the ionic composition of the liquid and the addition of surfactants changes the behaviour of nanoparticles between thermophobic and thermophilic (see Figure 1.11.b)[106]. In the first case the thermophoretic force adds to the optical one generated in the hotspot of plasmonic nanotweezers and enhances the plasmonic trapping effect whilst in the second case it is in the opposite direction of the optical gradient force and thus, it suppresses the trapping effect on the nanoparticle. This result shows that plasmonic trapping is a combination of optical, thermal and fluidic phenomena. Other groups worked on the control of electrostatic interactions, steric forces and electrokinetic phenomena, such as electrophoresis and electro-osmosis, to trap biomolecules and nanoparticles inside the sensing area of the nanopore. For instance, Willems et al. showed that it is possible to trap a small protein, dihydrofolate reductase, in a biological nanopore, Cytolysin A (ClyA), for several seconds by controlling the positive and negative charge distribution on the surface of the protein[107]. An interesting approach was proposed by Lam et al., who built an entropic cage made of a nanoporous silicon nitride membrane acting as a capping nanofilter layer above a larger nanopore in order to trap polymers, such as DNA for a long time[108]. In fact, the presence of the nanoporous membrane (randomly distributed nanopores with density of 312 pores/$\mu m^2$, the diameter ranges from 24 to 31 nm) acts as an entropic barrier for the DNA molecules due to the limited escape routes (see Figure 1.11.c). Interestingly, Cadinu et al. developed an electrical sensing platform based on a multi-pore architecture, namely four adjacent channels in parallel located at the tip of a glass nanopipette forming a multibarrelled nanopipette, in order to control at the same time the sensing, motion and trapping of DNA molecules (see Figure 1.11.d)[109]. In fact, they showed that were able to use independent channels: one at a negative potential to deliver a DNA molecule and another one at a positive potential to transfer it. By changing this patch pair every second it is possible to induce the exclusion from one barrel and then transfer to the adjacent one.

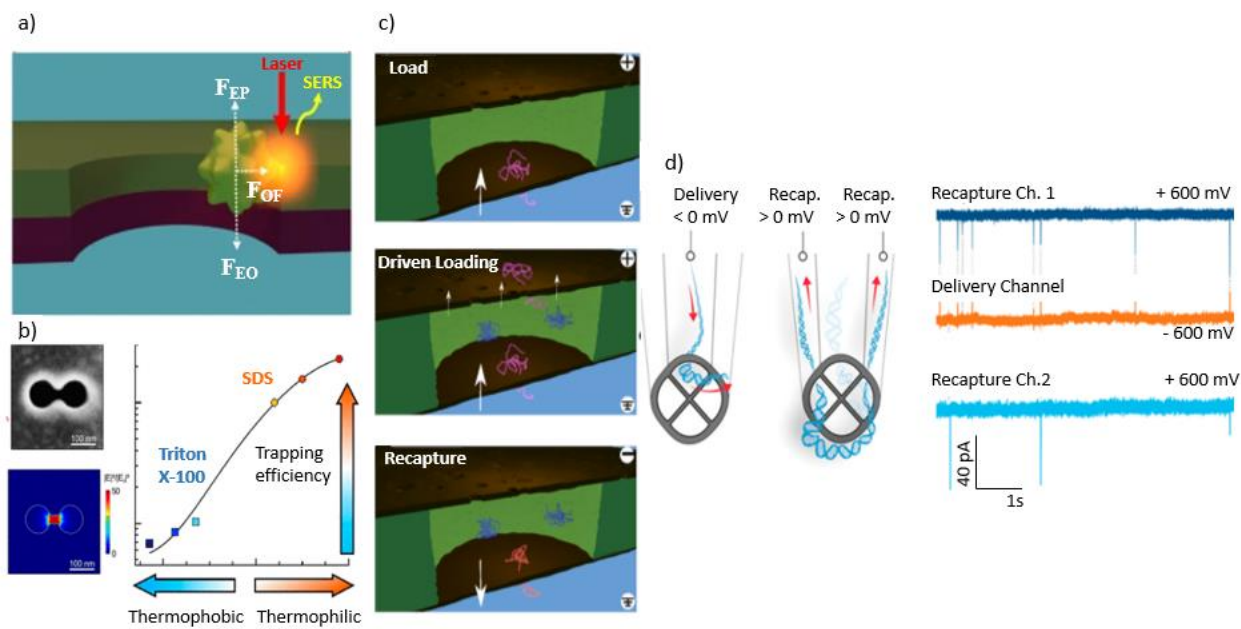

**Figure 1.11.** a) Examples of trapping strategies reported in literature. Electro-plasmonic trapping of molecules adsorbed on the surface of a plasmonic nanoparticle. The electrophoretic, electroosmotic and optical forces are represented by $F_{EP}$, $F_{EO}$ and $F_{OF}$ respectively. The nanoparticle is delivered close to the wall of a Au nanohole and trapped there for several minutes. Reproduced with permission from reference[104]. b) Nanoparticles are delivered in a plasmonic double nanohole to study the thermal contribution on the plasmonic trapping effect as a function of the surfactant added in the liquid medium. Reproduced with permission from reference [106]. c) An entropic cage made of a nanoporous membrane is built on a solid-state nanopore to create an entropic trapping effect. The molecule is driven through the nanopore with an applied bias, then is free to diffuse around the cage and finally, re-directed through the nanopore by reversing the external bias. Reproduced with permission from reference [108]. d) A multi-nanopore enables the manipulation of the motion of DNA molecules and in combination with applied electric fields across each nanopore induces a trapping mechanism. Reproduced with permission from reference [109].

Even if the limited temporal resolution of optical detection techniques and the lack of molecular selectivity of plasmonic nanopores are two completely different problems, some of the strategies found to improve the molecular selectivity of the nanopore can improve also the optical detection because they cause a significant lockdown of the molecule during the translocation. As example, Cai et al. functionalized a nanopore with a molecular beacon integrated in a DNA carrier for the recognition of complementary DNA sequences. The molecular beacon, namely an oligonucleotide with an internally quenched fluorophore, started to fluoresce when the target molecule bound to it. They combined both electrical and fluorescence detection showing synchronized events from ionic and photon time traces. In fact, the coupling of the information from the optical and the ionic time traces is useful for the analysis to exclude false positive events, such as folded DNA carriers. Furthermore, the average residence time of the target molecule in the detection volume increased when they were bound to the molecular beacon giving rise to higher fluorescence signals and thus, improving the signal to noise ratio. Indeed, various approaches to modify the walls of the nanopore and selectively bind the target molecules have been explored in various nanopore-based devices for electrical readout and thus they could be extended also for electro-optical experiments which are limited by the temporal resolution. For instance, Anderson et al. made a polymeric cushion of 3-(aminopropyl)trimethoxysilane (APTMS) between the pore's walls and DNA strands[110]. By varying the pH of the solution it was possible to change the effective positive charges distribution on the nanopore walls and this, in turn, was a way to engineer the DNA-nanopore interactions, strongly affecting the translocation time of DNA. This idea to use polymeric coatings to functionalize the walls of nanopores has been investigated also for plasmonic nanopores. In particular, Emilsson et al. showed that plasmonic nanopores sealed by poly(ethylene glycol) brushes were able to switch from this close-state, in which the nanopores are clogged by the entropic barrier formed by the polymeric brushes, to an open state after the introduction of antibodies in solution[111].

## 1.3 Molecular sensing with arrays of nanopores

One of the limitations of resistive-pulse sensing is the fact that the signal from a target molecule passing through a pore decreases if the number of the nanopores is increased. In general, this is not the case for optical sensing from nanopore-based devices since it is still possible to achieve the single-molecule resolution from a platform formed by an array of nanopores. The use of arrays of nanopores can be useful for applications which require high throughput, which is usually not easy to guarantee with single nanopore-based devices. For instance, the ion flux of $Ca^{2+}$ ions, initially placed in the ci compartment and free to diffuse into the trans compartment has been monitored checking the fluorescence signal of a $Ca^{2+}$ indicator situated in the trans compartment. By using this approach it is possible to quantify the ion flux of each nanopore and localize them from the fluorescence signal with a spatial resolution below the diffraction limit[112–115]. In particular, fluorescence imaging has provided an alternative way to detect events of translocation with the possibility to check at the same time hundreds of parallel channels. In this case, biological nanopores are limited by the thermal drift and the fact that it is not easy to organize biological nanopores in highly ordered periodic structures. Their diffusion in the lipid layer can be studied by monitoring the $Ca^{2+}$ flux from the channels[116]. Indeed, nanopore arrays have been tested for both electrical and optical sensing, enabling the parallel detection of DNA molecules[117,118]. As shown for the first time by McNally et al., solid-state ordered arrays of nanopores enabled parallel sensing by using a custom optical method based on total internal reflection and a two-color readout obtained by converting the nucleotides of a DNA molecule to a sequence of oligonucleotides, which is hybridized with molecular beacons with two different fluorophores (see Figure 1.12.a)[119]. In this way, the information corresponding to the four nucleotides is converted to a binary code formed by the four possible combinations of the two fluorophores. Furthermore, Im et al. integrated a pore-spanning lipid membrane on a plasmonic array of nanopores in order to monitor the binding of biomolecules on the lipid membrane from the shift of the EOT peak of the plasmonic nanopores array[120]. Recently, Wu et al. have developed a fabrication protocol based on the etching of a resist to cover with Au only the walls of ordered nanopores array on a silicon nitride membrane (see Figure 1.12.b-c)[121]. This strategy provided an opportunity to perform a site-selective functionalization exclusively of the walls of the nanopores by using alkanethiol chemistry. The authors showed that a selective functionalization of the walls of the nanopores (a 3 x 3 array) enables the detection of a cancer related biomarker, prostate-specific antigen (PSA) at concentrations as low as 80 aM by using the device for nanopore blocked sensing in combination with functionalized magnetic nanoparticles.

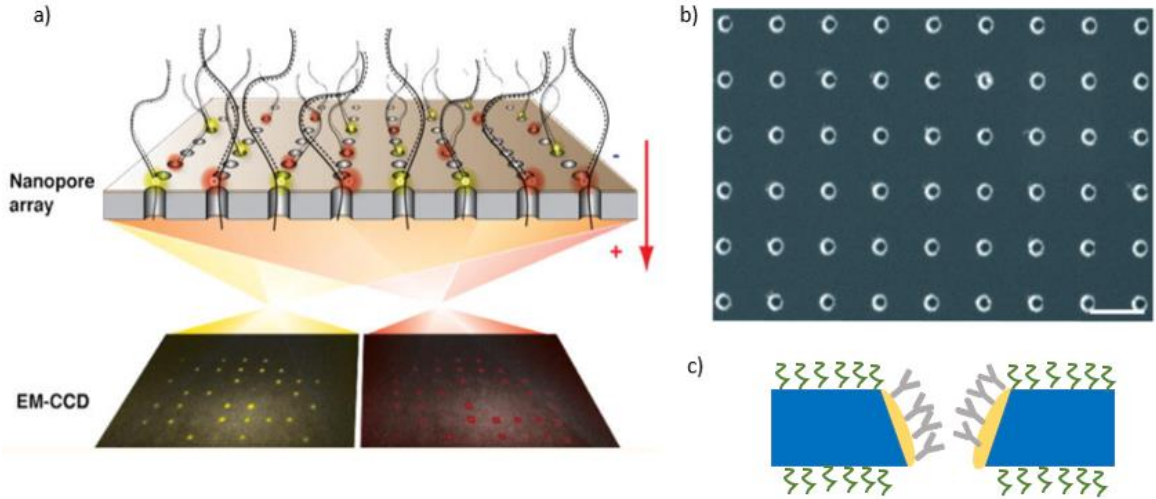

**Figure 1.12.** Examples of biosensing platforms based on arrays of nanopores. a) A strategy to convert the four nucleotides in a binary code generated by four possible combination of two fluorophores by using a process of conversion of the nucleotides in specific nucleotides, that are subsequently hybridized with molecular beacons. Reproduced with permission from reference [119]. b) SEM image of an array of nanopores which have only the internal walls covered with a layer of Au (scale bar: 500 nm). c) Scheme of the selective functionalization of the walls of the nanopore by following a procedure to attach antibodies only on the walls of the nanopores. Reproduced with permission from reference [121].

# 2 Fabrication of Nanostructured Hyperbolic Metamaterials

*In this Chapter it is described how to implement a low-cost and high throughput fabrication route to realize nanohole arrays in a hyperbolic metamaterial (HMM) multilayer. As will be discussed in more detail in Chapter 2, HMMs support highly confined electromagnetic modes and thus, the electromagnetic fields inside the nanoholes/nanopores are enhanced and confined close to the metallic thin layers creating small sensing volumes.*

Among other classes of materials, HMMs have received great attention from both a fundamental and an applicative point of view due to their unique anisotropic optical properties. HMMs are artificial materials obtained from proper subwavelength nanostructuring processes, enabled by the recent advances of nanofabrication techniques, with unexpected optical properties that have only rarely been found in nature. In Chapter 1 it was briefly described how the illumination of light enables the excitation of a propagating surface plasmon (SPP) at the surface of a metallic film in contact with a dielectric under proper conditions and a localized surface plasmon (LSP) on the surface of a nanoparticle. In some materials it is possible to excite coupled surface plasmons and give rise to a collective response that in the case of HMMs results in their anisotropic dispersion. Typically, these materials have been described by introducing the permittivity **ε** and permeability **μ** tensors in the frame of reference oriented along the principal axes of the crystals, since they exhibit their principal components with an opposite sign to the other two[122–124]. Here, we do not consider magnetic media and focus the attention only on optically anisotropic HMM that are described by the permittivity **ε:**

$$\boldsymbol{\varepsilon} = \begin{pmatrix} \varepsilon_\perp & 0 & 0 \\ 0 & \varepsilon_\perp & 0 \\ 0 & 0 & \varepsilon_\parallel \end{pmatrix}, \quad (2.1)$$

where the symbols ∥ and ⊥ are used to define the component parallel and perpendicular to the anisotropic axis respectively. Therefore HMMs are characterized by one of the two following conditions: the first one is $\varepsilon_\perp > 0$ and $\varepsilon_\parallel < 0$ and the HMM is called Type I HMM or dielectric HMM referring to its optical behaviour in the plane, and the second one is $\varepsilon_\perp < 0$ and $\varepsilon_\parallel > 0$ and the HMM is called Type II HMM or metallic HMM[125,126]. It is possible to show that if the permittivity tensors have opposite signs, the isofrequency surface for the extraordinary polarization, namely the component polarized in a plane containing the optical axis, is a hyperboloidal surface (see Figure 2.1a), which by considering the optical axis oriented along the **z** direction is given by:

$$\frac{k_x^2 + k_y^2}{\varepsilon_\parallel} + \frac{k_z^2}{\varepsilon_\perp} = \left(\frac{\omega}{c}\right)^2, \quad (2.2)$$

where $k_x$, $k_y$ and $k_z$ are the three components of the wave vector, ω is the frequency and $c$ the speed of light. From equation (2.2) it results that in HMMs arbitrarily large wave vectors can propagate whilst in common isotropic media they have an evanescent nature.

From their definition, it is clear that the way to engineer these materials requires that the motion of free electrons has to be restricted in one or two spatial directions. From a fabrication point of view, the two main nanostructured systems which have exhibited a permittivity with the properties of a HMM are subwavelength alternating metallic and dielectric layers[127–129], and nanowire arrays embedded in a dielectric medium[130–132] (Figure 2.1b-c). In the case of stacked metallic/dielectric layers various materials composing the metallic (such as Au and Ag) and the dielectric units (such as SiO$_2$, Al$_2$O$_3$ and TiO$_2$) have been investigated both theoretically and experimentally to understand how to control the propagation and the dispersion in three layered materials. HMMs have shown interesting optical properties, such as splitting of absorpyion and scattering bands[133–135], negative refraction[136,137], nanoscale light confinement[138], super resolution imaging and superlensing[139], and extreme biosensing[140,141].

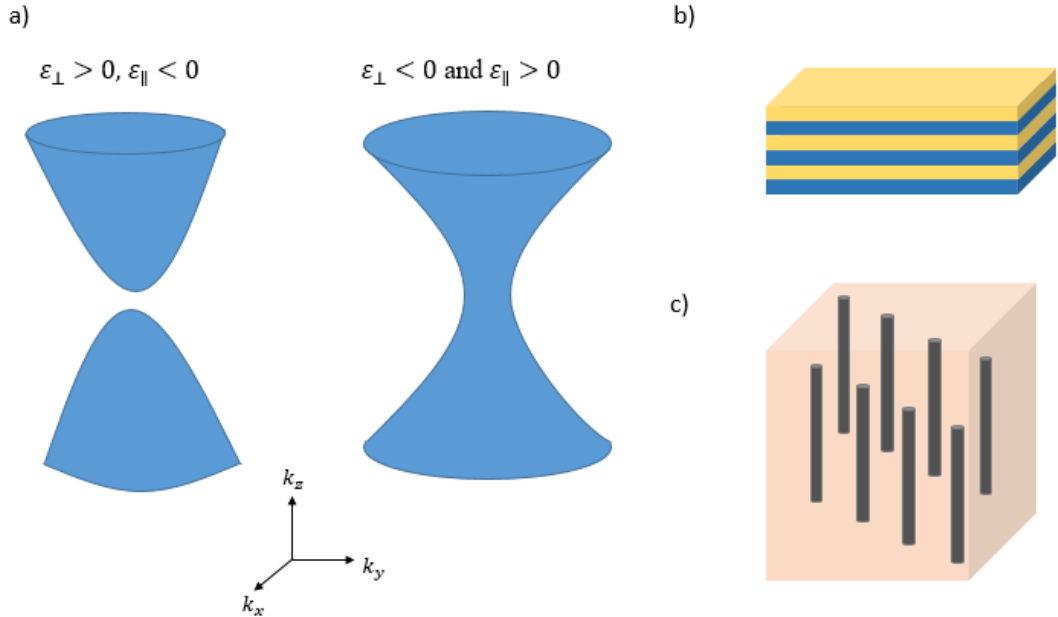

**Figure 2.1.** a) Isofrequency surfaces of extraordinary waves of Type I- and Type II-HMMs. b) Schematic illustration of a subwavelength stacked alternating metallic/dielectric pairs. c) Schematic illustration of nanowire arrays embedded in a dielectric matrix.

This Chapter is devoted to the discussion of the fabrication processes adopted to obtain nanostructured HMM. The prospective of engineering these metamaterials on the nanoscale is very interesting because the combination of their intrinsic optical properties with nanoscale systems could give rise to new properties that have not been explored yet, and potentially they could be employed for many relevant applications, such as biosensing, energy harvesting and nano optics. Here, we will focus the attention on HMMs obtained from sequential depositions of alternating metallic/dielectric units, which belong to the Type II-HMMs. The deposition system used to realise the layers composing the HMM multilayers, their material compositions and their thicknesses are described in Section 2.1. Since colloidal lithography is a high throughput and cost-effective nanofabrication technique, it was used to implement a fabrication route which enabled patterning of a large area and obtain at the same time a nanohole array on a HMM multilayer and particles coated with a HMM hat, which could be removed from the substrate and dispersed in a liquid phase. Indeed, HMM coatings could be added onto other types of nanoparticles in order to change their photothermal properties. As proof of concept, HMM coatings were realized with non-hexagonally close-packed (non-hcp) polystyrene (PS) nanospheres, detached from the substrate and dispersed in water as described in Section 2.2. Furthermore, a discussion on how to manipulate a hcp colloidal mask of PS spheres performing a novel oxygen plasma treatment combined with a vertical temperature gradient, which reshaped the array of spheres into a non-hcp array of PS nano mushrooms to create a natural undercut on the bottom of these structures that avoid the embedding of the particles after the deposition of the multilayers, is also reported in Section 2.2. On the other hand, this fabrication route creates directly a pattern on the HMM multilayer after the removal of the PS spheres that are used as mask. The realised patterns and how their features can be tuned are described in Section 2.3.

## 2.1 Deposition of HMM multilayers

HMM multilayers are fabricated by alternating the deposition of metallic and dielectric layers (see Figure 2.2a). In our case the metallic layer is composed by 10 nm of Au whilst the dielectric layer is composed by 20 nm of silicon dioxide ($SiO_2$) or aluminum dioxide ($Al_2O_3$), as specified each time in the next paragraphs.

Glass/Si substrates were cleaned by a three-step procedure consisting of 5 minutes each of sonication in acetone, isopropanol and deionized (DI) water. Then, the substrates were dried with a nitrogen gun.

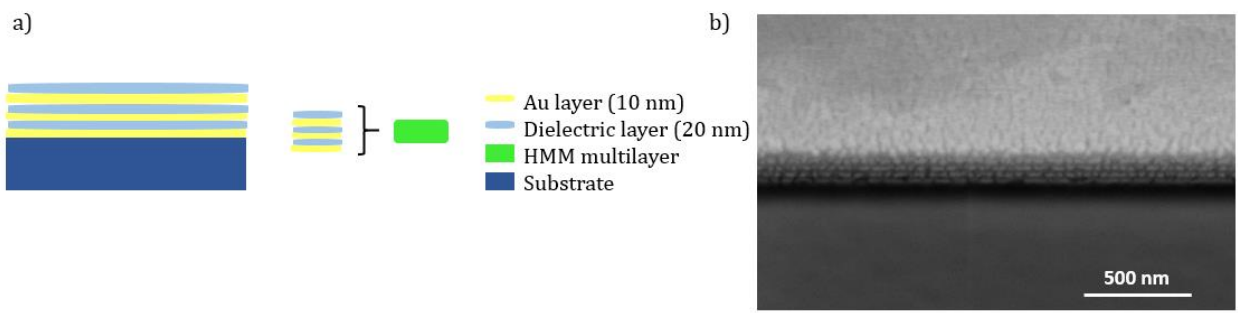

**Figure 2.2.** a) Sketch of a HMM multilayer composed of alternating units of Au/dielectric on a substrate. In order to make some schemes of the fabrication routes more clear for the reader the HMM multilayer will be represent with a green rectangle, meaning each time that it is composed by the alternating metal/dielectric layers. c) SEM image of the cross-section of a HMM multilayer made of 5 stacking bilayers of Au (10 nm) and $SiO_2$ (20 nm).

For the deposition of the Au/ $SiO_2$ and Au/ $Al_2O_3$ alternating units the substrates were loaded in the electron-beam evaporator chamber (E-beam, PVD75 Kurt J. Lesker Company), then it was evacuated to a pressure of about $2·10^{-7}$ Pa. An adhesion layer of 0.5 nm of Ti was evaporated with a rate of 0.1 Å/s between the substrate and the first layer of Au and between each Au/dielectric unit. Then, the Au and dielectric layers were sequentially deposited with rates of 0.1 Å/s and 0.4 Å/s respectively by moving the corresponding crucibles and waiting at least 10 min after each deposition to allow the materials to cool down before to start the next deposition.

As example, a SEM image of the cross-section of a HMM multilayer made of 5 bilayers of Au and $SiO_2$ is reported in Figure 2.2b. Even pumping the chamber for an overnight to achieve a better vacuum, the deposition of thin films (below 20 nm) with a high uniformity under the conditions described above is challenging. In fact, the deposited layers, especially the Au layers show some irregularities and a rough surface.

## 2.2 Nanostructured HMM: coatings on nanoparticles and nanohole arrays

The properties of the HMMs can be integrated with other nanostructured materials/nanoparticles. This possibility was explored by realizing HMM coatings on particles of different diameters. The fabrication was implemented on a hexagonally close-packed array of PS spheres by following the procedure described in Figure 2.3.

In order to fabricate periodic nanostructured arrays, an interfacial self-assembly lithographic technique was used to form a hcp close-packed monolayer of spheres on the substrate (Si or glass). Specifically, we used commercial PS spheres with average nominal diameters of 500 nm, 420 nm and 304 nm (5 wt % water suspension; Micro Particle GmbH).

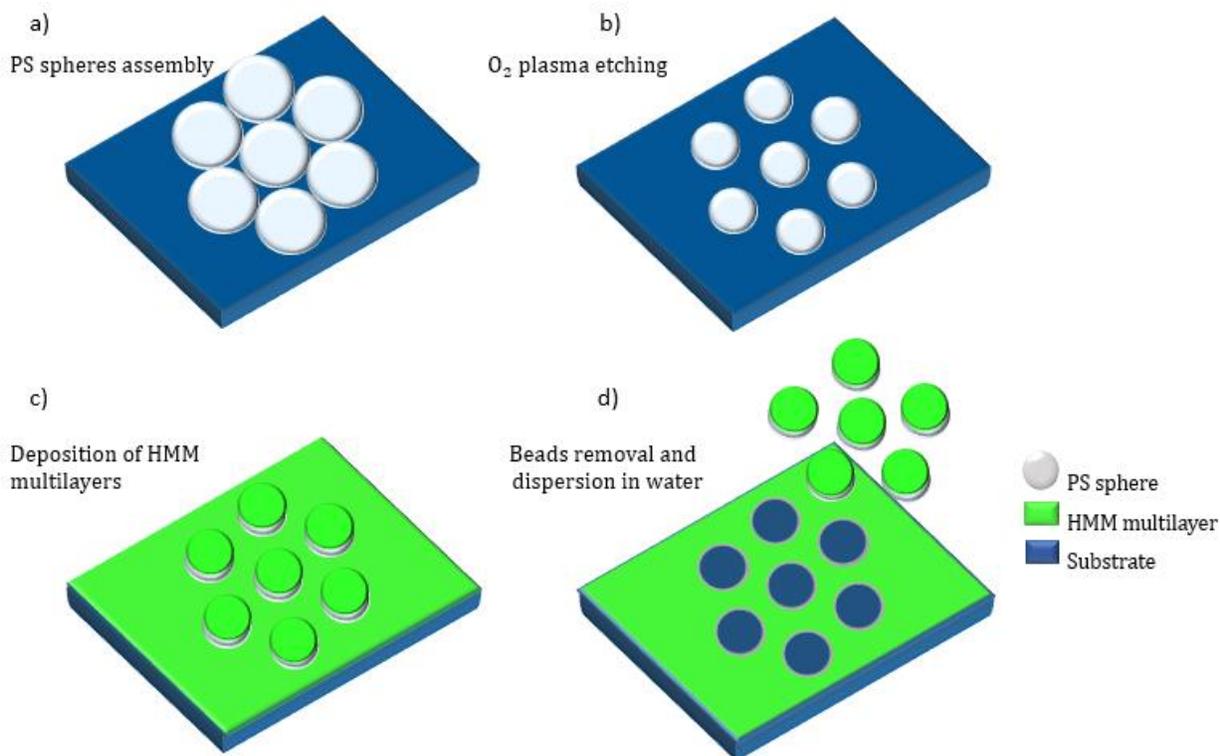

**Figure 2.3.** Scheme of the fabrication route of HMM coatings on PS spheres. a) A monolayer of hcp PS spheres is formed on the substrate by using an interfacial self-assembly lithographic technique. b) The PS spheres are exposed to a $O_2$ plasma treatment in order to separate the adjacent spheres. c) A HMM multilayer is deposited alternating the evaporation of metal/dielectric bilayers. d) The HMM-coated PS spheres are removed from the substrate by sonication and dispersed in water.

Then, the particles were exposed to an $O_2$ plasma treatment (power: 100 W, flow rate: 20 sccm; Gambetti) in order to etch the PS spheres with the aim to reduce their size whilst keeping them in their original positions. With this process the particles were no longer in contact and a hexagonally non-close-packed monolayer was obtained. The exposure time was adjusted taking into account the next step of fabrication which consists of the deposition of the HMM multilayer. In fact, if the interparticle distance is not large enough the PS spheres are embedded in the HMM film and their removal from the substrate is not possible. Furthermore even when the separation between the particles is enough to avoid this problem, if there are some junctions between the materials deposited on the substrate and the ones deposited on the particles the removal of the coated spheres is not completely successful. In order to avoid this problem the $O_2$ plasma treatment process was altered by adding in the presence of a vertical temperature gradient from the bottom to the top of the PS spheres (see Figure 2.4a). Practically, a glass petri dish was heated up at 65 °C in an oven, then the sample was rapidly placed on top of the petri and transferred in the bench $O_2$ plasma chamber (power: 100 W, rate flow: 20 sccm). Due to the presence of the temperature gradient during the etching process the bottom hemisphere of the PS spheres was etched faster than the top hemisphere. As a result, the hcp array of spheres was shaped into a non-close packed array of nano mushrooms[142]. The features of the mushrooms, such as their hat and their pillar, depend on the etching time and on the temperature gradient, since we found that placing the sample on a petri dish heated up at different temperatures strongly affects the morphology of the PS nano mushrooms produced (see the SEM images in Figure 2.4b-e). In fact, by varying the temperature of the glass petri from 35 °C to 80 °C, with a step of 15 °C and keeping unchanged the parameters of the $O_2$ plasma exposure (etching time: 120 s, power: 100 W, flow rate: 20 sccm), the size of the mushrooms' hat decreases whilst the length of the mushrooms' pillar increases, as shown in Figure 2.4f. Importantly, with this process it is possible to obtain thin pillars of about 20 nm meaning that the space underneath the mushrooms' hat is almost empty and can be filled with a thick layer of material, such as the HMM multilayers.

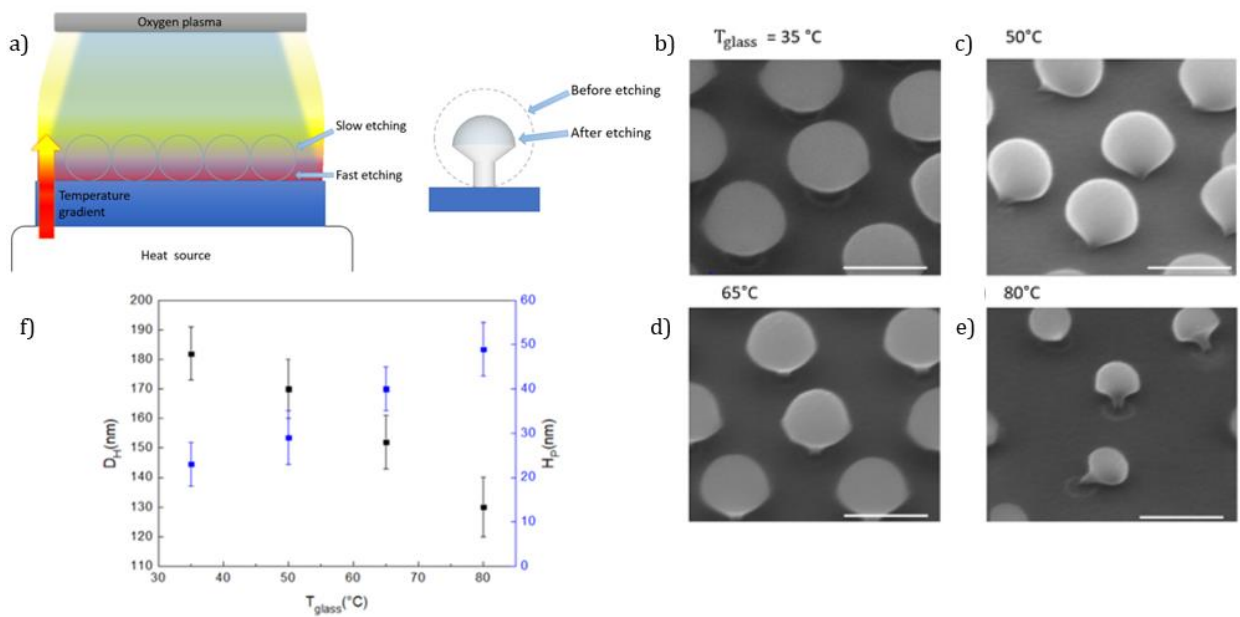

**Figure 2.4.** a) Scheme of the etching process implemented in the plasma cleaner chamber: a $O_2$ plasma treatment is performed on the hcp PS spheres array with the substrate in contact with a glass petri heated up to a certain temperature to create a vertical temperature gradient during the process. b-e) Tilted-view SEM images (scale bar: 200 nm) of 300 nm PS spheres shaped into nano mushrooms after the exposure for 120 s to the same plasma etching process whilst the substrate is in contact with a petri at a certain temperature: 35 °C, 50°C, 65°C and 80°C respectively. Some nano mushrooms are knocked over with the electron beam of the SEM system in order to reveal their shape. f) Average diameter of the hats and height of the pillars as a function of the temperature of the glass petri obtained from the analysis of SEM images with the ImageJ software.

Nano mushrooms with longer pillars can be obtained by using bigger PS spheres. For instance in Figure 2.5a shows the top-view SEM image of a hcp monolayer of spheres with a diameter of 420 nm after etching treatment was performed at 65 °C for 110 s (gas flow rate: 7.5 sccm, power: 100

W). The particles appear separated keeping the hcp order and an interparticle distance of about 200 nm. Moreover, as reported in the tilted-view SEM image presented in the inset of Figure 2.5b the spheres are etched anisotropically taking advantage of the temperature gradient applied vertically from the bottom of the monolayer, so that the bottom hemisphere of the particles shrink rapidly down to a pillar of 65 nm in length and 90 nm in height whilst the top hemisphere etches slower keeping the shape intact but reducing its size to 180 nm. This process is useful because it is possible to reduce the formation of junctions between the deposited material on top of the mushrooms' hat and the film deposited on the substrate. The fact that the particles are not completely embedded is useful for their removal and also to locally create hot spots between the tips of the mushrooms' cap and the film.

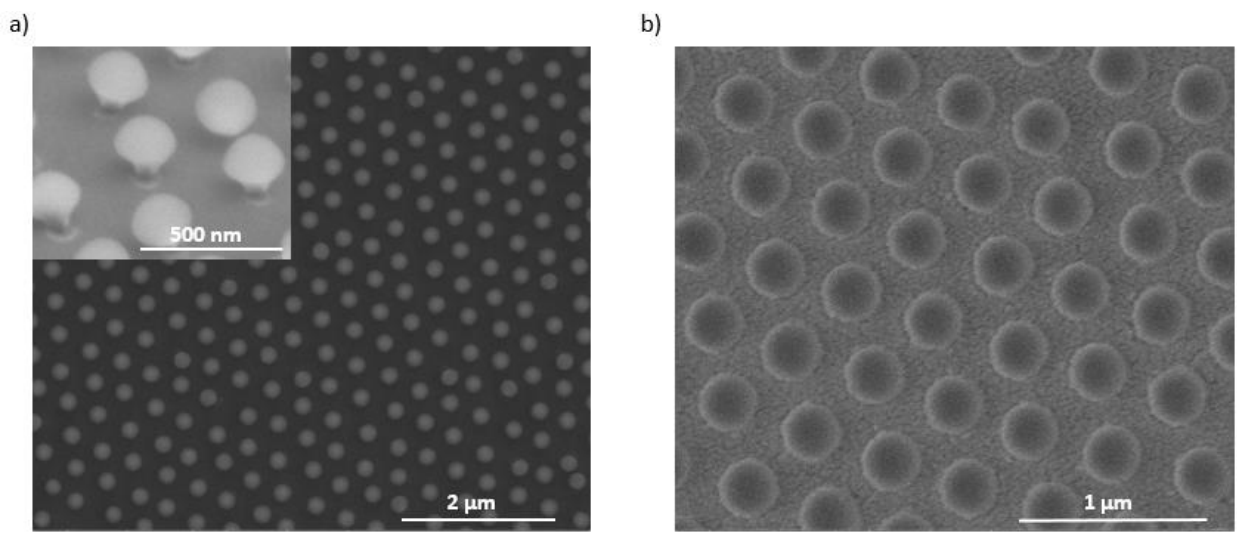

**Figure 2.5.** a) SEM image of hcp PS spheres with an initial diameter of 420 nm after the exposure to the temperature-assisted oxygen plasma treatment. The particles are separated and reshaped into nano mushrooms as is possible to visualize in the magnification of the tilted view of the sample reported in the inset. b) Top-view SEM image of the PS nano mushrooms after the deposition of the HMM multilayer. The presence of a coating on both the substrate and the particles is well visible.

After the temperature-assisted $O_2$ plasma treatment process the sample was loaded in the evaporator in order to deposit vertically the metal/dielectric units as described in the paragraph 2.1. Typically, 5 alternating bilayers made of 10 nm of Au and 20 nm of a dielectric material ($SiO_2$ or $Al_2O_3$) were deposited. A SEM image of the sample after these sequential depositions is reported in Figure 2.5b. It is clear that the particles are coated and that the space on the substrate between the particles is also filled with the deposited materials.

Finally, the sample was sonicated for 2 minutes in 1 mL of DI water, then the substrate was rinsed with a nitrogen gun. In this way the HMM-coated PS nanoparticles were removed from the substrate and transferred in water.

The morphological characterization of the HMM-coated nanoparticles was performed by SEM after drop casting an aliquot of the HMM-coated nanoparticles in water on a Si substrate and awaiting the evaporation of the solvent. In Figure 2.6a) is reported a SEM image of the random distributed HMM-coated nanoparticles after the drop casting. From the image it is clearly visible the presence of single nanoparticles and few aggregates composed of less than 10 nanoparticles each. Bigger aggregates of HMM nanoparticles were found only on circular areas of the substrates corresponding to the coffee ring effect occurring during the evaporation. Generally, it was not possible to find any residue of the HMM film which was a good indication of the fact that the sonication process did not damage the multilayer deposited on the first substrate. In Figure 2.6b) it is reported a SEM image of one of the larger aggregates, which shows that the nanoparticles are randomly oriented and thus, they are not connected between each other as a result of the deposition of the alternating layers. Furthermore, the layers composing the HMM coating, in this case 5 stacked units of $Au/Al_2O_3$, on top of each PS particle are well visible. From the analysis of this SEM image it is possible to give an estimation of the dimensions of the HMM-coated nanoparticles: the diameter of the PS sphere is 180 nm, the height of the HMM coating is about 150 nm, in accordance with the total thickness of the deposited materials, and diameter of the top hemisphere of the coating is about 330 nm. It is worth to point out that each layer forms a cap on the top hemisphere of the particle due to its spherical shape, meaning that the thickness of the tips is less than the nominal thickness of the deposited material. The increasing size of each layer is mainly due to the fact that the materials are deposited sequentially on nanospheres and not on a flat surface. However, the accumulation of small residual nanoparticles around the HMM coating on the PS particles is not due to the removal process because the presence of junctions between the HMM coating and the HMM film on the substrate should be significantly reduced by the undercut presented by the PS nano mushrooms. In order to prove this point a cross-section of the nanoparticles is shown in Figure 2.7a. At this purpose the particles were half milled with the FIB column and then imaged with the SEM system in order to show the quality of the layers far away from the surface. The SEM image reported in Figure 2.7a shows the cross-section of the HMM coating and the layers in the bulk appear slightly less indented than on the surface. It is reasonable to think that the major contribution to the roughness of the coating is still intrinsically due to the sequential depositions of the layers, as shown in Figure 2.2.

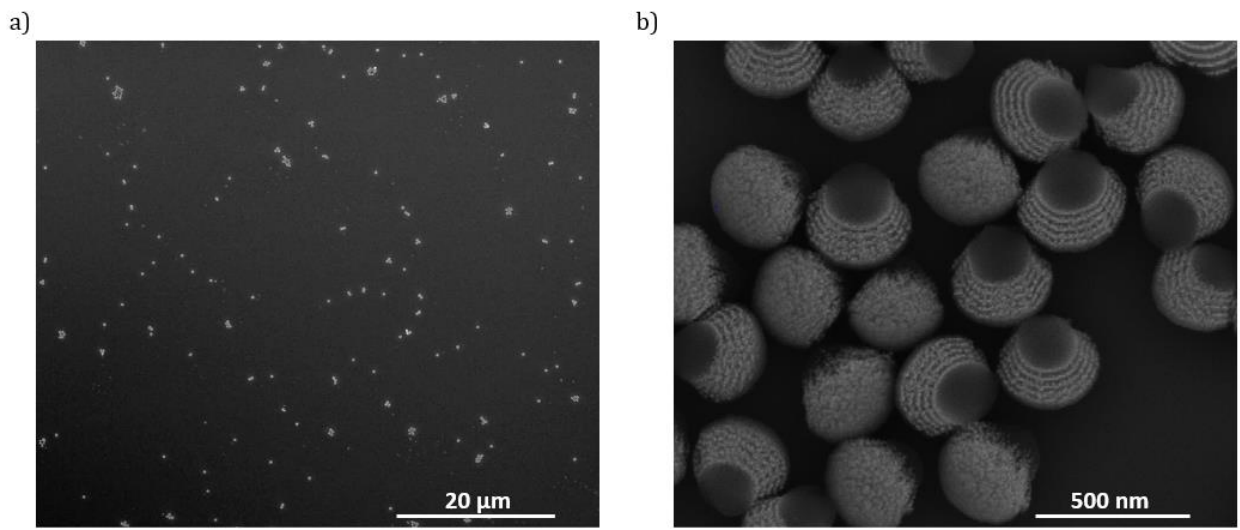

**Figure 2.6.** SEM images of the HMM-coated nanoparticles after drop casting on a Si substrate. a) The aqueous solution is made with a low concentration of HMM-coated nanoparticles in order to show that the particles are separated and that only small aggregates are present. b) A magnification of one of these agglomerates highlights that the particles are not attached as confirmed by their random orientation.

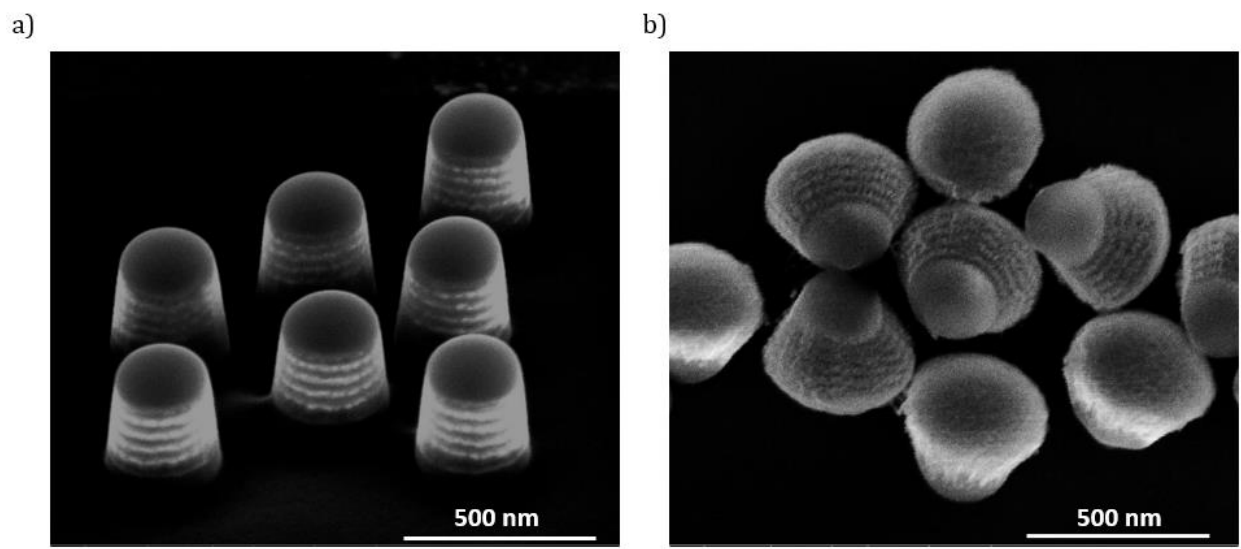

**Figure 2.7.** a) SEM image of the cross/section of the HMM-coated nanoparticles. The particles were milled by using the FIB column integrated in the SEM-FIB system. b) SEM image of nanoparticles coated with a HMM made of 5 units of Au/SiO$_2$.

The same fabrication process was repeated by changing the material composition of the HMM coating. In Figure 2.7b a SEM image of HMM coated nanoparticles in which the coating is made of 5 units of Au/Si O$_2$ is reported. These particles appear very similar to the ones coated with the Au/Al$_2$O$_3$ multilayer (shown in Figure 2.6b) and more in general, changing the materials composing the HMM did not affect the procedure implemented for the fabrication of the HMM-coated nanoparticles.

It is important to emphasize again that the fabrication route used to produce the HMM-coated nanoparticles is mainly based on the holey-mask colloidal lithography technique with a modification of the etching process of the PS particles which allows to reshape the spheres into nano mushrooms by introducing a temperature gradient as key role parameter during the $O_2$ plasma treatment. Conventional etching treatments of the PS spheres already reported in literature modify only the organization of the array in 2D, meaning that a hcp monolayer of spheres is converted into a non-hcp monolayer of spheres, which then is used as a colloidal mask (as sketched in Figure 2.2a-b.). The temperature-assisted $O_2$ plasma etching performed on hcp PS spheres does not only reduce the size of the particles increasing at the same time their reciprocal distance, but also completely modifies their morphology in such a way that a new nanostructured colloidal mask is obtained. The top of this nanostructured mask is formed by the hats of the mushrooms that are spherical in good approximation and thus can be used as mask the same as conventional masks of spherical particles. The thin pillar that is present on the bottom of the nano mushroom creates a space of several tens of nanometers between the substrate and the mushrooms' hat, which can be tuned by the temperature (Figure 2.4b-f) or by the initial size of the spheres. As a result, the nanostructured colloidal mask exhibits a natural undercut that prevents the embedding of the particles in the deposited material, improves the removal of the particles without the damage of the film deposited on the substrate and reduces the formation of junctions between the material deposited on the film and the one deposited on the particles. As a result, with this approach it is possible to deposit thick materials (the total thickness of the HMM multilayer is 150 nm). Furthermore, the discussed fabrication route can be used not only for the fabrication of the HMM-coated nanoparticles but also to pattern large areas on the substrate, as typical of colloidal lithography. In fact, the same approach can be used to realize a nanostructured HMM film on the substrate as discussed in the next paragraph.

## 2.3 HMM nanostructured films

After the removal of the particles a nanostructured HMM film is left on the substrate and its features depend on the interparticle distance between the PS particles after the etching treatment. Since after each deposition of the layers composing the multilayer, the shadow effect of the coated mushrooms' hat extends in length due to a corresponding increase of its size, two neighboring particles have to be at least 130 nm far apart from each other in order to avoid that the HMM coatings deposited on their hat form junctions and that after their removal the particles are not connected together. This affect also the quality of the pattern that is left on the substrate. In general, if the particles are far enough from each other after the deposition of the HMM the sample appears as reported in Figure 2.5.b. It is clear that after removing the beads what will be left on the substrate is a nanostructured HMM film exhibiting an ordered array of concentric nanoholes. A top-view SEM image of the nanohole array obtained after the removal of the beads is reported in Figure 2.8a. With the used etching conditions ($O_2$ gas flow: 7.5 sccm, power 100 W, etching time: 110 s, temperature: 65 °C) and after the HMM deposition the average distance between two neighboring nanoholes is about 100 nm. Due to the masking effect of the PS mushrooms' hat the substrate is patterned with concentric nanoholes made of alternating layers of Au/$SiO_2$. The diameters of the nanoholes on the first deposited Au layer and the last deposited $SiO_2$ layer are about 200 nm and 350 nm respectively. In fact, as a result of the sequential depositions that increase the size of the mushrooms' hat the nanoholes present a conical shape in the z-direction as it is possible to see in Figure 2.8b. By changing the etching parameters of the PS spheres it is possible to change both the diameter of the HMM nanohole array and the reciprocal average distance between each hole. Decreasing the etching time the initial size of the mushrooms' hat is larger and the interparticle distance is smaller and thus, it is expected to obtain larger HMM nanoholes that are on average closer. In Figure 2.8c it is shown the HMM nanohole array obtained from a hcp monolayer of PS spheres of the same diameter (nominal diameter: 420 nm) etched in the same conditions of the previous example but for only 90 s. In this case the diameters of the nanoholes on the first Au layer and the last deposited $SiO_2$ layer are 265 nm and 415 nm respectively. The average distance between two adjacent nanoholes is about 25 nm.

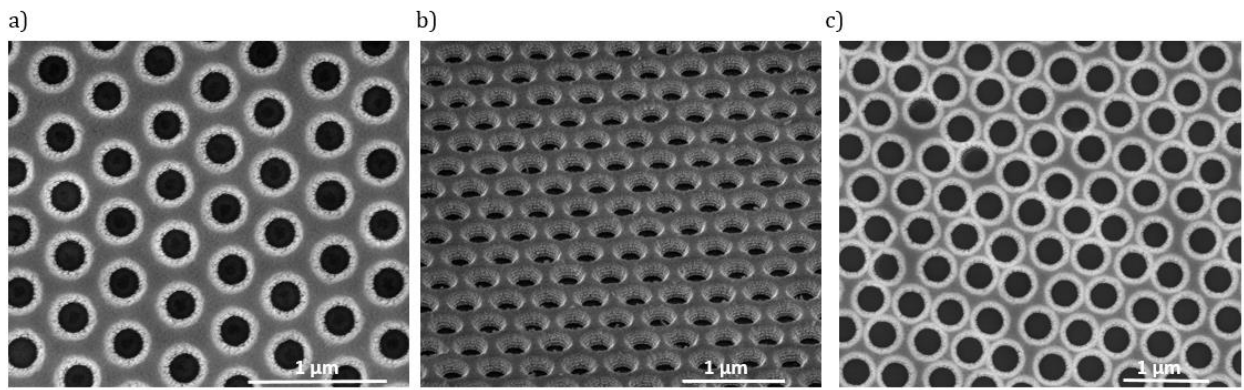

**Figure 2.8.** In a) and b) SEM images of the HMM nanohole array obtained after the removal of the PS particles: top-view and tilted-view respectively. In c) top-view SEM image of a nanohole array obtained on the HMM multilayer decreasing the etching time of the PS spheres from 110 s to 90 s.

By changing the initial diameter of the PS spheres it is possible to further tune the dimensions of the HMM nanohole array. Figure 2.9 shows SEM images of nanohole arrays on a HMM multilayer obtained from hcp PS spheres with diameters of 304 nm and 552 nm respectively. Both monolayers were exposed to the oxygen plasma treatment in the presence of the heat source at 65 °C for 110 s but the gas flow rate in the two cases was different: 20 sccm and 15 sccm respectively. From the 304 nm PS spheres and following the fabrication route explained in Section 2.2 concentric nanoholes with diameters of the first Au layer of 145 nm and of the last SiO2 layer of 285 nm were obtained (Figure 2.9a). Repeating the fabrication protocol on the 552 nm PS spheres the diameters of the first Au layer and of the last SiO$_2$ layer are on average 340 nm and 450 nm respectively (Figure 2.9b).

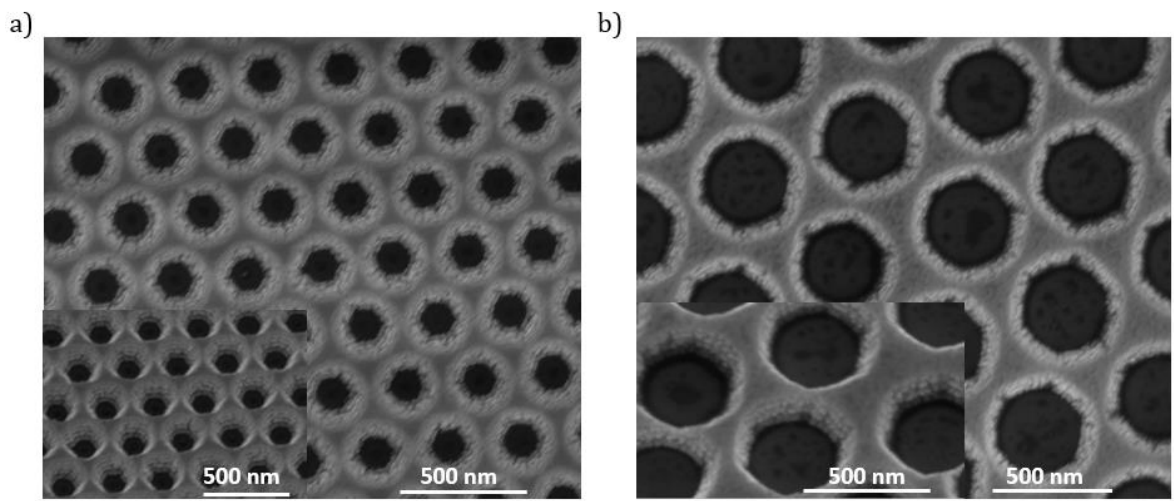

**Figure 2.9.** In a) and b) top-view SEM images of nanohole arrays of the HMM multilayer obtained from hcp PS monolayer with diameters of 304 and 552 nm respectively. Insets of tilted-view of the nanoholes arrays are also reported.

As a last example, a SEM image of another nanoholes array on a HMM multilayer obtained decreasing the etching time of a PS spheres with a diameter of 552 nm to 90 s is shown in Figure 2.10. As a result, the particles are less separated from each other and thus, the HMM film remaining on the substrate around each nanohole is organized in a snowflake structure. Generally, by increasing the etching time of the oxygen plasma treatment it is possible to increase the average distance between the nanoholes and decrease the diameter of the nanoholes as well. Furthermore, the diameter of the nanoholes can be further controlled by repeating the same process on a monolayer of PS spheres with a different diameter. Additionally, the oxygen flow rate during the oxygen plasma exposure affects the morphology of the PS mushroom template and thus, it can also be exploited to tailor the nanohole arrays. In particular, a higher flow rate gives rise to a faster etching and in combination with the applied temperature gradient to a higher piller underneath the PS mushrooms template. As already mentioned, this avoids the presence of junctions between the caps on the PS mushrooms and the film on the substrate after the multilayer deposition and facilitates the PS mushrooms removal.

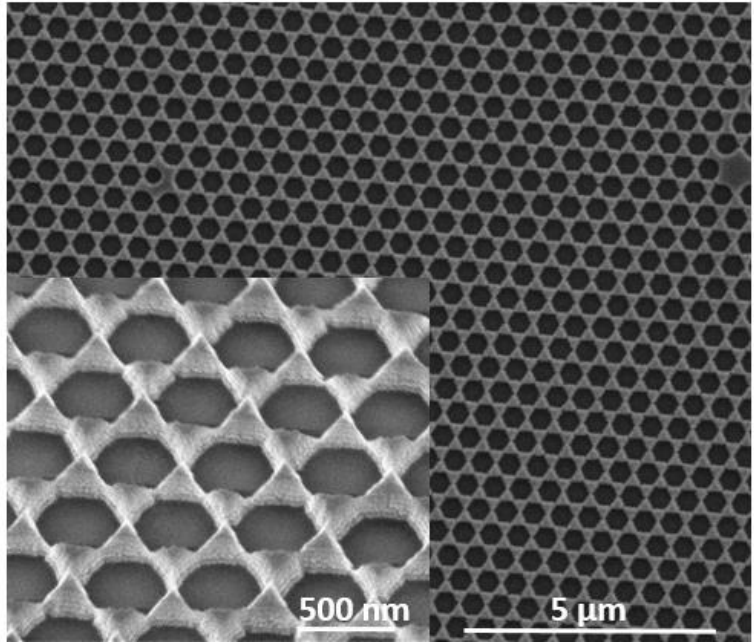

**Figure 2.10.** SEM image of nanohole arrays on a HMM multilayer obtained from a hcp PS spheres with a nominal diameter of 552 nm. The exposure time to the oxygen plasma treatment was reduced to only 90 s in order to slightly separate the PS spheres.

# 3 Results and discussion

*In this Chapter the optical properties of the nanoholes arrays in the HMM film and the HMM coated nanoparticles both obtained from the fabrication route described in Chapter 2 will be discussed. Various research groups have reported novel designs for both type I- and type II-HMMs exhibiting unconventional optical properties, as extreme sensitive optical modes, such as sharp surface plasmon polaritons (SPPs) and bulk plasmon polaritons (BPPs) modes that are excited satisfying the momentum matching condition with different coupling mechanisms and that can be used for biosensing applications with high sensitivity*[141]. For instance, extreme sensitivity has been demonstrated for a type II-HMM, made of 16 alternating layers of Au and $Al_2O_3$, which was integrated with a 2D subwavelength Au diffraction grating in order to excite the BPP modes via a grating coupling mechanism[140,143]. The effectiveness of this type II-HMM coupled with a Au-coated periodic grating was tested by changing the refractive index of the medium in contact with the HMM by the addition of a glycerol solution with different concentrations and monitoring the resulting shifts of the peaks positions of the reflectance spectra arising the excitation of the BPP modes, for both the spectral and angular configurations (as described in Section 1.2.3 of Chapter 1). Furthermore, this sensing platform was tested also in the presence of biotin, which is a small molecule with low molecular weight (244 Da) showing a limit of detection of 10 pM. Indeed, also type I-HMMs, made of Au nanopillars aligned vertically with respect to the glass substrate and separated from each other by few tens of nanometers have provided high-sensitivity as SPR-sensors[144]. By exciting the modes via prism coupling and changing the refractive index of the surrounding medium a sensitivity two order of magnitude higher than the one achieved with common LSPR-based sensors has been achieved[145]. Recently, Sreekanth et al. demonstrated the high sensitivity of the BPP modes excited in a type-I HMM, made of alternating layers of TiN and $Sb_2S_3$, via prism coupling achieving a limit of detection of 1 pM for the detection of biotin[146]. Therefore the BPP modes excited with various mechanism of coupling are very promising for biosensing applications of small analytes at low concentrations. Nevertheless, the already developed biosensing platforms based on HMMs require a coupling mechanism to excite BPP modes and thus, need the integration of a prism or a diffraction grating. This is one of the reasons why HMMs have not been investigated yet for biosensing applications in real biological media or in other configurations, such as for flow-through sensing. For what concern the nanostructured HMM presented in Chapter 2, a nanohole array in a HMM is promising because the molecules can contact directly with the inner multilayer structure and not only with its surface and could improve the sensitivity. Furthermore, the integration of a HMM nanohole array with a solid-state nanopore in a thin insulating membrane could be interesting to explore this class of material for electro-

optical sensing[147]. Therefore, the optical properties of the HMM coated nanoparticles and the nanohole arrays in HMM films are discussed in Section 3.1 and 3.2 respectively.

## 3.1 HMM coatings on nanoparticles re-dispersed in water

In Chapter 2 it is discussed how the HMM-coated PS nanoparticles were detached by sonication and suspended in 1 mL of DI water. The resultant solution appeared to have a black colour and since the particles are few hundreds nanometers in size they tended to sediment at the bottom of the vial after time but could be easily re-suspended with a pipette for several weeks. After this time, aggregation occurred since large aggregates visible by eyes were found in the vial.

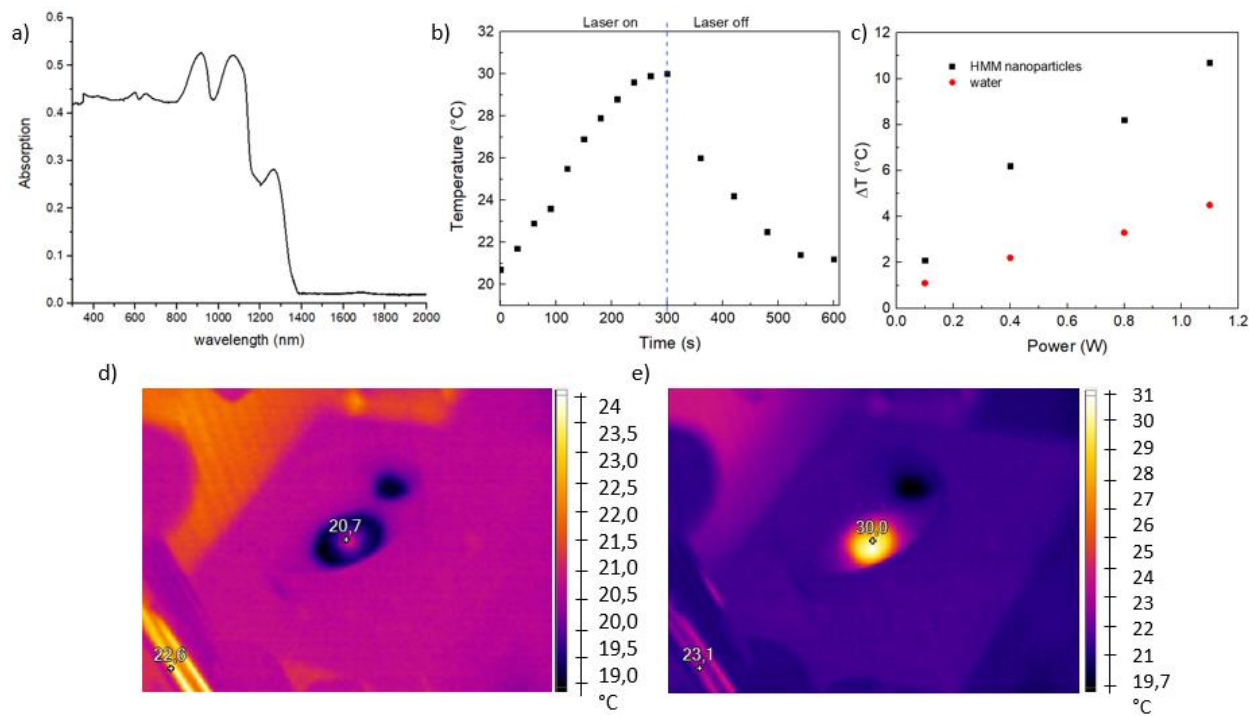

**Figure 3.1.** a) Absorption spectrum of the HMM coated PS nanoparticles in water. b) Temperature profile of the HMM-coated PS nanoparticles solution as a function of time under continuous irradiation (CW) with a 1064 nm laser. After 5 min the laser was switched off and the temperature of the solution decreased with time to the starting temperature value. c) Temperature increase as a function of power of the laser for the HMM nanoparticles in water (in black) and for a control sample made of water (in red). d)-e) Thermo-images of the HMM nanoparticles in solution drop casted in a circular petri dish before and after 5 min of laser illumination respectively.

The optical properties of the HMM-coated nanoparticles in water were characterized by pipetting the solution in a quartz cuvette and acquiring the absorption spectrum in the 300-2000 nm wavelength range with a Cary 5000 UV-Vis-NIR spectrophotometer. In Figure 3.1.a the absorption spectrum of the HMM coated nanoparticles, made of PS spheres coated with 5 alternating layers of Au/SiO$_2$ units as described in Chapter 2, is reported. These particles exhibit absorption peaks in both the first (NIR-I, 650-950 nm) and second (NIR-II, 1000-1350) nm near infrared windows, specifically at 910, 1064 and 1280 nm. Indeed, similar features in the NIR range from the absorption spectrum of spherical nanoparticles made of alternating metal/dielectric layer and obtained with a bottom-up approach have already been reported. The number of the peaks depends on the number of the metal/dielectric units while the position of the peaks on the materials composition and the thicknesses of the layers. Due to the high absorption in the NIR range the HMM coated PS nanoparticles were tested for photothermal experiments. In fact, it is well known that there is a linear proportion between the local thermal increase generated by nanoparticles and their optical absorption cross-section. It means that plasmonic nanoparticles can be used as local nano heaters upon illumination at the resonant wavelength. However, gold/silver nanoparticles exhibit resonances in the visible range. This is not suitable for some applications, such as for cancer treatment through hyperthermia, because the laser illumination in the visible in the visible range can cause many problems, such as the damage of the surrounding tissues. To overcome these limitations, plasmonic nanoparticles with resonances in the NIR range are of great interest and have great potential for this type of applications. Therefore, the increase of the temperature generated by the solution of HMM nanoparticles (concentration of about $1 \times 10^{10}$ particles/mL) was measured with a thermal camera (Fluke thermos camera) upon illumination with a 1064 nm laser (power: 200 mW). In Figure 3.1.b it is reported the measured temperature of the solution of HMM-coated nanoparticles as a function of time. When he sample was under laser illumination for 5 min the temperature increases as a function of time until the saturation value of 30 °C was reached. The, the laser was switched off and the temperature decreases exponentially in time to the starting temperature value of about 20.5 °C. In Figure 3.1.c it is shown the temperature increase as a function of the laser power for both the HMM nanoparticles solution and a control sample containing water. The data confirmed that the HMM nanoparticles play a role in the local thermal heating of the solution due to the absorption of the laser light in the II-NIR and that the thermal increase depends linearly on the laser power as expected. As example, in Figure 3.1.d-e two thermal images of the solution of HMM-coated nanoparticles before to switch the laser on and after 5 min of illumination respectively are reported. It is important to note that the HMM-coated nanoparticles exhibit an absorption peak at an even longer wavelength in the NIR-II centered at 1280 nm which with a proper re-design of the HMM and of the choice of the materials could be further shifted even in the NIR-III (1600-1870 nm). However, one limitation of the proposed structure and, as well as of the one reported in literature consists of the size of the particles (hundreds of nanometers) while, at least for biological hyperthermia applications involving in vitro and/or in vivo studies smaller particles are usually employed, also to avoid toxicity effects.

## 3.2 Nanohole arrays in HMM films

Hexagonally-non close packed nanohole arrays in HMM films were obtained after the removal of PS beads following the approach described in Chapter 2. In Figure 3.2.a. the optical simulation of the reflectivity of a nanohole array in multilayer films composed of an increasing number of $Au/Al_2O_3$ alternating units are reported. Firstly it is clear that the modes are sharper by increasing the number of bilayers and thus, the nanohole arrays in a HMM multilayer should have a higher sensitivity than the ones in a simple metal-insulator-metal (MIM) film. Furthermore, in the case of a nanohole in a HMM composed of three bilayers of $Au/Al_2O_3$ and an extra unit of Au (10 nm) and $Al_2O_3$ (20 nm) the electric field distributions of the electromagnetic fields of the modes at 850 nm and 950 nm are reported in Figure 3.2.b-c respectively. Here, it is possible to note how the distribution of the electromagnetic field for both the selected modes is highly confined at the metallic interfaces inside the nanohole. It means this layered nanohole has potential to detect molecules that are located close to the metallic layers of the HMM. Furthermore, the distribution of the electromagnetic fields for the two selected modes is different and thus, can be tailored depending on the selected mode.

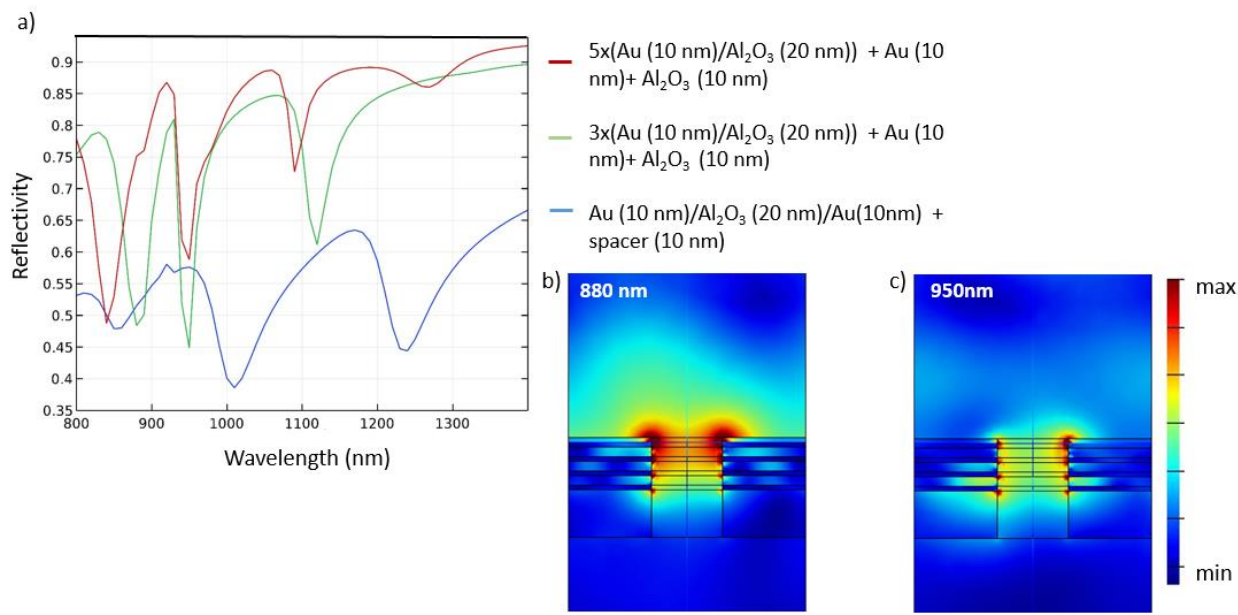

**Figure 3.2.** a) Optical simulations of the reflectivity as a function of the wavelength measured from a nanohole of 200 nm in diameter in three different multilayer systems as specified in the legend at the right of the plot. b)-c) Simulated electric field distribution inside a nanohole in the multilayer made of three units of Au (10 nm) and $Al_2O_3$ (20 nm) with an extra film of Au (10 nm) and $Al_2O_3$ (10 nm) for the modes at 880 nm and 950 nm respectively.

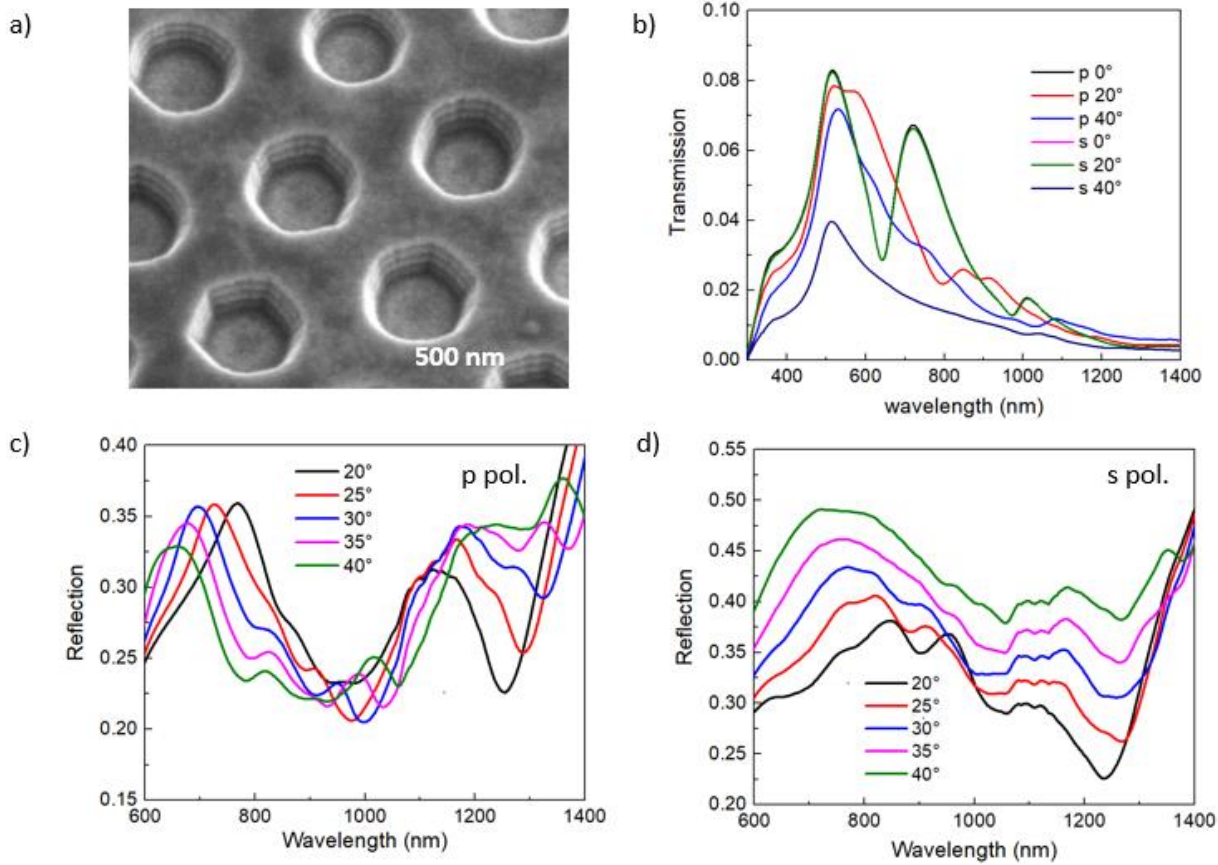

**Figure 3.3.** a) SEM image of a nanohole array in a HMM film made of 4 alternating units of Au (20 nm) and $Al_2O_3$ (20 nm) on a glass substrate. b) Transmission spectra acquired at different angles for both p an s polarized incoming light. c-d) Reflection as a function of the wavelength at various angles for p and s polarized incoming light respectively.

Due to intrinsic limitations of the available evaporator used for the deposition of the metallic films, a layer of 10 nm of Au was not a uniform, continuous layer but rather exhibited granularities as shown in Figure 2.2.b. This was the reason why no features were observed in the reflection spectra of nanohole arrays in HMM films made of Au/dielectric alternating units with 10 nm thickness of the Au layers. In fact, it was not possible to excite SPPs or BPPs as confirmed by reflection measurements performed for both p and s polarized light at various angles of incidence. In order to overcome this problem the thickness of the Au layers of the HMM multilayer was increased to 20 nm for each unit. As a result the quality of the films was improved and the nanohole arrays showed less indentations, as it is possible to see from the SEM image reported in Figure 3.3a. In this case the sample was fabricated on glass and four stacking units of Au (20 nm)/$Al_2O_3$ (20 nm) were deposited on the etched monolayer of PS spheres as described in Chapter 2. After the removal of the particles, a nanohole array in the HMM multilayer was left on the substrate. The diameter of the nanoholes is about 400 nm and the average distance between the centers of two of them is about 700 nm. The transmission spectra of the nanohole array in the HMM multilayer for both p and s polarized light (defined as the components of the incoming light which are parallel and perpendicular to the plane of incidence respectively) at various angles of incidence, from 20° to 60° are reported in Figure 3.3b. Here the peaks found in the transmission spectra are similar to the extraordinary optical transmission peaks observed from nanohole arrays as described in Section 1.2. In particular the peak at 780 nm was observed for both p and s polarized light at normal incidence. For s polarized light and for angle of incidence larger than 20° no features, apart from the peak at 510 nm, were observed. Instead for p polarized light many additional peaks were found in the transmission spectra by increasing the angle of incidence. In fact, the polarization of the incident light plays an important role for the excitation of the modes inside the HMMs since it is well known that they require p polarized incident light to be excited. The reflection spectra were acquired at various angles of incidence from 20° to 40° with a step of 5° for both p and s polarized light, as reported in Figure 3.3c-d. Multiple dips in the reflection spectra were observed for p polarized incident light whilst three main dips were observed in the reflection spectra for s polarized incident light only at the angles of incidence smaller than 25°. In fact, at angles of incidence smaller than 25° the reflection spectra for p and s polarized light are similar, then increasing the angle of incidence above 25° the reflection spectra are different for the two polarizations. The three main dips are centered at about 880 nm, 1000 nm and 1240 nm. Comparing this result with the simulated reflectance there is a significant red shift of the experimental dips that can be explained with the different thickness of the deposited Au films, which was 20 nm in the experiment whilst only 10 nm in the simulation. Moreover, the position of the peaks is affected by angle of incidence since they red-shift by increasing the angle of incidence. However, the possibility that the modes inside the HMMs are not properly excited by simple illumination with polarized light can not be completely excluded. It is important to note that the number of the dips in the experimental reflection spectra increases as a function of the angle of incidence. Perhaps this is due to the fact that the nanoholes in the HMM film are not cylindrical since the size of the nanohole in the first layer and the one in the last dielectric layer differ of about 90 nm. Indeed, the simulated reflection of the nanostructured HMM shows sharper dips compared to the ones observed experimentally. This discrepancy was caused by multiple factors. As already discussed, the uniformity of the deposited metallic and dielectric layers plays an important role because if the quality of the films is not good enough the propagation of the modes inside them is undermined. Furthermore, it is possible that a different geometrical design of the nanoholes in the HMM multilayer, namely the diameter/pitch ratio could facilitate the excitation/propagation of the modes inside the HMM. Nevertheless, even if the nanohole array requires further investigation, the nanostructured HMMs are still promising since the electromagnetic field is enhanced and confined close to the surface of the metallic films inside the holes the nanohole array in the HMM film could be used for biosensing applications probing the molecules inside the nanoholes. In the case of long molecules, such as DNA molecules, different portions of the same molecule residing in the channel could be detected at a given time. For this reason, one advantage of the presented nanostructured HMM compared to similar systems based on HMMs investigated in literature is the fact that does not require prism coupling or nanostructured metallic gratings to excite the modes of the HMM. On the other side, nanohole arrays in the HMM films still require further optimization of the geometrical parameters both for what concerns improvements of the thicknesses and uniformity of the deposited metallic/ dielectric units, and for the dimensions of the nanoholes and their periodicity. In this way, it could be possible to achieve the full excitation of the sharp modes in the HMM without any grating coupling mechanism, as predicted from the simulations.

# 4 Fabrication of ordered arrays of V-shaped nanopores

*In this Chapter it is presented the fabrication route of the second nanostructure investigated in this thesis, namely a free-standing Au coated nanoassembly of colloidal particles. The interstices of the template were modified in order to obtain an ordered array of plasmonic nanopores transferred on a supporting holey silicon nitride membrane.*

Historically, the first types of artificial nanopores were single solid-state nanopores on thin insulating free-standing membranes. Typically, these nanopores have been fabricated by drilling the pore with highly energetic electron or ion beams in a transmission electron microscope (TEM) or a focused ion beam (FIB). Various ions, including $He^+$, $Ne^+$, $Ar^+$ and $Ga^+$ have been used to produce energetic ion beams in FIB systems. Stein et al. implemented for the first time a technique, referred to as ion-beam sculpting, which is based on a bowl-shaped cavity realized by reactive ion etching (RIE) followed by exposure to an $Ar^+$ beam to remove material from the top of the membrane[148]. With the ion-beam sculpting technique symmetrical nanopores as small as 5 nm in diameter have been reported[149]. More in general, FIB systems have been widely employed to directly drill nanopores in free-standing membranes, which means that a specific location of the membrane is targeted avoiding the exposure of a large area. Drilling nanopores with FIB requires highly energetic ion beams and the tuning of the beam current according to the desired diameter and eccentricity of the pore. Nanopores drilled with FIB systems are usually conical with the top opening that is larger as a result of the continuous irradiation with the ion beam whose current has a Gaussian distribution. In general, the morphology of the nanopores can be controlled but small variations in the thickness of the free-standing layer can affect the diameter of the nanopore. The lower limit for the diameter is around 10 nm unless the source is a $Ga^+$ or $He^+$ ion beam, in which case it is possible to create smaller nanopores. As example, Gierak et al. used a 30 keV $Ga^+$ ion beam to fabricate nanopores with diameters of about 5 nm[150]. With the sub-nanometer probe of a He ion microscope nanopores with minimum diameters of 1.3 nm can be obtained. However, tuning some parameters of the ion beam, such as the energy, the temperature and the velocity, it is possible to control the transport of membrane material leading to the shrinkage of the nanopore. In this way, Li et al. proved that a nanopore with an initial size of 61 nm can shrink to 1.8 nm[148]. Moreover, other processes, such as thermal annealing can be performed to further shrink the size of the nanopore[151]. One constraint of FIB drilling is the integration of ions into the membrane during the ion beam exposure and as a result, a possible modification of the surface charge of the nanopore walls. For the fabrication of nanopores below 10 nm drilling techniques based on electron beams have become very popular. In fact, the focused electron beam (FEB) of a TEM has been used to manufacture nanopores as small as 2 nm with different shapes in both free-standing membranes $Si_3N_4$ and 2D materials, such as graphene and boron nitride (BN). Storm et al. developed a method to fabricate $SiO_2$ nanopores with single nanometer precision by combining standard silicon processing and a TEM microscope at an accelerating voltage of 300 kV[152]. A free-standing silicon membrane is thermally oxidized with a 40 nm layer of $SiO_2$. Then, electron beam lithography and RIE are used to create pyramid-shaped holes. The authors showed that an electron beam of high intensity in the range $10^5 - 10^7$ Am$^{-2}$ provides a way to shrink nanopores with an initial maximum diameter of 50 nm. For larger nanopores, the irradiation with the electron beam has the opposite effect since it grows in size. Therefore, it is possible to tune the diameter of the pore with the electron beam exposure by exploiting the surface tension effect in the $SiO_2$ layer and, simultaneously to check in real-time the fabrication. The shrinkage of nanopores with highly accelerated electron beams has been achieved in both TEM and SEM systems depositing specific materials, such as hydrocarbon compounds[153–155]. Electron beam lithography (EBL) combined with etching methods based on reactive ion etching (RIE) provides a strategy to fabricate an ordered array of nanopores with a diameter below 20 nm and high precision regarding the quality of the mask obtaining high quality shape and size distribution of the final nanopores array. Typically, a photoresist layer of poly(methyl methacrylate) is spin coated on the top of a $Si_3N_4$ layer, the pattern is defined by impinging the photoresist layer with EBL and then it is transferred to the membrane with an etching performed by RIE exposure. EBL masks of various materials and with different shapes, such as square or hexagonal shapes, have been successfully used by introducing additional lithographic steps[156–158]. Although the quality of the resulting masks is very high, the combination of EBL with RIE for the fabrication on nanopores relies on expensive lab equipment. Moreover, EBL patterning requires several hours for the realization of a mask and thus, is time-consuming for large scale manufacturing.

However, even if the fabrication of nanopores with ion/electron beams is well-established there are still some challenges. Firstly, the equipment in which such ions/electron beams operate are expensive and require trained operators. Moreover, it is time-consuming for large-area patterning and thus, other strategies have been explored in order to develop methods that are cost-effective and suitable for the fabrication of nanopores array. In fact, some applications, such as filtering and parallel sensing require the development of devices based on nanopores array on a large area of the substrate and not on a single nanopore. With this scope, the realization of nanopores with FIB or TEM technologies is not convenient because they are time-consuming and additional manufacturing steps are often required to complete the fabrication process. Various approaches, based on chemical solution etching, have been successfully implemented for the fabrication of nanopores on a large scale. Among others, a simple method is feedback chemical etching, which is based on the combination of a wet etching with a KOH solution and a mechanism of feedback, such as ionic current feedback or color feedback. As example, Chen et al. used a chemical etching to obtain a 14 x 14 nanopores array with an average diameter of 30 nm and a truncated-pyramidal shape[159].

Another method, known as ion-track etching, allows one to fabricate nanopore arrays on polymeric substrates[160,161]. Briefly, by accelerating heavy metal ions through a polymeric membrane it is possible to create tracks, which are etched with chemical solutions more rapidly than the remaining portions of the membrane. With this approach conical shaped nanopore arrays are usually obtained and thus, ion-track etching is a convenient method for the realization of devices which use properties typical of asymmetric channels, such as ion current rectification. The main limitations of this method are the impossibility to control the distribution of the nanopores array because it is not possible to predict the positions of the tracks, and the fact that a polymeric membrane is required, i.e. it is not possible to extend this fabrication process to other types of substrates. For the preparation of nanopores arrays in metal dioxide films the most popular method is based on electrochemical anodization[162]. In fact, self-organized nanoporous anodic alumina[163,164] or titania ($TiO_2$) films[165,166] can be easily obtained by using an electrochemical anodisation cell. Some parameters, such as the applied voltage and the concentration of the electrolyte solution can be tuned to change the diameter of the nanopores and their distribution. One of the main advantages of this approach is the high throughput since the nanopore arrays' density is around $10^9$-$10^{10}$ cm$^{-2}$. Metal-assisted chemical etching is another method to fabricate nanopores in Si substrate covered with metal particles[167–170]. The chemical etching is faster in the areas of the Si substrate where the metal particles are located resulting in the sink of the metal particles and the formation of the nanopores. In general, chemical etching solutions methods offer the possibility to fabricate nanopore arrays with high density in a cost-effective way, but present the problem that it is difficult to control the distribution of the nanopores array and thus, the production of ordered nanopore arrays is a challenge.

Another approach to fabricate nanopore arrays with short-range or long-range hexagonal order relies on nanosphere lithography (NSL). This method is based on various techniques, such as dip coating, interface assembly-techniques and spin coating, and produces a self-assembled hexagonally close-packed (hcp) 2D monolayer of colloidal nanospheres, or by direct drop casting of a solution you can obtain nanosphere arrays with short-range ordering. These arrays of particles are used as shadow masks or their interstices as windows for etching processes, such as RIE, to produce pores on the substrate. In the first case, the principal lithographic processes performed on the colloidal array are the etching of the spheres through RIE or other oxygen plasma treatments to shrink their diameter and obtain non-close packed ordered arrays of spheres, the deposition of materials, such as metals and dielectrics in high/ultrahigh systems, such as electron beam, thermal evaporation and ALD, and the removal of the colloidal beads. Importantly, the fabrication of nanoholes array on Si/glass substrates with NSL has been widely explored and the geometrical parameters, such as the diameter of the hole, and their reciprocal distance can be tailored controlling the etching process of the beads. In fact, Dahlin et al. implemented a protocol to realize nanopore arrays with a diameter below 50 on metal/insulator/metal films (MIM) and showed that the nanopores provide a way to excite surface plasmon polariton modes of MIMs[98]. Furthermore, the walls of these nanopores, which are made of three layers, could be functionalized selectively by using specific chemistry strategies to attach molecules or polymers only on a portion of their walls.

Although NSL has provided an effective fabrication method to realise ordered plasmonic nanostructures on a large area, it is difficult to produce nanopore arrays smaller than 50 nm and to perform the lift-off of the mask whilst preventing membrane cracks, apart from a few successful examples. To avoid problems related to the membrane breakage, 2D free-standing nanoassemblies have great potential because they can be suspended on "holey" substrates. In fact, 2D nanoassemblies are composed of particles of various materials, which are the elemental building blocks of the assembly. Then, the particles are self-assembled in periodic arrays by using interfacial self-assembly, spin coating and layer-by-layer self-assembly. Finally, neighboring particles are connected together by using a strategy to create reciprocal junctions, often by using a sintering process at a certain temperature or ligands attached on their surface. In this Chapter a method to prepare 2D Au-coated free-standing nanoassemblies based on the manipulation of the interstices of a hcp monolayer of PS spheres is presented. In this case the building block of the nanoassembly are PS spheres which are assembled with an interfacial self-assembly technique and connected together by using a thermal annealing. In Section 4.1 the procedure adopted for the realization of the Au-coated free-standing nanoassembly is described in detail. In particular, the technique used to form a self-assembly monolayer of PS spheres at the interface between the water-air interface of a water bath is discussed in the section 4.1.1. Then, a thermal annealing is performed to create junctions between the contact points of adjacent spheres. By controlling the temperature and the duration of the process it is possible to control the size of the junctions and induce a shrinkage of the interstices of the array as discussed in section 4.1.2. In fact, the manipulation of the interstices of the arrays allows one to obtain an array of circular nanopores. A Au layer was deposited to increase the robustness of the nanoassembly which, as a result of the deposition, exhibits plasmonic properties. The Au-coated nanoassembly is transferred on a "holey" insulating membrane by following the procedure described in section 4.1.3. Finally, a detailed morphological characterization of the Au-coated nanoassembly before and after the transfer on the holey membrane is reported in Section 4.2. From SEM images of the sample it is possible to understand how to tune the parameters of the thermal annealing to control the final shape and size of the nanopores. Furthermore, it gives an indication of the quality of the transfer of the Au-coated nanosheet on the holey membrane.

## 4.1 Preparation of V-shaped nanopores from a 2D free-standing nanoassembly

The fabrication of V-shaped nanopores directly integrated on a 2D Au-coated free-standing nanoassembly is based on the approach that is sketched in Figure 4.1.

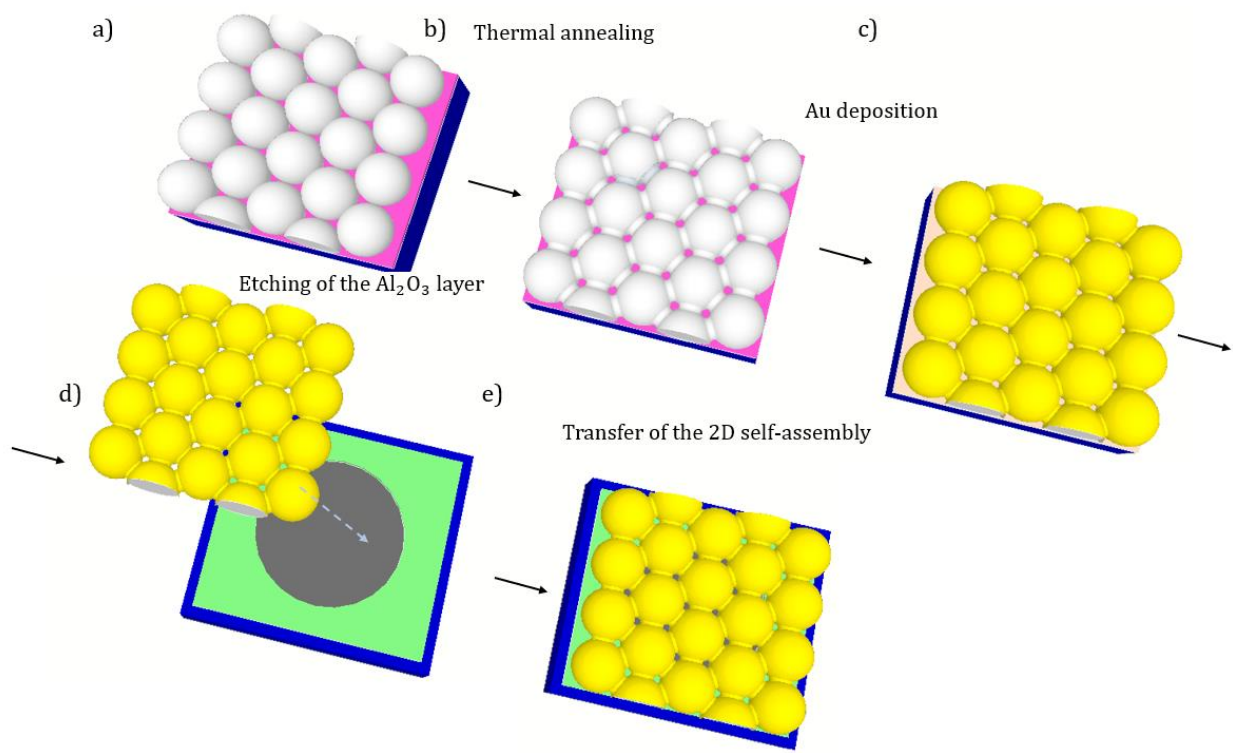

**Figure 4.1.** Sketch of the preparation of a 2D Au-coated free-standing nanoassembly. a) Firstly, an hcp monolayer of PS spheres is formed on a substrate covered with a sacrificial layer by using an interfacial self-assembly technique. b) A thermal annealing is performed to cross-link the junctions between adjacent spheres. c) A Au layer is deposited with a sputtering system. d) The Au-coated nanossembly is detached from the sacrificial layer by performing a chemical etching. e) Then, the free-standing nanoassembly is transferred on a "holey" $SiN_x$ membrane, namely a membrane patterned with holes of few microns.

The idea is to build a transferrable nanoassembly in which PS spheres, that are organised in a hcp monolayer, are the building blocks. After their assembly each PS sphere has six contact points with as many spheres and thus, with a controlled thermal annealing it is possible to induce the cross-linking of these contact points and form junctions between neighboring particles. Due to the melting of the junctions of the spheres, the interstices of the array are reduced in size and re-shaped into circular apertures. This is a way to connect the spheres of the monolayer and form a single compact nanoassembly, whose mechanical robustness can be improved by depositing a film, in this case a Au film, on top of it. The nanoassembly is easily detached because the PS spheres can be assembled on any kind of substrate including one with a sacrificial layer. Then, it is transferred at the air/water interface of a water bath, so that the floating nanosheet made of the Au-coated nanoassembly of PS spheres can be re-transferred on any type of substrate. In this case, the substrate is a holey free-standing $SiN_x$ membrane, namely a membrane patterned with an array of holes of few microns, so that after the transfer a free-standing Au-coated nanoassembly covers all the membrane including the micron-sized holes that are present in the membrane. The preparation of the free-standing nanoassembly is described in detail below.

Firstly, silicon (Si) substrates were cleaned with a triple-step procedure of 10 min each sonication in acetone, isopropanol and water, dried with a nitrogen gun. Then, a sacrificial layer of $Al_2O_3$ was deposited by an atomic layer deposition (ALD) system (final thickness: 30 nm, temperature: 80 °C, Oxford Instruments FlexAL). Then, an interfacial self-assembly technique was used to create a hcp array of PS spheres on top of the sacrificial layer.

### 4.1.1 Preparation of a hcp monolayer PS spheres with an interfacial self-assembly technique

A stock solution of negatively charged PS spheres (monodispersed in 5 wt % aqueous solution) with a nominal diameter of 140 nm were purchased from microParticles GmbH. The technique used to assembly the PS spheres in a hcp array is an interfacial self-assembly technique, which allows one to create monolayer of PS spheres at the air-water interface of a water bath, compact the ordered grains and transfer them on any kind of substrate. At this purpose, a clean 4x4 $cm^2$ Si wafer was exposed to an $O_2$ plasma treatment for 10 min (power: 100 W, gas flow rate: 20 sccm) to make the surface completely hydrophilic. Then, the substrate was suspended into a glass water bath with an angle of about 80° and the bottom edge of the substrate below the water surface. The PS spheres stock solution was diluted in ethanol (ratio 1:1) and dropped with a pipette on the Si wafer. In this way, the solution containing the particles spreads on the Si wafer forming small grains of ordered monolayers of PS spheres and the excess flows into the water bath. The solvent was dried naturally by evaporation, then the substrate was removed from the clamp used to suspend the wafer on the first water bath and slowly immersed into a second plastic bath. Doing so, the PS monolayers on the Si wafer are transferred at the water-air interface of the plastic bath. Previously, the pH of this bath was increased from 7 to 9 by the addition of a sodium hydroxide solution. The higher pH increases the compression between adjacent monolayers and encourages the formation of larger grains of ordered spheres. Repeating this procedure many times, i.e. pipetting the particles solution on the Si wafer, which is suspended above the glass bath with a clamp as described before and transferring the small ordered grains of particles that are formed on top of it into the water-air interface of the second plastic bath, a floating PS sphere monolayer covering all the area at the interface was formed. The compression between adjacent grains of hcp PS spheres was further increased by injecting at the interface 2 μL of a diluted solution of sodium dodecyl sulfate (20% in DI water). Then, Si substrates coated with the sacrificial layer were hydrophylized with an $O_2$ plasma treatment for 3 min (power: 100 W, gas flow rate: 20 sccm), immersed vertically in the plastic water bath, slowly placed horizontally beneath the water-air interface and carefully lifted in order to collect the PS monolayer. Finally, the sample was dried naturally at an angle. After the drying a hcp monolayer of PS spheres was formed on top of the sacrificial layer as represented in Figure 4.1.a).

### 4.1.2 Thermal annealing on a hcp monolayer of PS spheres

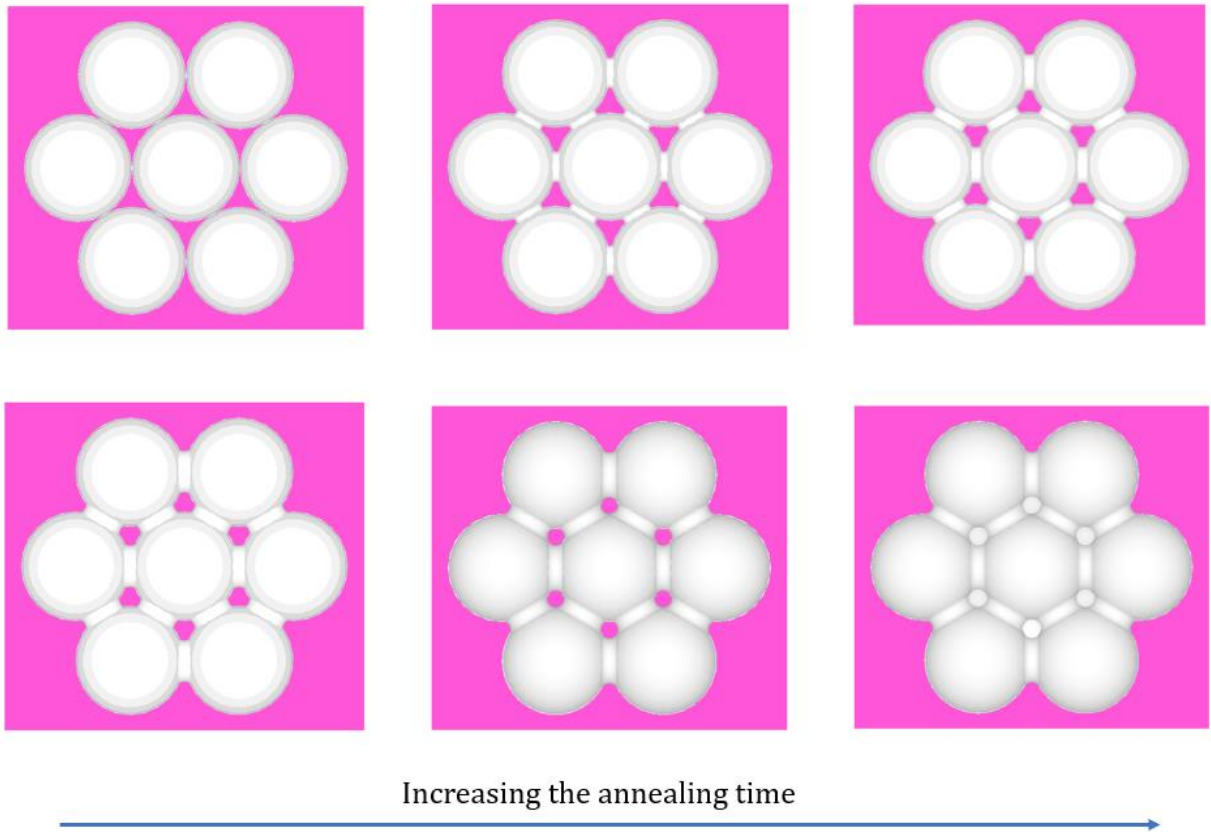

Increasing the annealing time

**Figure 4.2.** Sketch of the structural modification of the hcp array of PS spheres over time during the thermal annealing. The process is performed at the fixed temperature of 117 °C: at the beginning it induces the formation of cross-linked junctions between the contact points of neighboring particles. Then, the junctions become larger and, as a result the interstices reduce in size until they lose even their triangular shape and become circular. If the time is increased even

further the interstices are clogged by fused polystyrene since the junctions expand until the point that adjacent junctions also cross-link together.

Then, a thermal annealing on the hcp array of PS spheres was performed by placing the sample on a hotplate at 117 °C, which is higher than 105 °C, approximately the typical glass transition temperature of polystyrene. The duration of the thermal annealing was a critical parameter because after the first ten seconds the contact points between adjacent particles cross-linked forming junctions acting as bridges between the particles (see Figure 4.1.b). Prolonging the heating time for few seconds more it was possible to tune the length of these junctions, and as a result, also the reduction in size of the interstices of the array. However, prolonging the heating time even longer caused the melting of the spheres and a complete clogging of the interstices as well (see Figure 4.2). If not explicitly specified in this Chapter, the set heating time was 12 s, which allowed the formation of the junctions between the particles and the shrinkage of the triangular interstices of the array into circular apertures of about 13 nm. Then, the sample was rapidly removed from the hotplate and placed on an Al cube, which was previously cooled down in a fridge at 4 °C. The rapid cooling of the sample guaranteed that the monolayer of PS spheres was not affected by any other thermal effects due to residual heat.

### 4.1.3 Coating and transfer of the 2D nanoassembly of PS spheres

After the thermal annealing, the sample was covered with a Au layer (see Figure 4.1.c) from a sputtering deposition system (thickness: 40 nm, rate: 22 nm/min, Quorum Sputter coater Q150T ES). The deposition of Au was fundamental from two points of view: firstly, it increased the mechanical robustness and rigidity of the nanoassembly preventing the breaking of the junctions during the transfer process from the substrate; secondly, it was responsible for the optical properties exhibited by the free-standing nanoassembly.

The detachment of the Au-coated nanoassembly from the sacrificial layer was accomplished by soaking the sample in a chemical etchant solution. In fact, the $Al_2O_3$ layer was etched in an aqueous potassium hydroxide solution (6% w/v in DI water; KOH pellet from Sigma-Aldrich). Then, the sample was immersed vertically and slowly in a plastic water bath in order to release the Au-coated nanoassembly to the air-water interface.

Then, a $Si_3N_4$ membrane with a thickness of 100 nm was prepared and loaded in the SEM-FIB system to pattern in the center of the membrane periodic holes with a diameter of 2 µm (voltage: 30 kV, current: 25 nA; FEI Helios NanoLab 650) in order to realise a "holey" membrane as second substrate on which to transfer the Au-coated nanoassembly. In order to make the surface of the membrane completely hydrophilic a long oxygen plasma treatment was performed (exposure time:

7 min, power: 100 W, gas flow rate: 25 sccm). Then, the membrane was vertically immersed in the plastic water bath and carefully rotated until it was parallel to the water-air interface of the bath. Then, the Au coated nanoassembly was collected (see Figure 4.1.d) and the excess of water gently removed preventing any displacement of the nanoassembly with respect to the center of membrane. In this way the membrane is covered by the macroscopic nanosheet formed by the Au-coated nanoassembly and in particular, the micron-sized holes on the membrane are covered with the free-standing nanoassembly as well (see Figure 4.1.e).

## 4.2 Morphological characterization of the interstices of the hcp monolayer

The parameters for the thermal annealing, namely the temperature and the duration of the process, were adjusted by imaging with the SEM-FIB system the monolayer of PS spheres afterwards. In this way, it was possible to monitor any modification of the interstices of the array and the contact points between neighboring particles and tune the parameters of the thermal annealing until the interstices have the desired size and shape.

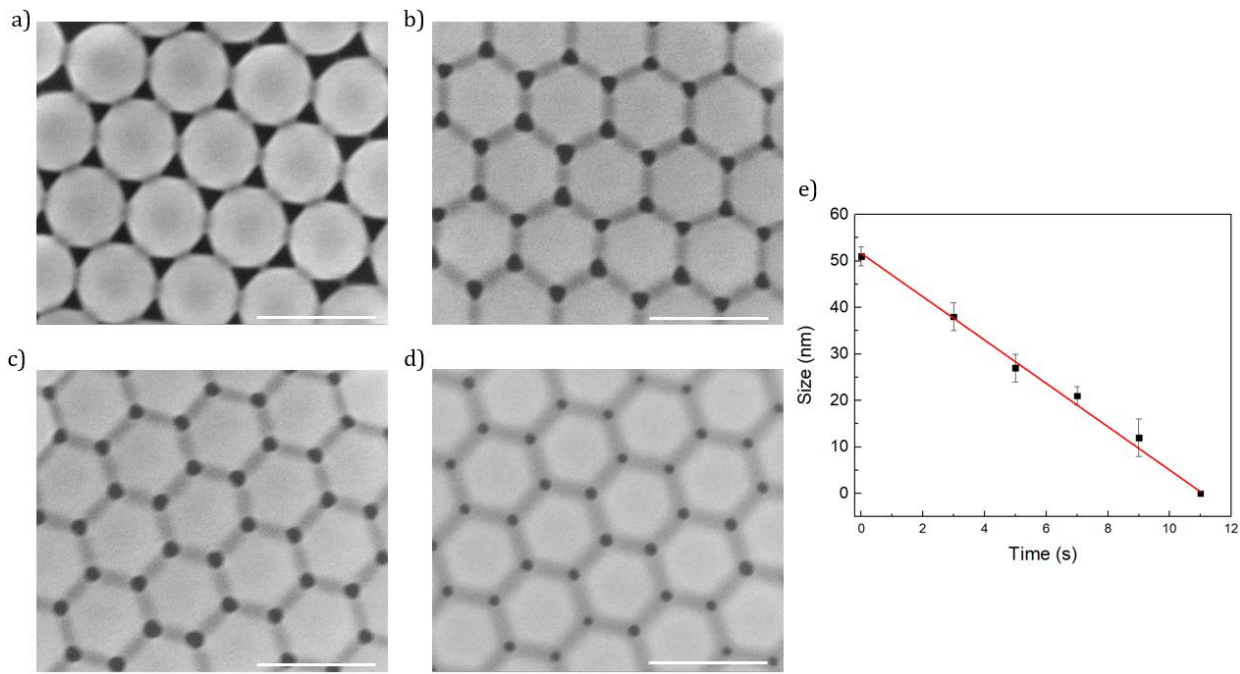

**Figure 4.3.** SEM images (scale bar: 200 nm) of hcp arrays of PS sphere after the thermal annealing at 117°C performed increasing the duration of the process: a) 3 s, b) 5s, c) 7s and d) 9s respectively. e) Plot of the average size of the interstices as a function of the duration of the thermal annealing.

In Figure 4.3.a-d SEM images of a hcp PS monolayer treated with a thermal annealing at 117 °C for different durations of the process are reported. After the first 3 s the contact points of the particles melt together forming junctions, which become more extended when the treatment is performed for longer. As a result, the triangular interstices of neighboring particles are reduced in size and also their shape is modified, as is well visible from the SEM images. In Figure 4.3.e a plot of the size of the interstices as a function of the annealing time is shown. The PS monolayer before any treatment exhibits triangular interstices of about 51 nm, then increasing the annealing time they shrink into circular apertures. The rate of the shrinkage estimated from the analysis of the SEM images reported in Figure 4.3.a-d is 4.6 nm·s$^{-1}$. In particular, after 9s of treatment the interstices are circular in good approximation and reduced on average to 12 nm. Performing the annealing for longer causes the total blockage of the interstices since adjacent PS spheres appear completely connected.

In order to realise the Au nanoassembly the thermal annealing was performed for 8 s at 117 °C in order to modify the interstices to circular apertures of about 15 nm. Then, 40 nm of Au were deposited as described in Section 4.2. After the deposition the top hemisphere of the PS spheres shows a Au cap and the junctions between adjacent spheres are strengthen by the Au coating as well. SEM images of the top-view and the cross-section of the Au-coated nanoassembly before the transfer are reported in Figure 4.4.a-b respectively. As a result of the thermal annealing and the Au deposition the array exhibits an array of periodic Au nanoholes (see Figure 4.4.a). From the analysis of Figure 4.4.b the estimated height of the Au caps is about 50 nm. The accumulation of Au particles on the sacrificial layer exactly where the interstices of the array are located proves that after the thermal annealing the interstices are completely open and that there is not melted polystyrene at the bottom of the array.

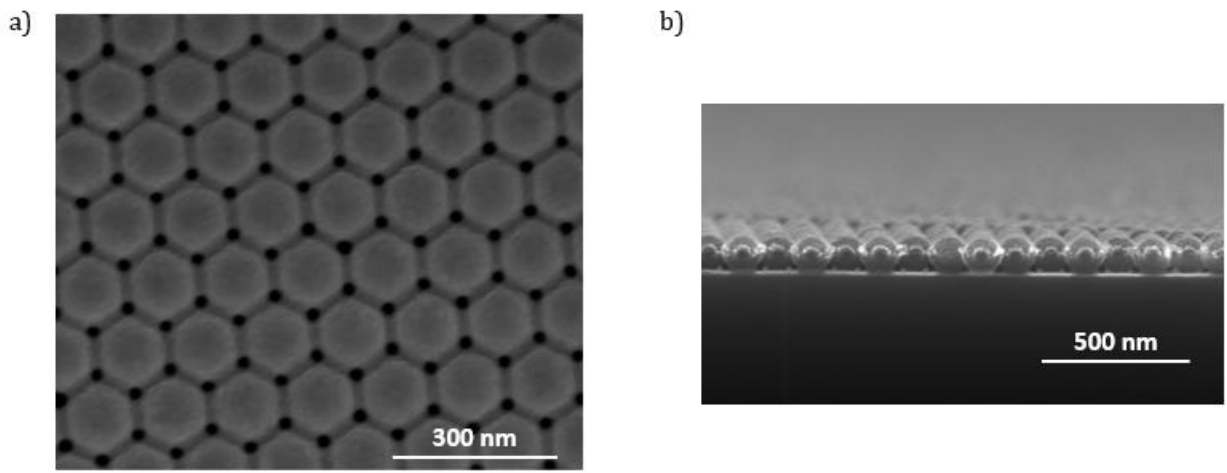

**Figure 4.4.** SEM images of the Au-coated nanoassembly before the transfer onto the holey membrane: a) top-view and b) cross-section of the same sample.

After the transfer the membrane is covered with the Au nanosheet. Since the membrane is patterned with holes of about 2 µm the Au-coated nanoassembly is suspended on top of them. In this way it is possible to use the modified interstices as an ordered array of nanopores. In order to investigate the appearance of the Au-coated nanoassembly suspended on the holes of the membrane a SEM image of the flipped membrane is shown in Figure 4.5a.

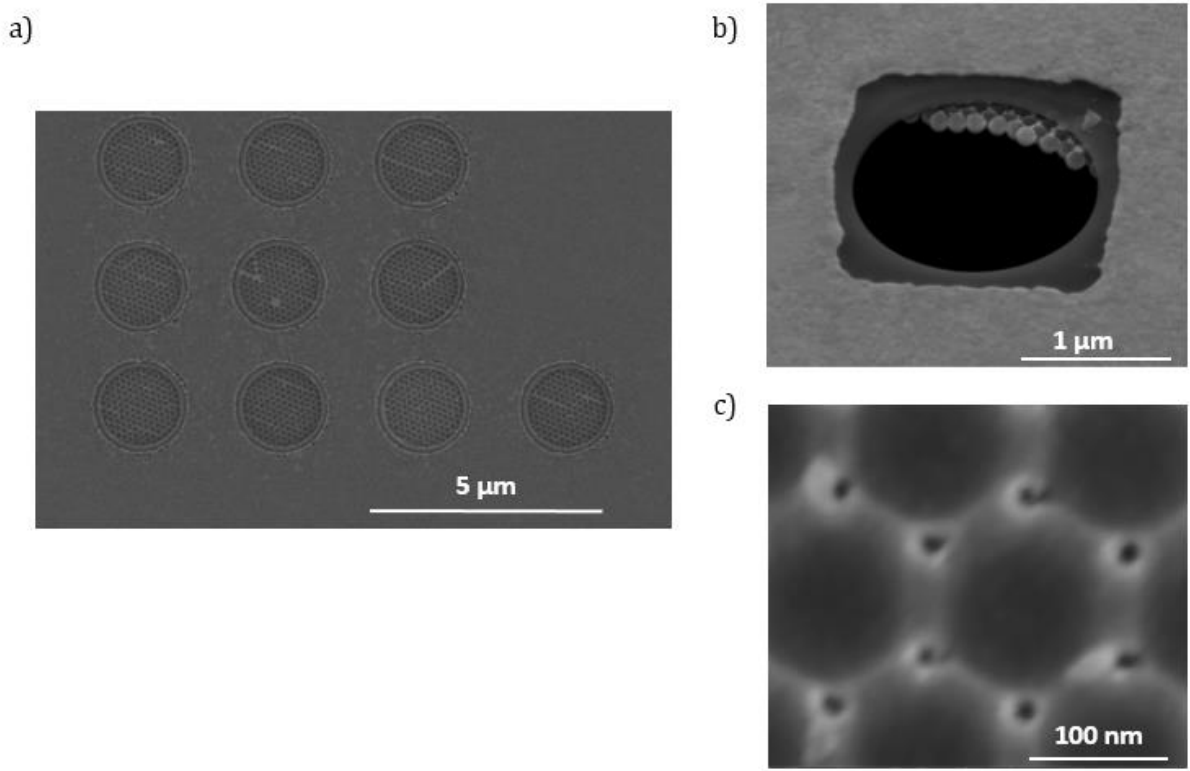

**Figure 4.5.** a) SEM image of the Au-coated nanoassembly after the transfer on the holey membrane. The image is acquired loading a flipped membrane in the SEM chamber. In this way it is possible to prove that the nanosheet covers also the patterned holes on the membrane. b) Tilted view-of a micron-sized hole partly covered with the free-standing nanoassembly to show clearly the transfer process. c) A magnification of the Au-coated nanoassembly suspended on one of the holes shows the quality of the free-standing modified interstices.

The presence of the Au nanosheet is clear, as well as the fact that it is not damaged after the transfer and that there are no empty areas. A tilted-view SEM image of a micron-sized hole in the membrane only partly covered with the free-standing nanoassembly is reported to clarify the concept beyond the fabrication method and the transfer process. In Figure 4.5c is shown a magnification of the Au nanosheet suspended on one of the micron-sized holes of the insulating membrane. In this way it is possible to investigate the features of the nanoassembly from a bottom-view. The presence of the PS spheres is evident and it is also possible to appreciate the presence of a Au crown around each interstice of the array. Here, since the nanoassembly is suspended on the membrane these modified interstices can be used as plasmonic nanopores. Due to the fact that the nanopores are V-shaped because they are formed from the annealing of PS spheres, the size of the nanopores at the top of the array is larger than the one at the bottom. In fact, the diameter estimated for the nanopores at the bottom of the array is about 10 nm.

# 5 Results and Discussion

*In this Chapter the optical properties of the plasmonic nanopores supported on the free-standing Au-coated nanoassembly are characterized. Furthermore, the device is tested as biosensing platform for SERS measurements in flow-through configuration.*

As discussed in Chapter 2 and Chapter 4, a hcp monolayer of PS spheres can be manipulated performing various processes, such as etching, deposition of materials, thermal annealing and surface decorations, in order to modify the geometrical parameters of the template, such as the shape of the PS spheres and the interstices formed by neighboring particles, and/or to provide additional functional properties, such as by the addition of metallic/dielectric coatings or by surface functionalization of the particles. In the case of the Au-coated nanoassembly fabricated following the procedure described in Chapter 2, after the thermal annealing and the Au deposition, the PS spheres were connected by the junctions formed by the contact points of adjacent spheres and were coated with a Au cap and metallic bridges on the junctions. The optical properties of the 2D Au-coated nanoassembly were measured after the transfer onto a $Si_3N_4$ membrane patterned with micron-sized holes. In this way, it is possible to eliminate the contribution to the optical response coming from the Au nanoparticles accumulated on the substrate in correspondence with the interstices of the array, as a result of the Au deposition. In fact, after the transfer onto the "holey" membrane, a plasmonic free-standing 2D nanoassembly exhibiting periodic nanopores was obtained. Therefore, the free-standing 2D nanoassembly covered the micron-sized holes patterned on the $Si_3N_4$ membrane. In this way, it was possible to collect the optical transmission at normal incidence directly from the plasmonic free-standing 2D nanoassembly. At this purpose, the transmission was measured with an ellipsometer (J.A. Woolam Co. V-VASE ellipsometer) in the wavelength range 300-1200 nm at the normal incidence. Furthermore, it was possible to also measure the transmission through the nanopore arrays by illuminating the sample in the opposite direction, namely the light was sent one time to the front side of the sample and another time to the back side. As shown in Figure 5.1.a, the resulting transmission spectra are identical for what concerns the spectral positions of the peaks and differs only for the intensity since the transmission collected illuminating the back side is lower than the one measured illuminating the front side. Indeed, the optical properties of a metal coated monolayers obtained by depositing a metallic film on a template of self-assembled spheres have already been reported in literature[171,172]. In the transmission spectra there are three bands: the main one centered at 515 nm and the other two supplementary bands centered at longer wavelengths, 600 nm and 890 nm respectively. The origin of the EOT bands and the dips observed in the transmission spectra of Figure 5.1.a have already been observed on similar metal coated micro/nanosphere templates and their origin has been theoretically predicted taking into account the periodicity of the template and the other main parameters, such as the thickness of the metallic layer and the material composition of both the spheres and the coating[171,173–175].

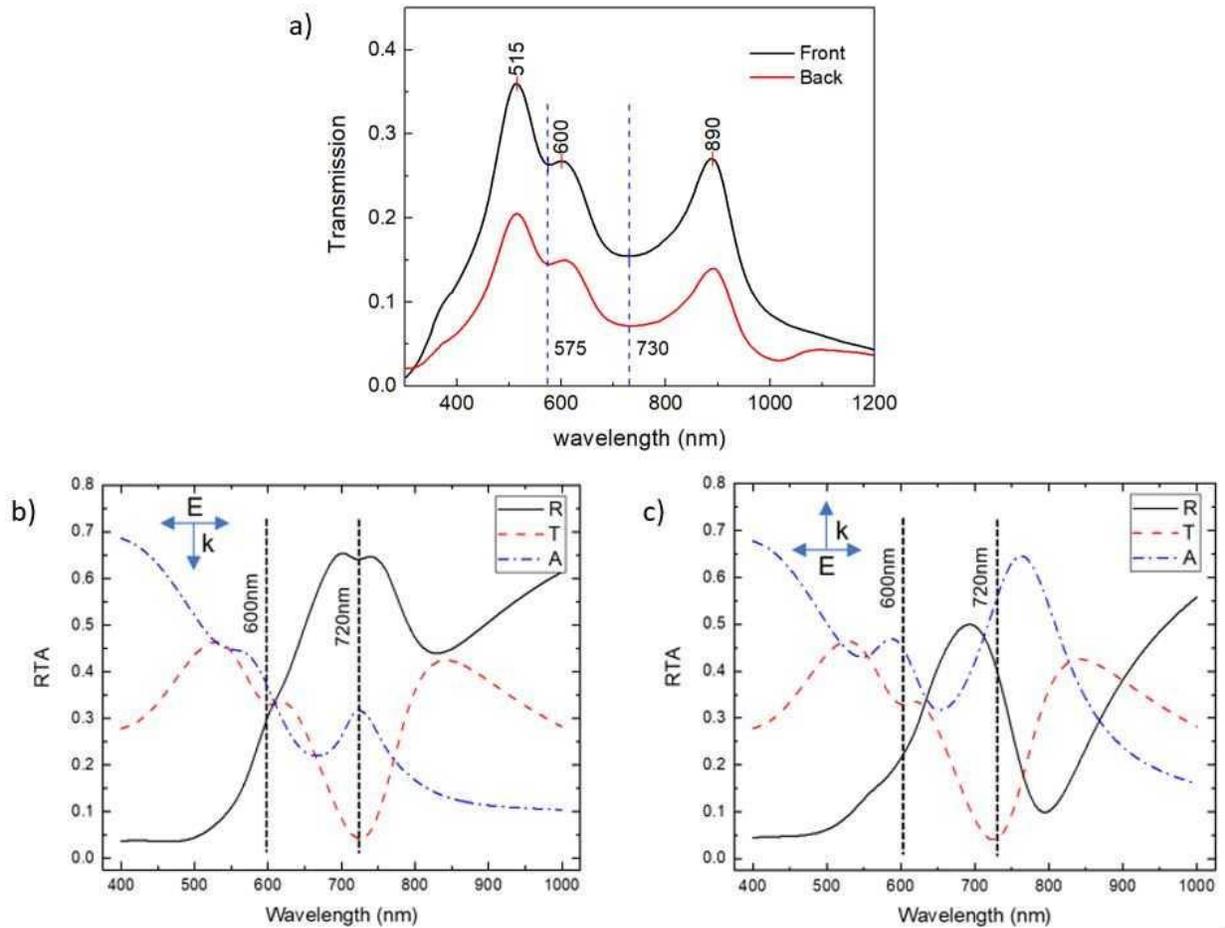

**Figure 5.1.** a) Experimental transmission spectra of the 2D free-standing plasmonic nanoassembly with the incident light impinging at normal incidence to the front side and the back side of the sample (black curve and red curves respectively). b)-c) Simulations of the optical properties of the Au-coated free-standing nanoassembly for the light impinging at the normal incidence to the front and back side of the sample respectively.

In Figure 5.1.b-c simulations of the optical properties of the Au-coated nanoassembly considering both cases of light impinging on the front and back side of the sample are reported respectively. The main features of the experimental transmission spectra are clearly reproduced, apart from small shifts of the peak and dip positons that are attributed to possible small discrepancies between the real thickness of the tips of the Au caps and the value used for the simulations and to some defects of the fabrication. The fact that the transmission intensity is lower when the light illuminated on the back side of the sample could be related to the absorption of light from the PS spheres of the 2D free-standing nanoassembly. In fact, their contribution is not included in the optical simulations, that takes into account only the nanostructured metallic template, namely the periodic Au caps connected by the Au bridges between adjacent spheres and the circular apertures, without the presence of the underneath PS template. In particular, the optical simulations were carried out taking into account the experimental geometrical parameters of the Au-coated nanoassembly (length and height of the Au cap: 140 nm and 50 nm respectively, diameter of the pores: 15 nm and thickness of the deposited Au film: 40 nm). It is interesting to observe the simulated electric field distributions of the Au-coated nanoassembly at 720 nm, the wavelength of the second minimum in transmission, for the light impinging the Au surface from the front side and the back side, as shown in Figure 5.2.a-b. In fact, if the light comes from the front side of the nanostructure the maximum enhancement of the electric field enhancement at the 720 nm is localized on the top of the metallic cap whilst if the light comes from the back side the electric field is highly enhanced and localized in the tips of the metallic cap.

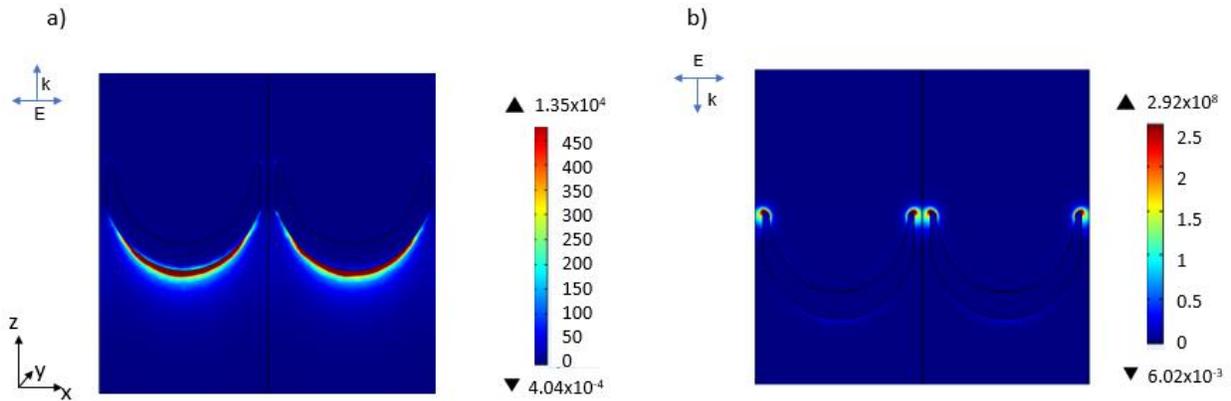

**Figure 5.2.** Simulated electric field distribution of the Au-coated nanoassembly at the wavelength of the second transmittance minimum. In a) and b) the incoming light is impinging the Au surface from the front side and the back side respectively.

## 5.1 Self-assembly of 4-ATP on the Au surface to test the SERS activity

The 2D plasmonic free-standing nanoassembly can be used as a sensing platform for optical detection based on SERS. It is possible to use this plasmonic nanomembrane for the detection of biomolecules that are adsorbed or bound on the surface, that is a widely used approach for sensing biomolecules from SERS substrates, or use the nanopores to deliver molecules through them and detect them in a flow-through configuration. To test the effectiveness of the 2D plasmonic free-standing nanoassembly as a SERS substrate a functionalization with 4-ATP molecules was performed. Specifically, a solution of 1 mM of 4-amino-thiophenol (4-ATP) in ethanol was incubated on the sample for 12 hours. As a result of the incubation, a self-assembled monolayer (SAM) of 4-ATP molecules was formed on the Au surface and the excess was washed away rinsing the sample several times in ethanol. Finally, the sample was dried with a nitrogen gun. The formation of a SAM on the surface of the 2D plasmonic free-standing nanoassembly is realized through covalent bonds between the Au surface and the thiol head group of the 4-ATP molecules. SERS spectra were collected with a Renishaw inVia Raman spectrometer with a 50x objective illuminating the sample with a 785 nm laser (power: 10%, exposure time: 1 s). In Figure 5.3a it is reported the SERS spectrum of 4-ATP obtained from the average of 120 spectra. The main Raman features of the 4-ATP are well-visible from the collected spectra and comparable with the ones reported in literature. In fact, 4-ATP molecules have been widely used in Raman Spectroscopy to characterize SERS substrate and thus, the vibrational fingerprints are well known[176,177]. In particular, a strong enhancement of the main vibrational bands of 4-ATP centered at 1004, 1076, 1142, 1169 and 1579 $cm^{-1}$ was observed. These peaks are related to specific vibrational modes of 4-ATP: for instance, the main spectral peaks at 1076 $cm^{-1}$ and at 1579 $cm^{-1}$ correspond to the S-C stretching vibration and to the aromatic ring chain vibrations respectively. In order to check the reproducibility of the Raman signals of 4-ATP molecules adsorbed on the 2D free-standing nanoassembly, SERS spectra were collected from each plasmonic well, namely from each micron sized hole in the membrane covered with the 2D plasmonic nanoassembly. Therefore, Figure 5.3.b) shows the average SERS spectra obtained from the analysis of the collected SERS signals (120 spectra collected from each well) from three micron-sized holes labelled as (0,0), (0,1) and (0,2). The main features of the spectra, such as the spectral peak positions, the width of the peaks and their intensity are all similar and no significant differences are noticed, meaning that the signals from different areas of the sample are reproducible. This is an important aspect because this platform has potential for parallel sensing since it is composed of independent plasmonic wells that, in prospective, could be functionalized with different receptors, for example with light activation of antibodies, and enable the discrimination of various molecules from the same biological sample. Furthermore, since the modified interstices of the 2D plasmonic nanoassembly provide an array of nanopores, this platform can work in flow-through configuration as will be described in the next section.

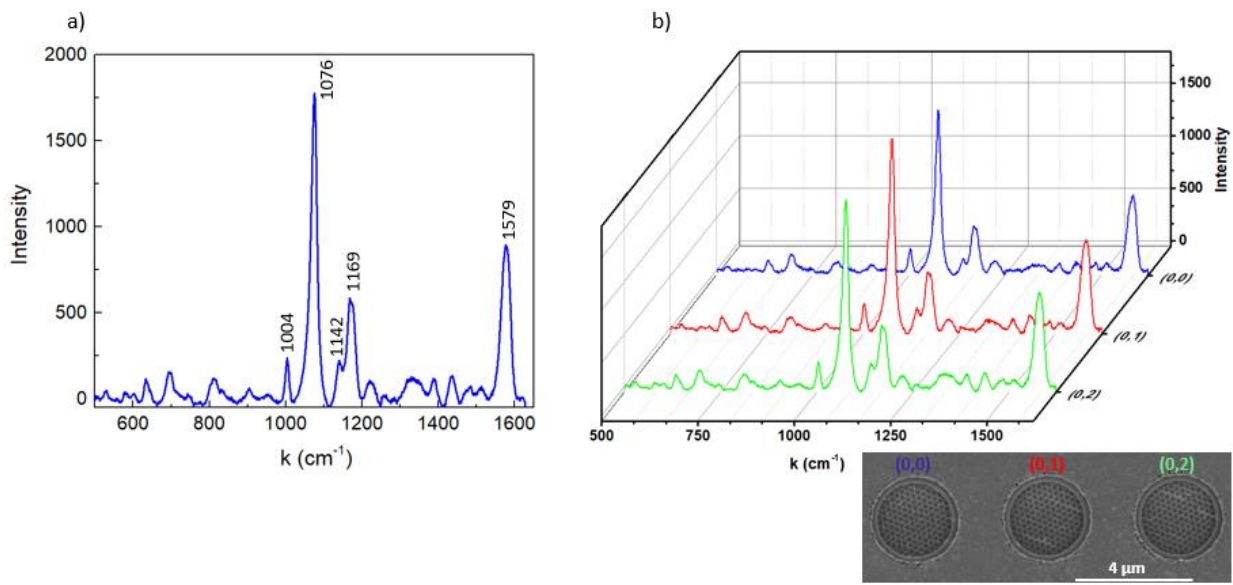

**Figure 5.3.** a) SERS spectrum of 4-ATP adsorbed on the Au surface of the free-standing nanoassembly. It results from the average of 120 collected spectra that have been processed, subtracting the baseline. b) Average SERS spectra obtained from different areas of the sample, namely from three micron-sized holes covered with the 2D free-standing nanoassembly and labelled as (0,0), (0,1) and (0,2) respectively. The inset refers to a SEM image of the three micron-sized holes in the membrane patterned with the plasmonic nanoassembly.

## 5.2 Detection of λ-DNA molecules in a flow-through configuration

In order to perform flow-through measurements, the chip with the 2D free-standing nanoassembly needs to be properly encapsulated and sealed to form two separated compartments. In this way, the molecules are injected in the cis compartment, usually encouraged to pass through the pores of the template by applying an external electric field and detected from the trans compartment when they reside in the plasmonic hotspots through the enhancement of the Raman signal. The procedure used to encapsulate the sample, the setup used to detect the molecules in flow-through and the analysis of the collected signals is reported below.

### 5.2.1 Encapsulation of the chip and preparation of the cis/trans compartments

The chip with the plasmonic free-standing nanoassembly was encapsulated in a microfluidic chamber made of polydimethylsiloxane (PDMS, Dow Corning SYLGARD 184 silicone elastomer) and cured at 65 °C for 2 h. In this way it was possible to build two fluidic compartments filled with an electrolyte solution of lithium chloride (LiCl, 1 M).

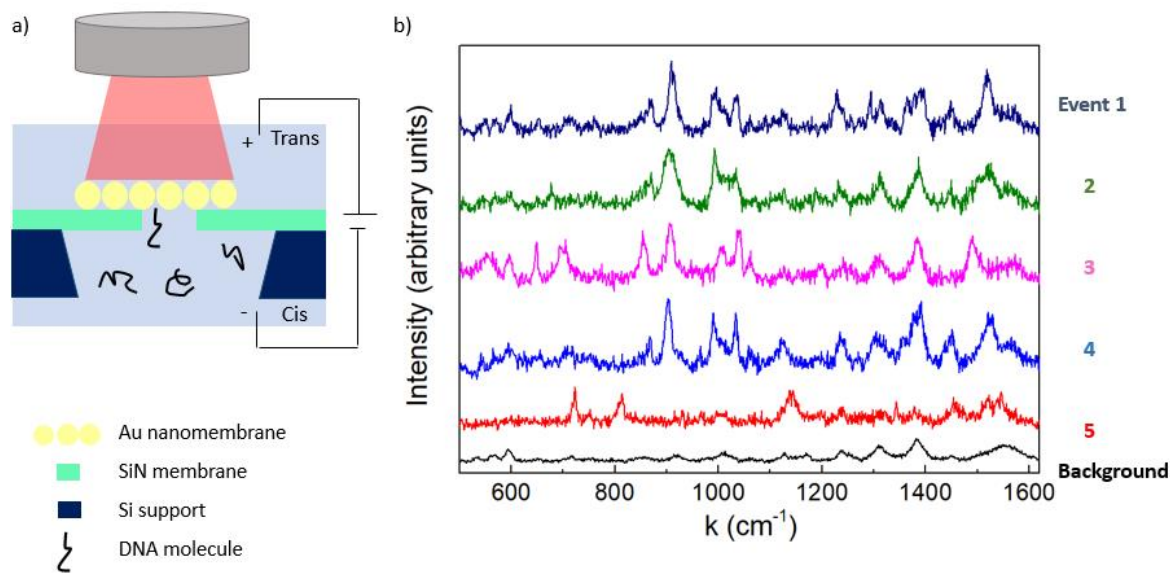

**Figure 5.4.** a) Sketch of the setup used to collect the SERS signals of λ-DNA molecules introduced in the cis chamber and passing through the plasmonic nanopores. The molecules are delivered to the nanopores with an applied electric field. b) Examples of collected SERS spectra over time after the introduction of λ-DNA in the cis compartment in the presence of the external bias of 30 mV. The black curve represents the background signal and its features of low intensity are related to the presence of glycerol in the cis chamber and the polystyrene beads underneath the Au caps of the array. The other spectra, labelled from 1 to 5, show extra features that are related to the passage of the molecules through the pores.

In order to increase the wettability of the surface and in particular of the nanopores, both the front side and the back side of the samples were exposed to an $O_2$ plasma treatment for three minutes (power: 100 W, gas flow rate: 25 sccm). Then, a solution of glycerol (1% in concentration in 1xTris-boratr-EDTA buffer) was heated up to 80 °C until it turned liquid, left to cool down only for few minutes and then, 10 μL of this viscous solution were injected in the cis chamber. After its complete cooling a gel with a thickness of about 0.5 mm was formed in the cis compartment underneath the membrane. The glycerol solution was injected into the cis compartment to increase the viscosity of the media forming a gel with randomly distributed pores of different sizes.

However, the solution of glycerol should not be added in the cis compartment when it is completely liquid because it could penetrate through the plasmonic nanopores of the free-standing nanoassembly and interfere with the SERS signal due to the DNA-Au interactions. Indeed, the glycerol can come in contact with the back side of the Au nanocaps and their interaction can give rise to some features with low intensity that represent a constant background during the collection of SERS signals over time, as discussed in more detail below.

The target molecules were λ-DNA molecules (5 µg/mL in Tris-EDTA solution) which were introduced in the cis compartment. Therefore, in order to be detected by SERS the molecules had to move towards the glycerol solution, which was a viscous media added to slow down the motion of the DNA molecule. In fact, the SERS signal from the molecules translocating through the pores was collected for 6 min with an integration time of 0.5 s, a 785 nm excitation wavelength and a 60 x 0.95 NA water immersion objective (see the sketch in Figure 5.4.a). In order to encourage DNA molecules to pass through the nanopores two Ag/AgCl electrodes were placed in the cis and trans compartment and an electric bias of 30 mV was applied for a specific time. In fact, it has been reported in literature that the application of an electric bias not only enables us to drive DNA molecules through the pores but also induce the linear state of the molecule preventing the formation of folded states[178,179]. Therefore, during the collection of Raman signals over time the bias was switched on and off each 90 s in order to study both the electrophoretic and the diffusion regime.

### 5.2.2 Analysis of the collected spectra

The collected spectra were analyzed both for the electrophoretic regime and the diffusion regime. For what concerns the measurements collected during the application of the external bias significant variations of the spectra, namely the appearance of extra vibrational bands in the spectra that are due to the passage of DNA molecules through the nanopores close to the Au surface. However, due to the presence of the glycerol gel near the nanopores openings some peaks with low intensity were observed in all the collected spectra (as shown in the black curve of Figure 5.4.b). These features were not caused by the passage of molecules through the nanopores but rather by vibrational modes of the glycerol[180,181]. From the assignment of the Raman peaks of glycerol reported in literature, the peaks at 590, 1011, 1240, 1310 and 1384 cm$^{-1}$ correspond to vibrational modes of glycerol, which was injected in the cis compartment when still warm to form a thin layer of gel underneath the plasmonic 2D free-standing nanoassembly. However, a minor contribution to the spectral bands at 600, 1011 and 1130 cm$^{-1}$ could also have arisen due to residues of the PS template in contact with the Au caps[182]. Therefore, the passage of the DNA molecules through the nanopores was detected monitoring the SERS signals over time due to the appearance of new SERS bands and/or significant modifications of the ones corresponding to the background signal. As represented in Figure 5.4.b some SERS spectra collected over time show features that are not observed in the background signal (black curve) and that correspond to vibrational bands related to the secondary structure of λ-DNA molecules that are encouraged to translocate through the nanopore under an applied voltage of 30 mV with a duration of 90 s. In fact, many peaks found in the collected spectra show features that correspond to the main vibrational modes of λ-DNA molecules[183–185]. The external structure of the λ-DNA, namely the deoxyribose and DNA backbone give rise to the main spectral features of the collected spectra with peaks centered at 670, 910, 1045, 1380 and 1456 cm$^{-1}$. In particular, the peak at 820 and 1060 cm$^{-1}$ is due to the phosphate group interaction. Moreover, some spectra show some features that are not assigned to the DNA backbone, but rather to the nucleobases. In fact, the peaks at 1260 cm$^{-1}$ is assigned to vibrational modes of thymine (T), adenine (A) and cytosine (C) and the one at 1320 cm$^{-1}$ to guanine (G). The feature observed at 1380 cm$^{-1}$ can be assigned to T, A and G whilst the one at 1534 cm$^{-1}$ is assigned to A. The spectra reported in Figure 5.4.b have been acquired while the electric bias was switched on for 90 s. Since the integration time was 0.5 s the total number of collected spectra during the electrophoresis was 180. Among them, 23 spectra show significant differences compared to the background signal. Moreover, switching the bias on and off as schematically described in Figure 5.5a, it is possible to study the detection events due to the passage of the DNA molecules through the nanopores under the electrophoretic pull and in the diffusion regime respectively. Therefore, after the first 90 s, the bias was switched off for another 90 s in order to study the diffusion regime. Again the total collected spectra were 180 but only 9 spectra show spectral features that can be assigned to the translocation of the DNA molecules through the plasmonic nanopores.

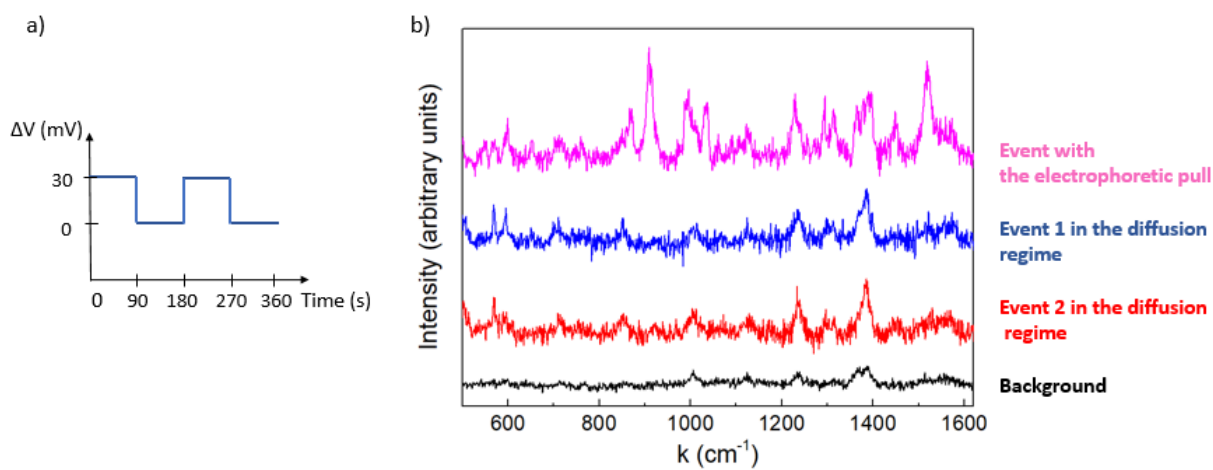

**Figure 5.5.** a) Schematic representation of the applied external voltage switched on and off each 90 s during the SERS measurements. b) Some examples of SERS spectra which represent the signal from the background (in black) and from the passage of DNA molecules through the nanopores detected in the diffusion regime (red and blue curves) and under the application of an external bias (pink curve).

Switching the bias on for another 90 s, we found 15 spectra which proved the passage of λ-DNA molecules through the nanopores and only 7 when the bias was switched off for the next 90 s. From Figure 5.5.b it is possible to note that the events corresponding to the detection of λ-DNA molecules in the diffusion regime are lower in intensity compared to the ones collected in the presence of an external bias and exhibit less spectral features. The reason could be the fact that the electrophoretic pull encourages the passage of a higher number of molecules through the nanopores within the integration time, whilst when no bias is applied only few molecules that are very close to the plasmonic nanopores pass through them. Furthermore, in the diffusion regime the molecules are not linearized but rather in a folded state and this could also be the reason why less spectral features appear in the SERS spectra.

Finally, it is worth to point out that the concentration of λ-DNA molecules injected in the cis compartment is around 30 pM that is almost three order of magnitude lower than the ones used for sensing experiments with single solid-state nanopore devices (a typical value is around 10 nM[1]). In fact, these sensors are often limited by the diffusion regime of the molecules, which affects the capture rate and in turns, also the minimum amount of molecules to inject in the cis compartment. In the plasmonic nanoassembly the presence of the nanopores array increases the capture rate enabling the detection of biomolecules even at low concentrations. Taking this into account, the free-standing nanoassembly has great potential as low-cost device for the separation and the optical detection of biomolecules.

# Conclusions and Perspectives

The recent advances of nanotechnology have paved way to the realization of new classes of functional systems on the nanoscale, which properties are unique and unexpected compared to their micro- or bulky counterparts. These nanosystems are of great interest for many applications and among others, biosensing performed on nanosized-sensors has been widely investigated due to the high sensitivity, even at the single molecule level that these nanoprobes can achieve. In particular, solid-state nanopores have received great attention since they mimic some transport properties of biological nanopores, which are fundamental in numerous biological processes, and enable the detection of biomolecules in liquid during their passage through the nanopore through the analysis of the ionic current trace collected over time. More recently, plasmonic nanostructures have been integrated with solid-state nanopores to improve their performances and couple the electronic readout with the optical readout boosted by plasmonic effects.

In this thesis two systems with plasmonic nanopores/nanoholes supporting highly confined electromagnetic fields have been explored in order to develop plasmonic nanopores with a reduced sensing area. In both cases, the strategy used for the fabrication of the nanopores is based on a scalable, low-cost approach that is more suitable for some applications that require multiple nanopores, such as high-throughput detection and parallel sensing.

The first design is based on a nanostructured HMM, made of a nanohole array in a multilayer composed by alternating metal/dielectric units. This system is promising because the electromagnetic field is highly confined inside the hole at the metallic interface of each metal/dielectric units. Firstly, a fabrication route based on colloidal lithography was implemented for the realization of the nanohole array in the HMM multilayer. Then, the optical properties of the nanostructured HMM were characterized with an ellipsometer. Although the experimental data are comparable with the simulations, it is clear that the nanostructured HMMs still require further optimization for what concerns the quality of the deposited multilayer and the geometrical parameters of the nanohole arrays, such as the diameter/pitch ratio. Therefore, in the future, further improvements on the fabrication are required to improve the optical response of the nanohole arrays in the HMM film. Moreover, the biosensing performance of this system on a glass substrate and in flow-through configuration will be investigated.

The second structure is based on a plasmonic nanomembrane formed by Au coated self-assembled nanoparticles supporting V-shaped nanopores in which the electromagnetic field is highly confined and enhanced at the tips of the nanopores. As a first result, the plasmonic nanomembrane was functionalized with a monolayer of a Raman active molecule in order to test its capabilities as SERS substrate. Then, the plasmonic nanomembrane was encapsulated in a fluidic chamber and tested for the detection of λ-DNA molecules in flow-through configuration by applying an external electric field to drive the molecules through the nanopores and performing optical detection in real time monitoring the SERS signals over time. From the analysis of the collected SERS spectra vibrational features of the λ-DNA molecules, which are its typical molecular fingerprint, were found in many of them, proving the potential of the plasmonic nanopores to be used for optical detection of biological molecules. The proposed device formed by a plasmonic nanomembrane has the advantage that its fabrication is based on a scalable and low-cost approach and that its transfer can be performed on any kind of substrate/support. Due to its plasmonic properties this nanomembrane enables the optical sensing of biomolecules passing through the nanopores. As proof of concept the device was tested for the detection of λ-DNA molecules with a concentration in the cis compartment of 30 pM, but in prospective it could be used for the identification of other biomolecules of interest, such as proteins. In the next future, this device will be used for the detection of proteins with the aim to develop a label-free sensing platform with high-throughput. In fact, one of the major limitations of the commonly used single solid-state nanopores is the low capture rate, which in turns determines that the required concentration of analytes in solution is typically higher than 10 nM.

In conclusion, for both the plasmonic nanostructures presented in this thesis a cheap and scalable fabrication approach was developed. Both of them exhibit highly localized electromagnetic fields inside a narrow sensing volume delimited by the nanoholes/pores and thus, have potential for flow-through sensing with electro-optical detection. In fact, the HMM nanohole array could be integrated with solid-state nanopore in an insulating thin membrane to enable flow-through detection. Furthermore, the metallic/dielectric multilayer composition of these nanoholes enables site-selective functionalization and in prospective, the possibility to target multiple and specific portions of the same molecule. For this reason, in combination with a single solid-state nanopore in a insulating membrane a single HMM nanohole could be suitable for molecular recognition of specific sequences and/or portions of a molecule. Instead the plasmonic nanoassembly is directly transferrable on a supporting membrane pre-patterned with micron-sized holes and thus, it does not require to be integrated with solid-state nanopores to realize flow-through sensing. Due to the large patterning area, the free-standing plasmonic nanoassembly has potential as sensing platform with large throughput for molecular sensing and filtration, which is usually not easily achieved with common single solid-state nanopore-based devices.

# Acknowledgements

Firstly I would like to thank my supervisor Dr. Francesco De Angelis who gave me the opportunity to work in his group, the "Plasmon Nanotechnologies" group at Istituto Italiano di Tecnologia (IIT) in Genoa, and meet people coming from many different countries, with different backgrounds and working in different research areas. As student this was a great opportunity not only from a scientific point of view but more in general, as personal life experience. Moreover, I would like to thank him for the time spent to discuss various projects and research topics, and all the advices he gave during these three years.

I would like to thank Dr. Darvill Price who taught me many different fabrication techniques which were available in the Cleanroom Facility of IIT. I thank you for your support, your suggestions and the ideas you shared with me during meetings. Your propensity for teaching and your passion for material science gave me energy and motivation during these years.

I thank you Dr. Nicolò Maccaferri who worked on the optical simulations and modelling of the hyperbolic metamaterials and Dr. Aliaksandr Hubarevich who worked on the optical properties of the plasmonic free-standing nanoassembly discussed in this work.

I thank you Dr. Jian-An Huang to support me with the Raman measurements and to give me many valuable advices to build fluidic cells.

Indeed, I would like to thank all the people, PhD student and Post-Docs, working in the "Plasmon Nanotechnologies" group during these years for the support and the scientific contributions you all shared with me during both formal meetings and unformal chats. You all contributed to improve and expand my knowledge in this research field. Not only, I want to thank you for the opportunity, even just during the coffee breaks, I had to listen about past experiences of your life or the places where you used to live or to work, because I feel this has expanded my prospective.

Finally, I would like to thank my family and friends for the love and support they gave me all my life.